\documentclass[fleqn,usenatbib]{mnras}

\usepackage{newtxtext,newtxmath}
\usepackage{amsmath}	

\usepackage[T1]{fontenc}

\DeclareRobustCommand{\VAN}[3]{#2}
\let\VANthebibliography\thebibliography
\def\thebibliography{\DeclareRobustCommand{\VAN}[3]{##3}\VANthebibliography}

\usepackage{graphicx}	
\usepackage{amsmath}	
\def\aap{AA}
\def\prl{Phys. Rev. Lett.}

\def\na{NewA}

\def\mnras{MNRAS}
\def\aaps{AAPS}
\def\apj{ApJ}
\def\apjs{ApJS}
\def\aj{AJ}

\def\prd{PhysRevD}
\def\jcap{JCAP}

\def\dgemail{gilman@astro.utoronto.ca}
\usepackage[titletoc]{appendix}
\usepackage{graphicx}
\usepackage{float}
\usepackage{amsmath} 
\usepackage{color}
\usepackage{wrapfig}
\usepackage{hyperref}
\usepackage{algorithm}
\usepackage{algpseudocode}
\usepackage{physics}
\def\data{{\bf{d}_{\rm{n}}}}

\def\msub{{\bf{m}_{\rm{sub}}}}

\def\qsub{{\bf{q}_{\rm{s}}}}
\def\qp{{\bf{q}_{\rm{p}}}}

\def\pkone{{0.0_{-0.4}^{+0.4}}}
\def\pktwo{{0.1_{-0.6}^{+0.7}}}
\def\pkthree{{0.2_{-0.9}^{+1.0}}}
\def\deltalphainferenceonesigma{{0.08}_{-0.34}^{+0.41}}
\def\mcconstraintonesigma{{1.3_{-0.5}^{+0.6}}}
\def\mcconstrainttwosigma{{1.3_{-1.1}^{+1.3}}}
\usepackage{natbib}

\usepackage[fleqn]{nccmath}

\title[strong lensing and the early Universe]{The primordial matter power spectrum on sub-galactic scales} 

\author[Gilman et al.]{\parbox{\textwidth}{
		Daniel Gilman$^{1}$\thanks{\dgemail}, Andrew Benson$^{2}$, Jo Bovy$^{1}$, Simon Birrer$^{3,4}$, Tommaso Treu$^{5}$, Anna Nierenberg$^{6}$
	} \\
	\\
	\\
	\parbox{\textwidth}{
		$^{1}$Department of Astronomy and Astrophysics, University of Toronto, 50 St. George Street, Toronto, ON, M5S 3H4, Canada\\
		$^{2}$Carnegie Observatories, 813 Santa Barbara Street, Pasadena, CA 91101, USA\\
		$^{3}$Kavli Institute for Particle Astrophysics and Cosmology and Department of Physics, Stanford University, Stanford, CA 94305, USA\\
		$^{4}$SLAC National Accelerator Laboratory, Menlo Park, CA, 94025\\
		$^{5}$Department of Physics and Astronomy, University of California,
		Los Angeles, CA 90095, USA\\
		$^{6}$Department of Physics, University of California Merced, 5200 North Lake Rd. Merced, CA 95343\\}
}

\pubyear{2015}

\begin{document}
\label{firstpage}
\pagerange{\pageref{firstpage}--\pageref{lastpage}}
\maketitle

\begin{abstract}
The primordial matter power spectrum quantifies fluctuations in the distribution of dark matter immediately following inflation. Over cosmic time, over-dense regions of the primordial density field grow and collapse into dark matter halos, whose abundance and density profiles retain memory of the initial conditions. By analyzing the image magnifications in eleven strongly-lensed and quadruply-imaged quasars, we infer the abundance and concentrations of low-mass halos, and cast the measurement in terms of the amplitude of the primordial matter power spectrum. We anchor the power spectrum on large scales, isolating the effect of small-scale deviations from the $\Lambda$CDM prediction. Assuming an analytic model for the power spectrum and accounting for several sources of potential systematic uncertainty, including three different models for the halo mass function, we obtain correlated inferences of $\log_{10}\left(P / P_{\Lambda \rm{CDM}}\right)$, the power spectrum amplitude relative to the predictions of the concordance cosmological model, of $0.0_{-0.4}^{+0.5}$, $0.1_{-0.6}^{+0.7}$, and $0.2_{-0.9}^{+1.0}$ at k = 10, 25 and 50 $\rm{Mpc^{-1}}$ at $68 \%$ confidence, consistent with cold dark matter and single-field slow-roll inflation. 
\end{abstract}




	\section{Introduction}
	The theory of inflation, first proposed by Alan Guth in 1981 \citep{Guth81}, simultaneously resolved outstanding challenges to the `Big Bang' paradigm while providing a predictive framework to characterize the initial conditions for structure formation. Soon after the original paper by Guth, several authors pointed out \citep{Starobinsky82,GuthPi82,Bardeen++83} that fluctuations of a scalar field driving inflation, the `inflaton', would seed inhomogeneities in the distribution of matter. The primordial matter power spectrum, $P\left(k\right)$, quantifies these inhomogeneities as a function of (inverse) length scale $k$. In addition to inflation, dark matter physics beyond the concordance model of cold dark matter plus a cosmological constant, $\Lambda$CDM, can alter the shape of the matter power spectrum on small scales \citep{Bode++01,Schneider++12,Vogelsberger++16}. The connection to the inflaton and dark matter physics makes $P\left(k\right)$ one of the most fundamental quantities in cosmology. 
	
	The largest scale constraints on $P\left(k\right)$ come from analyses of the Cosmic Microwave Background (CMB) radiation \citep{Planck2020,Planck2020a,Planck2020b}, and the clustering an internal structure of massive galaxies \citep{Fedeli++10,Reid++10,Troxel++2018}. Pushing to smaller scales, the Lyman-$\alpha$ forest uses the presence of neutral hydrogen as a proxy for the underlying dark matter density field, and delivers constraints on $P\left(k\right)$ on scales reaching $k \sim 5 \ \rm{Mpc^{-1}}$ \citep{Viel++04,Chabanier++19}, although other analyses of Lyman-$\alpha$ forest data suggest sensitivity to the linear matter power spectrum on somewhat smaller scales \citep[e.g.][]{Viel++13,Irsic++17,Rogers++21}. Recently, \citet{Sabti++21} used measurements of the ultra-violet luminosity function, which connects to $P\left(k\right)$ through the abundance of galaxies at redshifts $4-10$, to push to even smaller scales, reaching $k = 10 \ \rm{Mpc^{-1}}$. In combination the data agree with the predictions of $\Lambda$CDM, and suggest a common origin for the density perturbations in the early Universe described by power-law spectrum $P\left(k\right) \sim k^{n_s}$, with the spectral index $n_s = 0.9645 \pm 0.0044$ measured precisely from analyses of the CMB \citep{Planck2020}. The value of $n_s$ inferred from the data amounts to a success for slow-roll inflation \citep{Linde82,AlbrechtSteinhardt82,SteinhardtTurner84}, which predicts a value of $n_s$ close to unity.
	
	Constraints on the power spectrum on smaller length scales $k > 10 \ \rm{Mpc^{-1}}$, beyond the reach of existing measurements, could reveal departures from the form of $P\left(k\right)$ that the CMB, Lyman-$\alpha$ forest, and other probes have so precisely measured. Deviation from the power-law form of $P\left(k\right)$, particularly a scale-dependence, or `running', of the spectral index larger than $|n_s - 1|$, could falsify slow-roll inflation. Such a finding would have profound consequences for inflationary cosmology, and perhaps the nature of dark matter \citep{Chluba++12}. However, inferring $P\left(k\right)$ on smaller scales becomes increasingly challenging because structure formation becomes highly non-linear, complicating the mapping between $P\left(k\right)$ and observables. Physically, this non-linearity manifests in the formation of gravitationally-bound structures referred to as dark matter halos. 
	
	Strong gravitational lensing by galaxies provides a direct and elegant observational tool to determine the properties of otherwise-undetectable dark matter halos \citep{Dalal++02,Vegetti++14,Inoue++15,Hezaveh++16,Birrer++17,Gilman++20a,Hsueh++20}. Strong lensing refers to the deflection of light by the gravitational field of a massive foreground object, with the result that a single background source becomes multiply-imaged. In quadruply-imaged quasars (quads), a galaxy and its host dark matter halo, which together we refer to as the `main deflector', produce four highly magnified but unresolved images of a background quasar. The image magnifications in quads depend on second derivatives of the projected gravitational potential in the plane of the lens, $\Phi$, which depends on the projected mass density, $\kappa$, through the Poisson equation in two dimensions, $\nabla^2 \Phi \propto \kappa$. A compact dark halo that dominates the local mass density near an image can impart a large perturbation to $\nabla^2 \Phi$, and hence the image magnification, even if the halo has a mass orders of magnitude lower than the main deflector. Using existing datasets and analysis methods, lensing of compact, unresolved sources reveals the properties of halos with masses below $10^8 M_{\odot}$, with a minimum halo mass sensitivity around $10^7 M_{\odot}$ \citep{Gilman++19}. 
	
	To identify the corresponding $k$-scales relevant for an inference of $P\left(k\right)$, we can compute the Lagrangian radius $R_l$ for a halo of mass $m$ defined by $m = (4 \pi/3) \Omega_m \rho_{\rm{crit}} R_l^3$, where $\Omega_m \rho_{\rm{crit}}$ represents the contribution to the critical density of the Universe from dark matter, and evaluate the corresponding wavenumber $k = 2 \pi / R_l$. The minimum halo mass sensitivity of quad lenses depends on the size of the background source. Existing measurements of narrow-line flux ratios can reach $10^7 M_{\odot}$, while upcoming measurements of mid-IR flux ratios through JWST GO-02046 (PI Nierenberg) will reach $10^6 M_{\odot}$. Computing $R_l$ for the mass range probed by existing data $10^7 - 10^{10} M_{\odot}$, suggests that $k$ scales between $10 - 100 \ \rm{Mpc^{-1}}$ should contribute to the signal. Strong lenses perturbed by dark halos encode properties of the primordial matter power spectrum on scales two orders of magnitude smaller than those currently accessible by the CMB, pushing constraints on the primordial power spectrum to sub-galactic scales.
	
	In this work, we push constraints on $P\left(k\right)$ to scales $k > 10 \  \rm{Mpc^{-1}}$ by performing a simultaneous inference of the concentration and abundance of low-mass dark matter halos. We then recast this measurement in terms of the primordial matter power spectrum, using theoretical models in the literature to connect the power spectrum to the halo mass function and concentration-mass relation. This work builds upon earlier work in which we have analyzed the halo mass function \citep[e.g.]{Gilman++20a} and the concentration-mass relation \citep[e.g.][]{Gilman++20b} independently of one another. As we will demonstrate, the joint inference of both quantities simultaneously adds crucial information that makes a strong lensing inference of $P\left(k\right)$ possible. 
	
	This paper is organized as follows. Section \ref{sec:inference} reviews the inference methodology and dataset used in this analysis. Section \ref{sec:pkandobservables} discusses how the primordial matter power spectrum affects the populations of dark matter halos that strong lensing can detect, and presents the analytic model for the primordial matter power spectrum used in our analysis. Section \ref{sec:models} describes the models implemented in the strong lensing analysis that parameterize the halo mass function, the concentration-mass relation, the main deflector mass profile, and the lensed background source. Section \ref{sec:constrainpk} presents our inference of small-scale dark matter structure, and describes how we interpret this measurement in terms of the power spectrum. We summarize our findings and give concluding remarks in Section \ref{sec:conclusions}. 
	
	Throughout this work, we used cosmological parameters from WMAP9 \citet{Hinshaw++13}. We perform strong lensing computations using the open source software package {\tt{lenstronomy}}\footnote{https://github.com/sibirrer/lenstronomy} \citep{BirrerAmara++18,Birrer++21}, and generate populations of dark matter halos for the lensing simulations using {\tt{pyHalo}}\footnote{https://github.com/dangilman/pyHalo} \citep{Gilman++21}. {\tt{pyHalo}} makes use of {\tt{astropy}} \citep{Astropy}, and the open source software {\tt{colossus}}\footnote{https://bdiemer.bitbucket.io/colossus/cosmology\_cosmology.html}\citep{Diemer18} for computations involving the halo mass function and the concentration-mass relation. We also used {\tt{galacticus}} \citep{Benson++12} for computations of the halo mass function and concentration-mass relation with a varying primordial matter power spectrum.

	\section{Inference methodology and dataset}
	\label{sec:inference}
	We begin by reviewing the Bayesian inference framework that we use to constrain a set of hyper-parameters describing the properties of dark matter in a sample of quads. This inference pipeline was developed and tested with simulated datasets by \citet{Gilman++18} and \citet{Gilman++19}. It was then applied to real datasets to constrain models of warm dark matter \citep{Gilman++20a}, the concentration-mass relation of CDM halos \citep{Gilman++20b}, and extended to accommodate models of self-interacting dark matter \citep{Gilman++21}. In Section \ref{ssec:bayesianinf}, we discuss the inference method, and in Section \ref{ssec:data} we discuss the sample of lenses used in our analysis, most of which we have analyzed in previous work. The material unique to this paper begins in Section \ref{sec:pkandobservables}. 
	
	\subsection{Bayesian inference in substructure lensing}
	\label{ssec:bayesianinf}
	Quad lens systems comprise a quasar situated behind a massive\footnote{Typically an early-type galaxy residing in a $\sim 10^{13} M_{\odot}$ host dark matter halo.} galaxy and its host dark matter halo, which together we refer to as the main deflector. With precise alignment  between the observer, source, and the main deflector, four paths through space connect the observer with the source, with the result that an observer sees four highly-magnified images of the quasar. Observables include the four image positions, and the flux of each image. As the source brightness is unknown, the relevant quantity associated with the image brightness is a magnification ratio, or a flux ratio, taken with respect to any of the four images. 
	
	The image positions and flux ratios for a sample of lenses form a data vector $\boldsymbol{D} = \left(\boldsymbol{d}_1, \boldsymbol{d}_2, \boldsymbol{d}_3, ...\right)$, with the dataset for the $i$th lens labeled $\boldsymbol{d}_i$. Our goal is to obtain samples from the posterior probability distribution
	\begin{ceqn}
	\begin{equation}
	\label{eqn:posterior}
	p\left(\qsub| \boldsymbol{D }\right) \propto \pi \left(\qsub\right) \prod_{n=1}^{N} \mathcal{L}\left(\data | \qsub \right).
	\end{equation}
	\end{ceqn}
	The quantity $\qsub$ specifies a set of hyper-parameters that describe the dark matter structure in the lens model, $\pi\left(\qsub\right)$ is the prior on the hyper-parameters, and $\mathcal{L}\left(\data | \qsub\right)$ is the likelihood of the $n$th lens, given the model. For the purpose of this analysis, $\qsub$ specifies the slope and amplitude of the halo mass function and concentration-mass relation. 
	\begin{figure*}
		\includegraphics[clip,trim=6cm 1cm 6cm
		2cm,width=.95\textwidth,keepaspectratio]{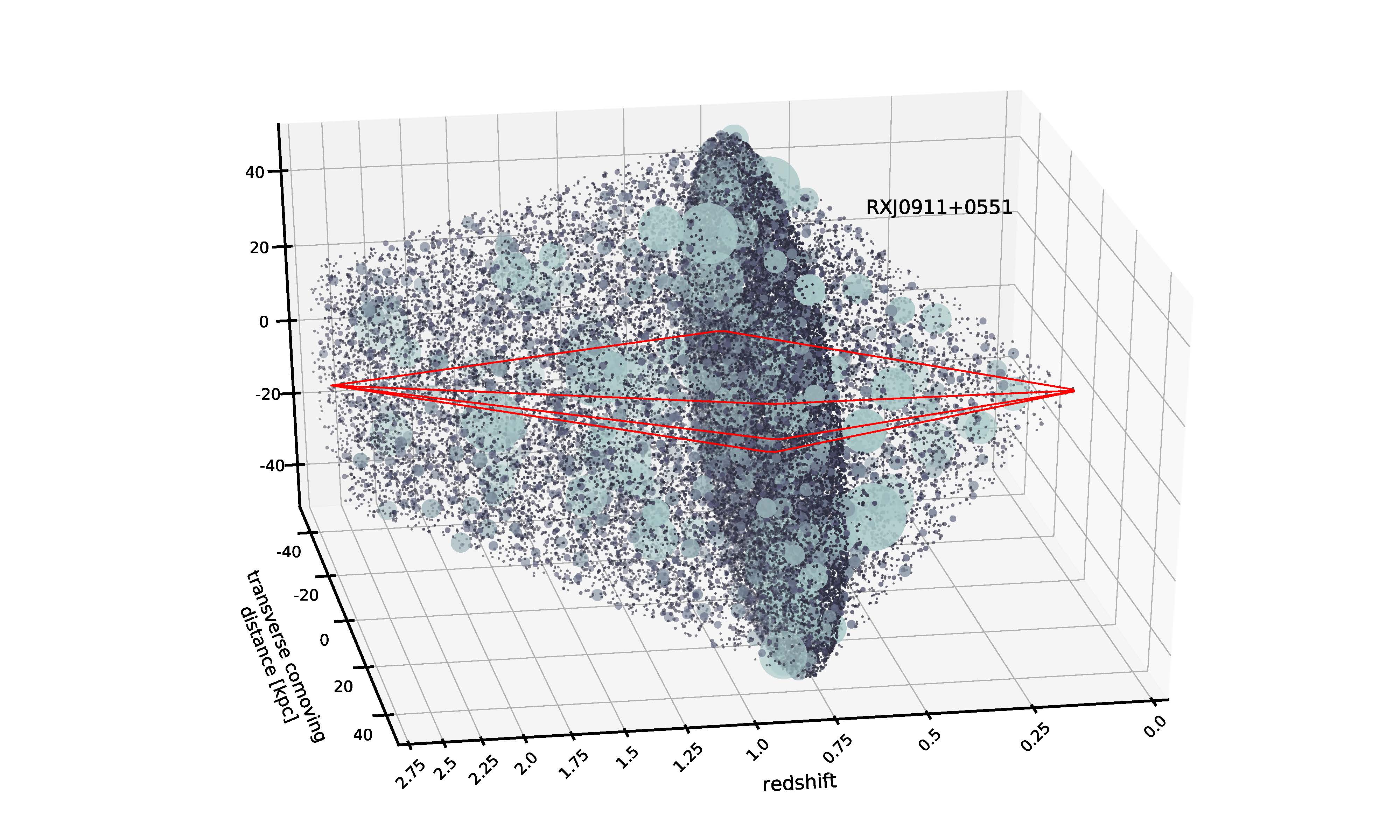}
		\caption{\label{fig:rxj0911realization} A schematic representation of the ray-tracing simulations we perform to infer the properties of dark matter structure in each lens system. The red lines show the path traversed by the light rays in the lens system RX J0911+0551. The source lies to the left, at the point of intersection of the rays. The x-axis represents the line of sight distance in redshift units between the observer and a quasar source at $z =2.76$. The y and z-axes show a physical distance scale expressed as the transverse comoving separation. Black dots distributed along the line of sight represent one possible configuration of halos in the field, while subhalos of the main deflector's host dark matter halo appear packed together at the lens redshift $z = 0.66$. The size and color of each dot varies proportionally with the mass of the halo and its concentration, respectively.}
	\end{figure*}
	The inference method accommodates any set of parameters $\qsub$, provided we can generate individual realizations\footnote{By a `realization' of halos, we refer to a set of coordinates, masses, and structural parameters such as the concentration, scale and truncation radii.} of dark matter halos, labeled $\msub$, from the model. The individual realizations connect the hyper-parameters to observables, so we can compute the likelihood by marginalizing over all possible configurations of $\msub$, and other nuisance parameters $\boldsymbol{\xi}$
	\begin{ceqn}
	\begin{equation}
	\label{eqn:likelihood}
	\mathcal{L}\left(\data | \qsub \right)  = \int p\left(\data | \msub, \boldsymbol{\xi} \right) p \left(\msub, \boldsymbol{\xi} | \qsub \right) d \msub d \boldsymbol{\xi}.
	\end{equation}
	\end{ceqn}
	Directly evaluating the integral in Equation \ref{eqn:likelihood}, however, presents an insurmountable challenge. Most configurations of $\msub$ and $\boldsymbol{\xi}$ produce a lens that looks nothing like the data, and hence only a small volume of parameter space contributes to the integral. 
	
	To make the evaluation of Equation 4 tractable, we focus computational resources on specific combinations of $\boldsymbol{\xi}$ and $\msub$ that, by construction, match the image positions of the observed dataset $\data$. We target the small volume of parameter space that contributes to the likelihood integral by solving for a set of parameters (contained in $\boldsymbol{\xi}$) describing the main lensing galaxies' mass profile that map the four observed image positions to a common source position in the presence of the full population of dark matter subhalos and line of sight halos specified by $\msub$. This task amounts to a non-linear optimization problem using the recursive form of the multi-plane lens equation \citep{BlanfordNarayan86} 
		\begin{ceqn}
	\begin{equation}
	\label{eqn:raytracing}
	\boldsymbol{\theta_K} = \boldsymbol{\theta} - \frac{1}{D_{\rm{s}}} \sum_{k=1}^{K-1} D_{\rm{ks}}{\boldsymbol{\alpha_{\rm{k}}}} \left(D_{\rm{k}} \boldsymbol{\theta_{\rm{k}}}\right).
	\end{equation}
	\end{ceqn} 
	
	In the preceding equation, $\boldsymbol{\theta_{\rm{k}}}$ is the angular coordinate of a light ray on the $k$th lens plane, $\boldsymbol{\theta}$ is a coordinate on the sky, $D_{\rm{k}}$ represents the angular diameter distance to the $k$th lens plane, $D_{\rm{ks}}$ represents the angular diameter distance from the $k$th lens plane to the source plane, and $\boldsymbol{\alpha}_k$ is the deflection field at the $k$th lens plane that includes the contributions from all dark matter halos at that redshift. The dimension of the optimization problem depends on the number of free parameters describing the main deflector mass profile. We return to this topic when describing the lens model for the main deflector in Section \ref{ssec:lensmodels}. We account for uncertainties in the measured image positions by adding astrometric perturbations to the image positions used in Equation \ref{eqn:raytracing}.
	
	With a combination of $\boldsymbol{\xi}$ and $\msub$ that produces a lens system with the same image positions as in the data, we can proceed to compute the model flux ratios at the correct image positions. At this stage, however, the task at hand still seems insurmountable because marginalizing over all possible configurations of halos specified by $\msub$ involves, for all intents and purposes, an infinite number of lensing computations. However, we can again reduce the computational expense by sampling from the likelihood function in Equation \ref{eqn:likelihood} with an Approximate Bayesian Computing approach \citep{Rubin1984}. Given the flux ratios of the observed dataset $\boldsymbol{f}_{\rm{obs}}$ and a set of model-predicted flux ratios $\boldsymbol{f}_{\rm{model}}$, we compute a summary statistic 
		\begin{ceqn}
	\begin{equation}
	\label{eqn:summary}
	S \equiv \sqrt{\sum_{i=1}^{3} \left({f}_{\rm{obs(i)}} - {f}_{\rm{model(i)}}\right)^2},
	\end{equation}
	\end{ceqn}
	and accept proposals of $\qsub$ with the condition $S < \epsilon$, where $\epsilon$ represents a tolerance threshold. As $\epsilon$ approaches zero, the ratio of the number of accepted samples between two models $\qsub_1$ and $\qsub_2$ approaches the relative likelihood of the models $\frac{\mathcal{L}\left(\data | \qsub_1 \right)}{\mathcal{L}\left(\data | \qsub_2 \right)}$, allowing us to approximate the intractable likelihood function in Equation \ref{eqn:likelihood}, up to a constant numerical factor. We can then multiply the likelihoods obtained from individual lenses to obtain the posterior distribution in Equation \ref{eqn:posterior}\footnote{Before combining the likelihoods for individual lenses, we apply Gaussian kernel density estimator (KDE) with a first-order correction to the likelihood at the edge of the prior to remove edge effects. Applying the KDE results in a smooth interpolation of the likelihood, reducing shot noise when multiplying several discretely-sampled, but high-dimensional, probability distributions.}. 
	
	Figure \ref{fig:rxj0911realization} shows a schematic example of one lens system RX J0911+0551, with a full population of subhalos and line of sight halos included in the lens model. The red lines depict the path of lensed light rays through lines of sight populated by dark matter halos. For this particular lens, the Einstein radius, lens and source redshifts, and $\Lambda$CDM-predicted (sub)halo mass function together result in approximately 10,000 halos appearing in the system. For the analysis presented in this paper, we generate populations of halos like the one shown in the figure $\mathcal{O}\left(10^6\right)$ times per lens, and retain the top 3,500 samples to compute the likelihood\footnote{The number of samples accepted in the posterior is the minimum number required to obtain a continuous approximately of the likelihood through a kernel density estimator, and is subject to convergence criteria discussed in \citet{Gilman++19} and \citet{Gilman++20a}}. 
	
	\subsection{Lens sample}
	\label{ssec:data}
	Our lens sample consists of eleven quadruply-imaged quasars. Each system meets two criteria designed to mitigate sources of systematic error in the flux ratio measurements and the lens modeling. 
	
	First, each lens in the sample has flux ratios computed from [OIII] doublet emission at 4960$\AA$ and 5007$\AA$ that emanates from the nuclear narrow line region (NLR), mid-infrared emission, or CO (11-10) emission from a more compact area surrounding the background quasar. For a typical source redshift, both the NLR and radio emission regions subtend angular scales greater than 1 milli-arcsecond on the sky, removing the contaminating effects of microlensing by stars in the main deflector but preserving sensitivity to milli-arcsecond scale deflections produced by dark matter halos. Exploiting differential magnification by finite-size sources eliminates microlensing as a source of systematic error, but in turn it requires explicit modeling, which we account for in our analysis. 
	
	The second criterion we impose demands that the main deflector show no evidence for stellar disks, as these structures require explicit treatment in the lens modeling \citep{Gilman++17,Hsueh++16,Hsueh++17}. Stellar disks appear prominently with current imaging data, allowing for the rapid identification and removal of problematic systems from the lens sample. 
	
	The lenses in our sample with narrow-line flux measurements, and the corresponding reference for the astrometry and image fluxes are RXJ 1131+0231 \citep{Sugai++07}, B1422+231 \citep{Nierenberg++14}, HE0435-1223 \citep{Nierenberg++17}, WGD J0405-3308, RX J0911+0551, PS J1606-2333, WFI 2026-4536, WFI 2033-4723, WGD 2038-4008 \citep{Nierenberg++21}. We use mid-IR flux measurements for PG 1115+080 \citep{Chiba++05}, and the flux ratios from compact CO emission for MG0414+0534 \citep{Stacey++18,Stacey++20}. The only quad in our system that we have not included in a previous work is RXJ 1131+0231. For this work, we expanded our lensing simulation pipeline to model the asymmetric narrow-line emission around the background quasar discussed by \citet{Sugai++07}. 
	
	For the lenses in our sample with fluxes measured from nuclear narrow-line emission presented by \citet{Nierenberg++14, Nierenberg++17, Nierenberg++21}, we use a forward modeling approach to measure the image fluxes and positions, while simultaneously accounting for possible variations in the point spread function (PSF). Our baseline model for the PSF was based on a combination of multiple Gaussians in the case of OSIRIS data (B1422+231), and the empirical effective PSF for lenses with WFC3 data \citep{Anderson++16}. We verified the sensitivity and accuracy of this method by measuring the properties of simulated lenses with characteristics similar to the observed lenses. We also tested for detectable extended emission in the narrow-emission and placed upper limits on the possible size of the narrow-emission region \citep{Nierenberg++14,Nierenberg++17}. The effects of additional systematic uncertainties due to e.g. differential dust absorption along the different quasar sight-lines are measured to be less than a few hundredths of a magnitude (much smaller than our flux measurement uncertainty) for early-type deflectors, given their very low dust content of early-type galaxies as well as the typical relative redshift of the sources and deflectors \citep{Fal++99, Ferrari++99}. The astrometric uncertainties of the image positions are on the order of a few milli-arcseconds \citep[e.g.][]{Nierenberg++21}.
	
	\section{Connecting the primordial matter power spectrum to dark matter structure}
	\label{sec:pkandobservables}
	A challenging aspect of inferring $P\left(k\right)$ from astronomical data -- whether from the CMB, the Lyman-$\alpha$ forest, or strong lensing -- stems from the simple fact that the power spectrum itself is not directly observable. In the case of strong lensing, the amount of perturbation to image magnifications in quads depend on the overall amount of structure, as determined by the mass function $n\left(m,z\right)$  \citep[e.g.][]{Gilman++20a}, and the central density, or lensing efficiency, of halos determined by the concentration mass relation $c\left(m,z\right)$ \citep[e.g.][]{Gilman++20b,Amorisco++21,Minor++21}, with the result that a strong lensing measurement connects to $P\left(k\right)$ through the $n\left(m,z\right)$ and $c\left(m,z\right)$ relations. In turn, both of these relations connect to the primordial matter power spectrum through the linear matter power spectrum $P_{\rm{lin}}\left(k\right) \equiv P\left(k\right) T\left(k\right)^2$, or the primordial spectrum multiplied by the square of the linear transfer function $T\left(k\right)$ presented by \citet{EisensteinHu98}. 
	
	To cast a strong lensing measurement in terms of $P\left(k\right)$, we require a specific model for $P\left(k\right)$, and a way to compute $n\left(m,z \right)$ and $c\left(m,z\right)$ from it in the mass range relevant for substructure lensing $10^7 - 10^{10} M_{\odot}$\footnote{See Section \ref{ssec:modelshmf} for a discussion on why this is the relevant mass range for substructure lensing.}. In Section \ref{ssec:theoreticalmodels}, we describe the theoretical models that connect $n\left(m,z \right)$ and $c\left(m,z\right)$ to $P\left(k\right)$. Next, in Section \ref{ssec:pkmodel}, we describe the parameterization for $P\left(k\right)$ used in this analysis, and in Section \ref{ssec:haloabundance} we illustrate how changes to the power spectrum on small scales manifest in the abundance and concentrations of halos. 
	
	\subsection{Theoretical models for halo abundance and internal structure}
	\label{ssec:theoreticalmodels}
	Before discussing in detail the technical aspects of how the primordial matter power spectrum determines the properties of dark matter halos, we can make qualitative predictions for the effects on dark matter structure that would result from an increase in power at a particular (but arbitrary) scale $\tilde{k}$, which corresponds to a halo mass scale $\tilde{m} \propto \tilde{k}^{-3}$. First, adding power at $\tilde{k}$ increases the fraction of volume elements in the Universe that collapse into halos of mass $\tilde{m}$. The mass function $n\left(m,z\right)$ therefore connects to $P\left(k\right)$ by enumerating how many peaks in the density field collapse into a halo of mass $m$ \citep{PressSchechter++74}. Second, increasing the amplitude of fluctuations at the scale $\tilde{k}$ causes fluctuations on this scale to collapse earlier, accelerating the formation of halos of mass $\tilde{m}$ relative to a scenario with no enhancement of power at $\tilde{k}$. The central density of a halo reflects the background density of the Universe when the halo formed, and because the expansion of space relentlessly dilutes the background density, halos that collapsed earlier will have higher central densities than halos that collapsed later. Thus, the concentration-mass relation $c\left(m,z\right)$, which predicts the median central density of a halo as a function of mass and redshift, connects to $P\left(k\right)$ through the timing of structure formation \citep{Navarro++97,Bullock++01,Eke++01,Wechsler++02}.
	
	\subsubsection{The halo mass function}

	Models for the mass function and the concentration-mass relation (discussed in the next section) establish a mapping between the initial conditions of the density field, quantified through the linear matter power spectrum, and the properties of collapsed halos at later times. The halo mass function has a concrete, analytic connection to $P\left(k\right)$ through the variance of the density field $\sigma\left(R_l, z\right)$
		\begin{ceqn}
	\begin{equation}
	\sigma^2\left(R_l, z\right) = \frac{D\left(z\right)^2}{2 \pi^2} \int_{-\infty}^{\infty} k^2 P_{\rm{lin}}\left(k\right) \tilde{W}^2\left(k R_l\right) dk,
	\end{equation}
	\end{ceqn}
	where $\tilde{W}\left(k R_l\right)$ is the Fourier transform of the spherical top-hat window function, $D\left(z\right)$ is the linear growth function for the perturbations, and $R_l = \left(\frac{3M}{4\pi \Omega_m \rho_{\rm{crit}}}\right)^{\frac{1}{3}}$ is the Lagrangian radius of the halo computed with the fractional contribution of matter, $\Omega_m$, to the critical density of the Universe $\rho_{\rm{crit}}$. 
	
	The number of halos per logarithmic mass interval $\log m$ per unit volume depends on $\sigma$ through
		\begin{ceqn}
	\begin{equation}
	\frac{d^2N}{d \log m \ dV} \equiv n\left(m,z\right) = f(\sigma, z) \frac{\rho_0}{m} \frac{d \log \sigma^{-1}}{d \log m}
	\end{equation}
	\end{ceqn}
	where $f\left(\sigma,z\right)$ represents the fraction of mass contained in halos for a given variance. 
	
	Over the last two decades, several mass function models have appeared in the literature with different parameterizations for $f\left(\sigma, z\right)$. We use as a baseline the model presented by \citet[][]{ST99}, hereafter referred to as Sheth-Tormen. The Sheth-Tormen model extended Press-Schechter formalism \citep{PressSchechter++74} to models of ellipsoidal collapse, with the inclusion of two free parameters, $a$ and $p$, calibrated against simulations. The Sheth-Tormen mass function has
		\begin{ceqn}
	\begin{equation}
	\nu f\left(\nu\right) = A \left(1 + \frac{1}{\nu^{\prime p}}\right) \left(\frac{\nu^{\prime}}{2}\right)^\frac{1}{2} \frac{e^{\frac{-\nu^{\prime}}{2}}}{\sqrt{\pi}},
	\end{equation}
	\end{ceqn}
	where $\nu^{\prime} \equiv a \nu\left(m, z\right)^2$, where $\nu\left(m, z\right) = \delta_c / \sigma$ is the peak height in terms of the variance $\sigma$, and $\delta_c = 1.686$ is the overdensity threshold for spherical collapse in an Einstein de-Sitter universe. \citet{Bohr++21} recently compared the predictions of Sheth-Tormen model on halo mass scales $\sim 10^7 M_{\odot}$ and found excellent agreement with their simulations, although the simulations only evolved structure until $z = 5$. 
	
	Most other mass function models presented to date either re-calibrate the Sheth-Tormen model to different cosmologies \citep[e.g.][]{Despali++16}, or empirically adjust them to match the high-mass end of the mass function to higher precision \citep[e.g.][]{Tinker++08}. Each model makes modestly different predictions for the slope and amplitude of the mass function on the mass scales of interest, which could in principle affect our results, so we perform the analysis described in the remainder of this paper using two other forms for $f\left(\sigma, z\right)$. In Appendix \ref{app:massfunctions}, we show that the effect of assuming a different model for the halo mass function has a smaller effect on our result than the statistical measurement uncertainties assuming any of the mass functions. 
	
	\begin{figure*}
		\includegraphics[clip,trim=0.2cm 4.5cm 0.25cm
		4cm,width=.95\textwidth,keepaspectratio]{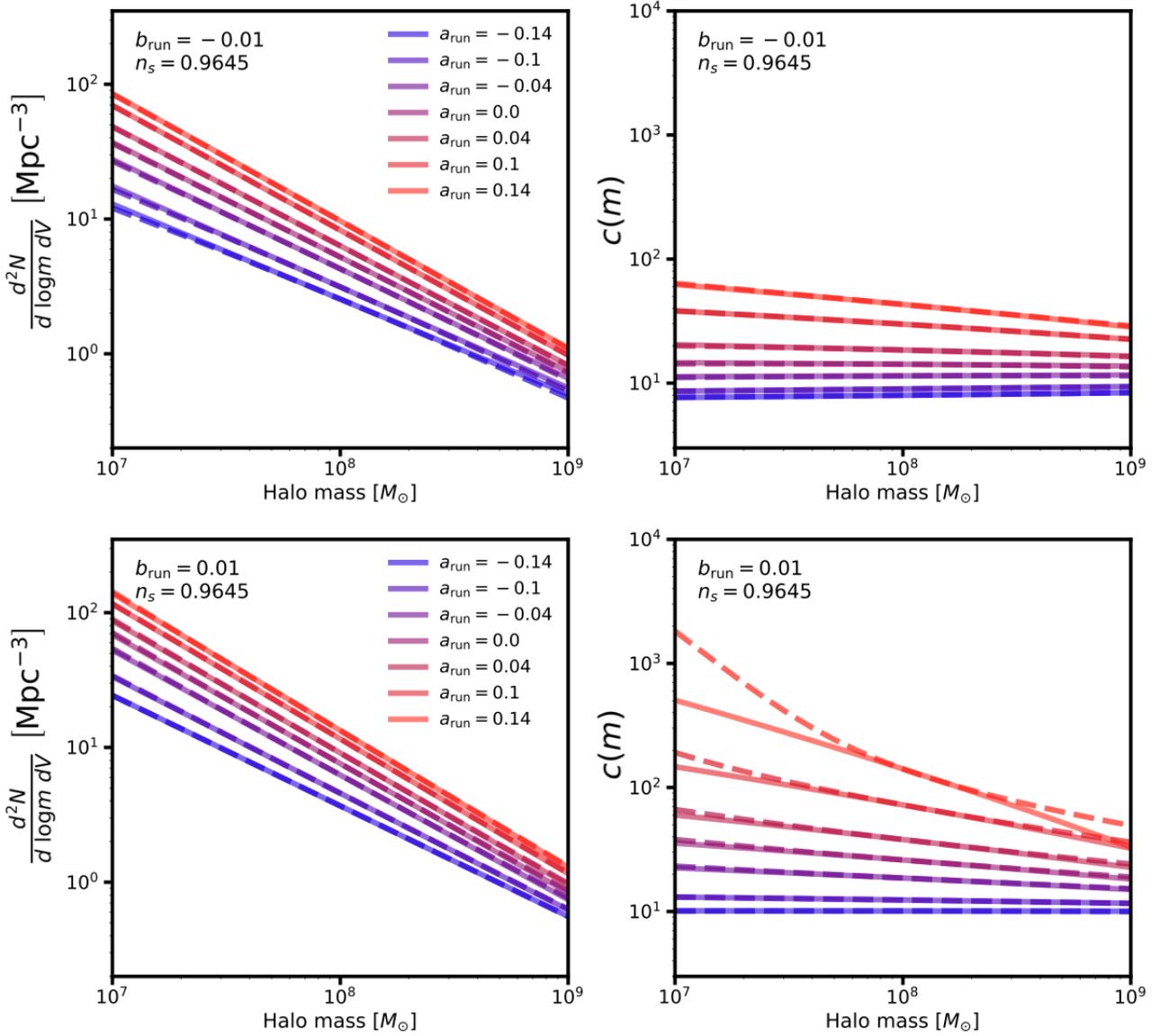}
		\caption{\label{fig:mcrel} Theoretical predictions (solid lines) and model fits (dashed lines, see Section \ref{ssec:recasting}) to the halo mass function (left) and concentration-mass relation (right) as a function of $a_{\rm{run}}$ and $b_{\rm{run}}$, the parameters that determine the scale dependence of the spectral index of the primordial matter power spectrum on scales $k > 1 \ \rm{Mpc^{-1}}$. Increasing small-scale power through positive values of $a_{\rm{run}}$ and $b_{\rm{run}}$ makes halos more concentrated and more abundant, while decreasing small-scale power makes them less abundant and less concentrated. The curves in the figures use a fixed value of the spectral index $n_s$. The effects of varying $n_s$ are qualitatively similar to the effects of adding power through $a_{\rm{run}}$ and $b_{\rm{run}}$. }
	\end{figure*}
	
	\subsubsection{The concentration-mass relation}
	To connect halo concentrations to the power spectrum, we use the concentration-mass relation model presented by \citet{DiemerJoyce19}, which provides a particularly accurate fit for the concentrations of low-mass halos relevant for a lensing analysis with quads. The model predicts halo concentration $c$ by solving
		\begin{ceqn}
	\begin{equation}
	c = F^{-1}\left[\frac{\Omega_m \left(z\right)}{\Omega_m \left(z_{\rm{pe}}\right)} \frac{1 + z_{\rm{pe}}}{1 + z} F\left(c_{\rm{pe}}\right)\right]
	\end{equation}
	\end{ceqn}
	where $F\left(x\right) = \log\left(1 + x\right) - \frac{x}{1 + x}$ for Navarro-Frenk-White (NFW) \citep{Navarro++97} profiles, $F^{-1} \left(F\left(x\right)\right) = x$, and $\Omega_m$ represents the fraction of the critical density of the Universe composed of matter.  The quantities $c_{\rm{pe}}$ and $z_{\rm{pe}}$ refer to the concentration and redshift at the time the halo concentration began changing through `psuedo-evolution', which refers to a change in halo concentration due to the evolution of the background density of the Universe, even though the physical structure of the halo remains fixed \citep{Diemer++13}. 
	
	Diemer and Joyce connect the power spectrum to halo concentrations by postulating that $c\left(m,z\right)$ depends on both the epoch of halo formation, and on the effective slope of the power spectrum evaluated near the Lagrangian radius of a halo $\partial \log P / \partial \log k \big \vert_{R = \kappa R_l}$. They assume a linear relationship between $c_{\rm{pe}} = a + b \left(\tilde{n}\left(k\right) + 3\right)$, and an effective logarithmic slope $\tilde{n}\left(k\right) \equiv -2 \frac{d \log \sigma}{d \log R} \big \vert_{R = \kappa R_l} - 3$. \citet{DiemerJoyce19} give additional details regarding the calibration of this model using simulations with a variety of power spectra, which the authors use to determine the best-fit values of $a$, $b$, and $\kappa$.

	\subsection{A model for the primordial matter spectrum $P\left(k\right)$}
	\label{ssec:pkmodel}
	In order to connect the primordial matter power spectrum $P\left(k\right)$ to observables, we must be able to integrate it to compute the variance $\sigma\left(R_l, z\right)$, and differentiate it to compute the logarithmic slope of the power spectrum $\frac{\partial \log P}{\partial \log k} \big \vert_{R = \kappa R_l}$. These practical considerations, together with the limiting constraining power of existing data, disfavor a completely free-form model for $P\left(k\right)$ where we vary the amplitude in different $k$ bins. Instead, given the precise measurements of the power spectrum on scales $k < 1 \ \rm{Mpc^{-1}}$ by a number of different methods, we anchor $P\left(k\right)$ on large scales, and allow the small-scale amplitude to vary assuming a functional form given by 
		\begin{ceqn}
	\begin{equation}
	\label{eqn:Pkequation}
	P\left(k\right) = \left\{
	\begin{array}{ll}
	P_{\rm{\Lambda CDM}}\left(k\right) & \quad k \leq k_0 \\
	P_{\rm{\Lambda CDM}}\left(k_0\right) \left(\frac{k}{k_0}\right)^{n\left(k\right)} & \quad k > k_0 \\
	\end{array}
	\right\}
	\end{equation}
	\end{ceqn}
	where we define $P_{\rm{\Lambda CDM}} \equiv P_0 \left(\frac{k}{k^{\star}}\right)^{0.9645}$ with the normalization $P_0$ and spectral index fixed to the latest analyses of CMB data \citep{Planck2020} with $k^{\star} = 0.05 \ \rm{Mpc^{-1}}$. The function $n\left(k\right)$ specifies a scale-dependent spectral index that we vary beyond the pivot scale, and expand to second order in $\log k$\footnote{$\log$ refers to a natural logarithm, unless it is explicitly written with base 10.}
		\begin{ceqn}
	\begin{equation}
	n\left(k\right) = n_s + a_{\rm{run}} \log \left(\frac{k}{k_0}\right) + b_{\rm{run}} \log \left(\frac{k}{k_0}\right)^2
	\end{equation}
	\end{ceqn}
	around a pivot scale $k_0 = 1 \ \rm{Mpc^{-1}}$. This parameterization is motivated in part by the predictions of slow roll inflation, which predicts a hierarchy of coefficients for the scale-dependent terms $\frac{b_{\rm{run}}}{a_{\rm{run}}} \sim |n_s - 1|$ \citep{LiddleLyth2000}. 
	
	The optimal choice for the pivot scale depends on the sensitivity of the dataset used to constrain the power spectrum. For the dataset relevant for this analysis, we estimate sensitivity to halos with masses down to $10^7 M_{\odot}$, corresponding to $k \sim 100 \ \rm{Mpc^{-1}}$. However, because halo abundance and structure depends on integrals and derivatives of $P\left(k\right)$, a range of $k$ scales affect the properties of halos of any given mass. Further, halos of different mass contribute in varying degree to the signal we extract from the data. We will examine the role of the pivot scale and discuss its physical interpretation in Section \ref{ssec:pivotscales}, when presenting our main results. 
	
	 In our analysis, we will treat $n_s$, $a_{\rm{run}}$, and $b_{\rm{run}}$ as free parameters, and anchor the amplitude $P_0$ to the value inferred from the CMB. As our analysis probes over an order of magnitude in $k$ scales, variation of parameters in the spectral index act on long lever arms. For this reason, the effect on the mass function and concentration-mass relation from varying $P_0$ within the constraints from large-scale measurements is completely negligible relative to the effect of varying $n_s$, $a_{\rm{run}}$, and $b_{\rm{run}}$ (see Appendix \ref{app:systematics}). We assign priors to $n_s$, $a_{\rm{run}}$, and $b_{\rm{run}}$ (summarized in Table \ref{tab:params}) that result in a large dynamic range of power spectrum amplitudes, subject to the constraint that the power spectra result in mass functions and concentration-mass relations that resemble power laws in mass and peak height, respectively, as these are the models we implemented in our inference made with strong lenses (see Section \ref{sec:models}).

	\subsection{Predicting halo abundance and concentration from $P\left(k\right)$}
	\label{ssec:haloabundance}
	Assuming the functional form for the power spectrum in Equation \ref{eqn:Pkequation} in terms of $n_s$, $a_{\rm{run}}$, and $b_{\rm{run}}$, the variance $\sigma$ and the effective logarithmic slope $\tilde{n}\left(k\right)$ are both implicitly functions of these parameters, i.e. $\sigma \equiv \sigma\left(R_l, z, n_s, a_{\rm{run}}, b_{\rm{run}}\right)$ and $\tilde{n}\left(k\right) \equiv \tilde{n}\left(k, n_s, a_{\rm{run}}, b_{\rm{run}}\right)$. Thus, we can compute the mass function and concentration-mass relation for any $n_s, a_{\rm{run}}, b_{\rm{run}}$. We perform these computations using the semi-analytic modeling software {\tt{galacticus}} \citep{Benson++12}. 
	
	Figure \ref{fig:mcrel} shows, in solid lines, the theoretically-predicted mass functions and concentration-mass relations for different combinations of $a_{\rm{run}}$ and $b_{\rm{run}}$ for the Sheth-Tormen mass function, using the model for $P\left(k\right)$ in Equation \ref{eqn:Pkequation}. The dashed lines show fits to these relations in terms of parameters we will constrain with a lensing analysis (see Section \ref{sec:constrainpk}). We find excellent agreement between the models and the theoretical predictions for the mass function and concentration-mass relation, with a slight breakdown occurring for models with $a_{\rm{run}} = 0.14$ and $b_{\rm{run}} = 0.01$. Appendix \ref{app:systematics} discusses how we quantify this source of systematic error, and propagate them through our analysis pipeline. The three mass function models we consider (see Appendix \ref{app:massfunctions} for results using models other than Sheth-Tormen) exhibit similar trends with small-scale changes to $P\left(k\right)$. 
	
	As discussed at the beginning of this section, and as Figure \ref{fig:mcrel} clearly demonstrates, increasing small-scale power leads to covariant changes in the amplitude of the halo mass function and concentration-mass relation. Since strong lensing data is sensitive to both halo abundance and concentration, we expect significant constraining power over the power spectrum parameters from a sample of quadruply-imaged quasars. To extract this signal, we perform a strong-lensing inference of the halo mass function and concentration-mass relation. We discuss the models implemented in the lensing analysis in the following section. 
	
	\section{Models implemented in the strong lensing analysis}
	\label{sec:models}
	In this section, we describe the structure formation and lens models used to infer the properties of dark matter halos with masses below $10^{10} M_{\odot}$. First, Section \ref{ssec:mfuncmcrelmodels} describes the parameterization of the halo and subhalo mass functions, and the concentration-mass relation. Section \ref{ssec:lensmodels} describes our model for the mass profile of the main lensing galaxy (the `main deflector'), and Section \ref{ssec:sourcemodel} discusses how we model the finite-size of lensed background source. Section \ref{ssec:priorchoices} comments on how we chose priors for the parameters introduced in this section that determine the average abundance and concentration of dark matter halos.
	
	\subsection{Parameterization of the halo mass function and concentration-mass relation}
	\label{ssec:mfuncmcrelmodels}
	Two distinct populations of dark matter halos perturb strongly lensed images: field halos along the line of sight, and subhalos around the host dark matter halo. In this section, we discuss the analytic expressions for the mass function and concentration-mass relation for each of these populations of halos, parameterized by hyper-parameters whose joint likelihood we will compute in the lensing analysis. Our goal will eventually be to connect the models described in this section to the theoretical predictions for the mass function and concentration-mass relation described in the previous section to cast the lensing inference in terms of $P\left(k\right)$. Notation used throughout this and the remaining sections is summarized in Table \ref{tab:notation}, and we summarize all model parameters discussed in this section in Table \ref{tab:params}. 
	
	\subsubsection{Model for the halo and subhalo mass functions}
	\label{ssec:modelshmf}
	On the low-mass end, the size of the lensed background source determines the minimum halo masses we can probe with our data. The fluxes in our sample measured from narrow-line, mid-IR, and CO 11-10 emission come from regions surrounding the background quasar spatially extended by $\mathcal{O}\left(10\right)$ parsecs \citep[e.g.][]{MullerSanchez++11}. Based on the deflection angle produced by a halo of a given mass relative to the size of the source, our data encodes measurable perturbation of halos down to roughly $10^7 M_{\odot}$. On the high-mass end, halos more massive than $\sim 10^{10} M_{\odot}$ tend to host a visible galaxy, in which case we would infer their presence and insert them into the lens model (see Section \ref{ssec:lensmodels}). In addition, the number density of halos more massive than $10^{10} M_{\odot}$ makes it unlikely that they would be found in the lens system. Thus, the population of dark matter halos massive enough to affect our data whose presence is not given away by a luminous galaxy is in the range $10^{7} - 10^{10} M_{\odot}$. To ensure we capture the full signal from low-mass objects, we render halos down to $10^6 M_{\odot}$, such that the full mass range in which we generate substructure is $10^6 - 10^{10} M_{\odot}$.
	
	We parameterize the mass function as
		\begin{ceqn}
	\begin{equation}
	\label{eqn:massfunction}
	\frac{d^2N}{d \log m dV} = \delta_{\rm{LOS}} \left(\frac{m}{m_0}\right)^{\Delta \alpha} \left[1+ \xi_{2\rm{halo}} \left(M_{\rm{host}}, z\right)\right] \frac{d^2N_{\rm{0}}}{d \log m  \ dV}
	\end{equation}
	\end{ceqn}
	where  $\frac{d^2N_{\rm{0}}}{dm dV}$ represents the Sheth-Tormen mass function model evaluated with $n_s = 0.9645$, $a_{\rm{run}} = 0$, and $b_{\rm{run}} = 0$. To account for how the mass function responds to changes in the power spectrum, we include a free normalization factor $\delta_{\rm{LOS}}$ and logarithmic slope $\Delta \alpha$. We also add a contribution from the two-halo term $\xi_{\rm{2\rm{halo}}}$\footnote{For details, see \citet{Gilman++19}.}, which accounts for the presence of correlated structure around the host dark matter halo with mass $M_{\rm{host}}$ \citep{Lazar++21}. 
	
	To model subhalos of the host dark matter halo, we sample masses from a subhalo mass function defined in projection, which we parameterize as 
		\begin{ceqn}
	\begin{equation}
	\frac{d^2N}{dm dA} = \Sigma_{\rm{sub}} \left(\frac{m}{m_0}\right)^{\alpha + q\Delta \alpha} \mathcal{F} \left(M_{\rm{halo}}, z\right),
	\end{equation}
	\end{ceqn}
	with a pivot scale $m_0 = 10^8 M_{\odot}$. The function $\mathcal{F} \left(M_{\rm{halo}}, z\right)$ accounts for the evolution of the projected mass in substructure with the host halo mass and redshift \citep{Gilman++20a}. Factoring the evolution with host halo mass and redshift out of $\Sigma_{\rm{sub}}$ allows us to combine inferences of the parameter from different lenses, in host halos with different masses at various redshifts. 
	
	The amplitude $\Sigma_{\rm{sub}}$ absorbs the effects of tidal stripping by the host halo and the central galaxy, which can impact the amplitude of the subhalo mass function by destroying subhalos, particularly if their orbits have small pericenters. By marginalizing over a flexible logarithmic slope $\alpha$, we can also account for mass-dependence in the tidal stripping, which would alter the slope of the subhalo mass function. We expect a similar amount of tidal stripping among the lenses in our sample because the host halos have similar masses of $\sim 10^{13} M_{\odot}$ with elliptical galaxies at their centers. 
	
	The subhalo mass function has a logarithmic slope $\alpha + q \Delta \alpha$, where $\alpha$ is the CDM prediction for the logarithmic slope $\alpha \sim -1.9$ \citep{Springel++08}, and $\Delta \alpha$ is the same variation in logarithmic slope we apply to the field halo mass function. As the host dark matter halo accretes its subhalo population from the field, we expect the subhalo mass function to have a similar logarithmic slope to the field halo mass function, but we allow for some flexibility in this connection by introducing the parameter $q$. For our analysis, we assume a uniform prior on $q \sim \mathcal{U}\left(0.7, 1.0\right)$, such that the logarithmic slopes of the subhalo and field halo mass functions vary nearly in tandem. 
	\begin{table*}
		\centering
		\caption{The notation for several quantities that appear frequently in the text.}
		\label{tab:notation}
		\begin{tabular}{l | c | r} 
			\hline
			notation & description\\
			\hline 
			\\
			$P\left(k\right)$ & the primordial matter power spectrum\\
			\\
			$P_{\rm{\Lambda CDM}}$ & the $\Lambda$CDM prediction for the power spectrum anchored \\
			& by measurements on large scales\\
			\\
			$\qsub$ & hyper-parameters sampled in the lensing analysis, including $\Sigma_{\rm{sub}}$, $\delta_{\rm{LOS}}$, $\Delta \alpha$, $c_8$, and $\beta$\\
			\\
			$\qp$ & parameters describing $P\left(k\right)$ beyond the pivot scale at $1 \ \rm{Mpc^{-1}}$, including $n_s$, $a_{\rm{run}}$, and $b_{\rm{run}}$\\
			\\
			$p\left(\qsub | \boldsymbol{D}\right)$ & posterior distribution of $\qsub$ given the data $\boldsymbol{D}$\\
			\\
			$p\left(\qp | \boldsymbol{D}\right)$ & posterior distribution of $\qp$ given the data $\boldsymbol{D}$\\
			\\
			$P_k$ & the power spectrum $P\left(k\right)$ evaluated at some scale $k$\\
			\\
			$\tilde{p}\left(P_k | \boldsymbol{D}\right)$ & posterior distribution of $P_k$ given the data, assuming a uniform prior on $\qp$\\
			\\
			$p\left(P_k | \boldsymbol{D}\right)$ & posterior distribution of $P_k$ given the data, assuming a log-uniform prior on $P_k$\\
			\\
		\end{tabular}
	\end{table*}

	\begin{table*}
		\centering
		\caption{The parameters sampled in this analysis that describe the halo mass function and concentration-mass relation, and the primordial matter power spectrum. }
		\label{tab:params}
		\begin{tabular}{l | c | r} 
			\hline
			parameter & description & prior\\
			\hline 
			\\
			$\delta_{\rm{LOS}} $ & amplitude of the line of sight halo mass function at $10^8 M_{\odot}$&  $\mathcal{U}$ $\left(0, 2.5\right) $\\
			& relative to Sheth-Tormen & \\
			\\
			$\beta $ & logartihmic slope of the concentration-mass relation & $\mathcal{U}$ $\left(-0.2, 15\right) $\\
			& pivoting around $10^8 M_{\odot}$ & \\
			\\
			$c_8 $ & amplitude of the concentration-mass relation at $10^8 M_{\odot}$& $\log_{10}\mathcal{U}$ $\left(0, 4\right) $\\
			\\ 
			$\Delta \alpha $ & modifies logarithmic slope of halo and subhalo mass functions & $\mathcal{U}$ $\left(-0.6, 0.9\right) $\\
			& pivoting around $10^8 M_{\odot}$ & \\
			\\
			$q $ & couples the logarithmic slopes of the& $\mathcal{U}$ $\left(0.7, 1.0\right) $\\
			& subhalo and field halo mass functions & \\
			\\
			$\Sigma_{\rm{sub}} \left[\rm{kpc}^{-2}\right]$ &subhalo mass function amplitude at $10^8 M_{\odot}$& $\mathcal{U}\left(0, 0.125\right)$ \\
			\\
			$\alpha$ & CDM prediction for the logarithmic slope of the&  $\mathcal{U}$ $\left(-1.95, -1.85\right) $\\&subhalo mass function pivoting around $10^8 M_{\odot}$& \\
			\\
			$n_s$ & spectral index of $P\left(k\right)$ at $k > 1 \ \rm{Mpc^{-1}}$ & $\mathcal{U}$ $\left(0.3645, 1.5645\right) $\\
			\\
			$a_{\rm{run}}$ & running of the spectral index at $k > 1 \ \rm{Mpc^{-1}}$ & $\mathcal{U}$ $\left(-0.2, 0.2\right) $\\
			\\
			$b_{\rm{run}}$ & running of the running of the spectral index & $\mathcal{U}$ $\left(-0.018, 0.018\right) $\\	
			\\
			$\sigma_{\rm{src}} \left[\rm{pc}\right]$ & background source size &  \\ &nuclear narrow-line emission&$\mathcal{U}\left(25, 60\right)$\\ 
			& mid-IR/CO (11-10) emission & $\mathcal{U}\left(1, 20\right)$ \\&
			\\
			$\gamma_{\rm{macro}}$ & logarithmic slope of main deflector mass profile  & $\mathcal{U}$ $\left(1.9, 2.2\right) $\\
			\\
			$\gamma_{\rm{ext}}$ & external shear across main lens plane  & (lens specific)\\
			\\
			$a_4$ & boxyness/diskyness of main& $\mathcal{N}$ $\left(0, 0.01\right) $ \\&deflector mass profile & \\
			\\
		\end{tabular}
	\end{table*}
	
	\subsubsection{Model for the concentration-mass relation}
	We model the mass-concentration relation as a power law in peak height, a parameterization that accurately reproduces concentration-mass relations predicted by simulations of structure formation \citep{Prada++12,Gilman++20b}. We assume a functional form given by
		\begin{ceqn}
	\begin{equation}
	\label{eqn:mcrelation}
	c\left(m, z\right) = c_8 \left(1 + z\right)^{\zeta} \left(\frac{\nu_0\left(m, z\right)}{\nu_0 \left(10^8, 0\right)} \right)^{-\beta}.
	\end{equation}
	\end{ceqn}
	where we have defined $\nu_0 \equiv \delta_c / \sigma\left(m, z, 0.9645, 0, 0\right)$ as the peak height evaluated for a power spectrum with spectral index $n_s = 0.9645$, $a_{\rm{run}} = 0$, and $b_{\rm{run}} = 0$ (see parameter definitions in Table \ref{tab:params}). We account for changes to halo concentrations from the power spectrum through a flexible normalization $c_8$, and logarithmic slope $\beta$. We introduce an additional term $\left(1 + z\right)^{\zeta}$ to slightly modify the redshift evolution in order to match the redshift evolution predicted by concentration-mass relations studied the literature. The results we present are marginalized over a uniform prior on $\zeta$ between -0.3 and -0.2. 
	
	We evaluate the concentration-mass relation at the halo redshift for field halos. For subhalos, we evaluate their concentrations at the infall redshift, which we predict using {\tt{galacticus}} \citep{Benson++12}. The distinction between subhalo and field halo concentrations arises because psuedo-evolution, or a changing concentration parameter due to the changing background density of the Universe, determines the redshift evolution of $c\left(m,z\right)$ for low-mass halos in the field \citep{DiemerJoyce19}. As soon as a field halo crosses the virial radius of its future host and becomes a subhalo, it follows an evolutionary track determined by the tidal field of the host halo, and the elliptical galaxy at its center \citep{GreenvandenBosch19};  the concept of `psuedo-growth' no longer applies. Finally, we note that evaluating subhalo concentrations at infall implies the mass definition of the $m_{200}$ evaluated at infall, rather than the time of lensing. We account for tidal evolution of subhalos between the time of infall and the time of lensing by truncating their density profiles.
	
	\subsubsection{Model for halo density profiles}
	We model the density profiles of both field halos and subhalos as truncated NFW profiles \citep{Baltz++09}
		\begin{ceqn}
	\begin{equation}
	\frac{\rho\left(c, x, \tau \right)}{\rho_{\rm{crit}} \left(z^{\prime}\right)} = \frac{200}{3}\frac{c^3}{\log\left(1+c\right) - \frac{c}{1+c}} 
	\left(\frac{\tau^2}{x \left(1 + x\right)^2 \left(x^2 + \tau^2\right)}\right)
	\end{equation}
	\end{ceqn}
	with $x = r / r_s$ and $\tau = r_t / r_s$, with truncation radius $r_t$. We have explicitly defined the profile in terms of the concentration $c$, and $\rho_{\rm{crit}}\left(z^{\prime}\right)$, the critical density of the Universe at redshift $z^{\prime}$, which represents either the halo redshift (for field halos), or the infall redshift (for subhalos). We tidally truncate subhalos according to their three dimensional position inside the host halo \citep[see][]{Gilman++20a}, and truncate field halos at $r_{50}$, comparable to the halo splashback radius \citep{More++15}. 
	
	\subsection{Main deflector lens model}
	\label{ssec:lensmodels}
	As we omit lens systems with stellar disks, the remaining systems have structural properties typical of the early-type galaxies that dominate the strong lensing cross section, with roughly elliptical isodensity contours and an approximately isothermal logarithmic profile slope $\gamma$ \citep{Auger++10}. These observations motivate the use of a power-law ellipsoid for the main mass profile with a logarithmic slope $\gamma$ sampled from a uniform prior ranging between $-1.9$ and $-2.2$. If the lens system has a luminous satellite detected in the imaging data, we model it as a singular isothermal sphere with astrometric uncertainties of $50$ m.a.s., and set the Einstein radius to the value determined by the discovery papers. Following common practice, we account for structure further away from the main deflector by embedding the power-law ellipsoid model in an external tidal field with shear strength $\gamma_{\rm{ext}}$. We assume a uniform prior on $\gamma_{\rm{ext}}$ that we determine on a lens-by-lens basis\footnote{We eventually constrain $\gamma_{\rm{ext}}$ itself when we select models the fit the data using the summary statistic computed from each flux ratio. To choose an appropriate prior, we intially sample a wide range of $\gamma_{\rm{ext}}$ to determine what values can fit the observed flux ratios. Based on this initial estimate, we than focus the sampling on a more restricted range of $\gamma_{\rm{ext}}$ for each system.}. 
	
	To account for deviations from purely elliptical isodensity contours in the main lens profile, we add an octopole mass moment with amplitude $a_4$ aligned with the position angle $\phi_0$ of the main deflector's mass profile. The octopole mass moment is given by
		\begin{ceqn}
	\begin{equation}
	\kappa_{\rm{oct}}\left(r\right) = \frac{a_4}{r} \cos \left(4 \left(\phi - \phi_0\right) \right).
	\end{equation}
	\end{ceqn} 
	
	For positive (negative) values of $a_4$, the inclusion of $\kappa_{\rm{oct}}$ produces disky (boxy) isodensity contours. The inclusion of this term increases the flexibility of the main deflector lens model, mitigating sources of systematic error that may arise from an overly-simplistic model for the main deflector \cite{Gilman++17,Hsueh++18}. For each system, we marginalize over a uniform prior on $a_4$ between $-0.01$ and  $0.01$. This range of $a_4$ reproduces the range of boxyness and diskyness observed in the surface brightness contours of elliptical galaxies \cite{Bender++89}. Because the boxyness or diskyness of the light profile should exceed the boxyness or diskyness of the projected mass profile after accounting for the contribution of the host dark matter halo, this prior introduces the maximum reasonable amount of uncertainty in the main deflector's boxyness or diskyness.  

	 When solving for a set of macromodel parameters that map the observed image coordinates to the source position with Equation \ref{eqn:raytracing}, we allow the Einstein radius, mass centroid, axis ratio, axis ratio position angle, the position angle of the external shear, and the source position, to vary freely. We sample the logarithmic profile slope, the strength of the external shear, and the amplitude of the octopole mass moment, from uniform priors (see Table \ref{tab:params}), and keep their values fixed during the lensing computations performed for each realization of dark matter halos. 
	
	\subsection{Background source model}
	\label{ssec:sourcemodel}
	The effect of a halo with a fixed mass profile on the magnification of a lensed image depends on the size of the background source \citep{Dobler++06,Inoue++05}. We account for finite-source effects in the data by computing flux ratios with a finite-source size in the forward model. For systems with flux ratio measurements from the nuclear narrow-line region, we marginalize over a uniform prior on the source size between $25-60 \rm{pc}$ \citep{MullerSanchez++11}, and for the radio emission we marginalize over a uniform prior between $1-20 \rm{pc}$ \citep{Stacey++20}. We model the source as a Gaussian, and when quoting its size refer to the full width at half maximum. We compute the image magnifications by ray-tracing through the lens model, and integrate the total flux from the source that appears in the image plane. 
	
	The only exception to the source modeling described in the previous paragraph applies to the lens system RXJ1131+1231, which has NLR flux ratio measurements presented by \citet{Sugai++07}. \citet{Sugai++07} show that the NLR surrounding the quasar in this lens system appears to have an asymmetric structure that extends to the north. We model this additional source component by adding a second Gaussian light profile to the north of the main NLR, with a spatial offset between it the main NLR that we allow to vary freely between 0-80 parsecs. In addition, we rescale both the size and surface brightness of the second source component relative to the size and surface brightness main NLR by random factors sampled between $0.3 - 1$. We marginalize over both the size of the main NLR, and the spatial offset, size, and brightness of the second source component in our analysis. 
	
	\subsection{Choice of priors}
	\label{ssec:priorchoices}
	We use uninformative (uniform) priors on the parameters sampled in the lensing analysis, imposing a minimal set of assumptions in the inference. The range of priors summarized in Table \ref{tab:params} is determined by the requirement that we be able to compute the likelihood of a particular power spectrum, described by parameters $n_s, a_{\rm{run}}, b_{\rm{run}}$, given the lensing data. For example, because the model we consider for the power spectrum predicts a halo concentration at $10^8 M_{\odot}$ between 1 and 10,000, we set a lower limit on the prior $\log_{10} \left(c_8\right) = 0.0$ and an upper limit of $\log_{10} c_8 = 4.0$. As discussed in the following section, our final results ultimately depend only on the relative likelihood between two different sets of $\qsub$ parameters, and therefore do not depend on the choice of prior. As we use uniform priors on each parameter, the posterior distribution of $\qsub$ parameters given the data is directly proportional to the required likelihood. 
	
	\section{Constraints on $P\left(k\right)$}
	\label{sec:constrainpk}
	This section presents our main result: an inference of the primordial matter power spectrum on small scales obtained from a measurement of halo abundance and internal structure using eleven quadruply-imaged quasars. Table \ref{tab:notation} summarizes the notation that appears frequently throughout this section. We divide the presentation of our main results into three subsections.
	
	First, Section \ref{ssec:lensingmeasurement} describes how we use the models for the halo mass function and concentration-mass relation described in the previous section to compute the probability of $\qsub = \left(\delta_{\rm{LOS}}, \beta, \log_{10} c_8, \Delta \alpha, \Sigma_{\rm{sub}}\right)$ given the data, or $p\left(\qsub | \boldsymbol{D}\right)$. We then briefly comment on the results, which have standalone value as a detailed inference of the properties of low-mass dark matter halos and subhalos. Second, Section \ref{ssec:recasting} describes how we translate the constraints on the $\qsub$ parameters sampled in the lensing analysis to constraints on the parameters $\qp \equiv \left(n_s, a_{\rm{run}}, b_{\rm{run}}\right)$ that describe the power spectrum. We then present an inference of $p\left(\qp| \boldsymbol{D}\right)$, or the joint constraints on the $\qp$ parameters given the data. Section \ref{ssec:inferencepk} describes how we use the probability distribution $p\left(\qp| \boldsymbol{D}\right)$ to infer the amplitude of $P\left(k\right)$ at different $k$ scales, assuming the model for the power spectrum in Equation \ref{eqn:Pkequation}. Finally, Section \ref{ssec:pivotscales} examines how our results depend on the choice of the pivot scale, and what scales are probed by our data.
	
	To check the robustness of the results presented in this section, we validate the methodology we use to infer $P\left(k\right)$ with simulated datasets. We present the results of these tests in Appendix \ref{app:simdatatest}, and discuss the adequacy of the models we use to interpret the data in Appendix \ref{app:goodnessoffit}.
	
	\subsection{Joint inference of the halo mass function and concentration-mass relation }
	\begin{figure*}
		\includegraphics[clip,trim=0cm 0cm 1cm
		1cm,width=.95\textwidth,keepaspectratio,angle=0]{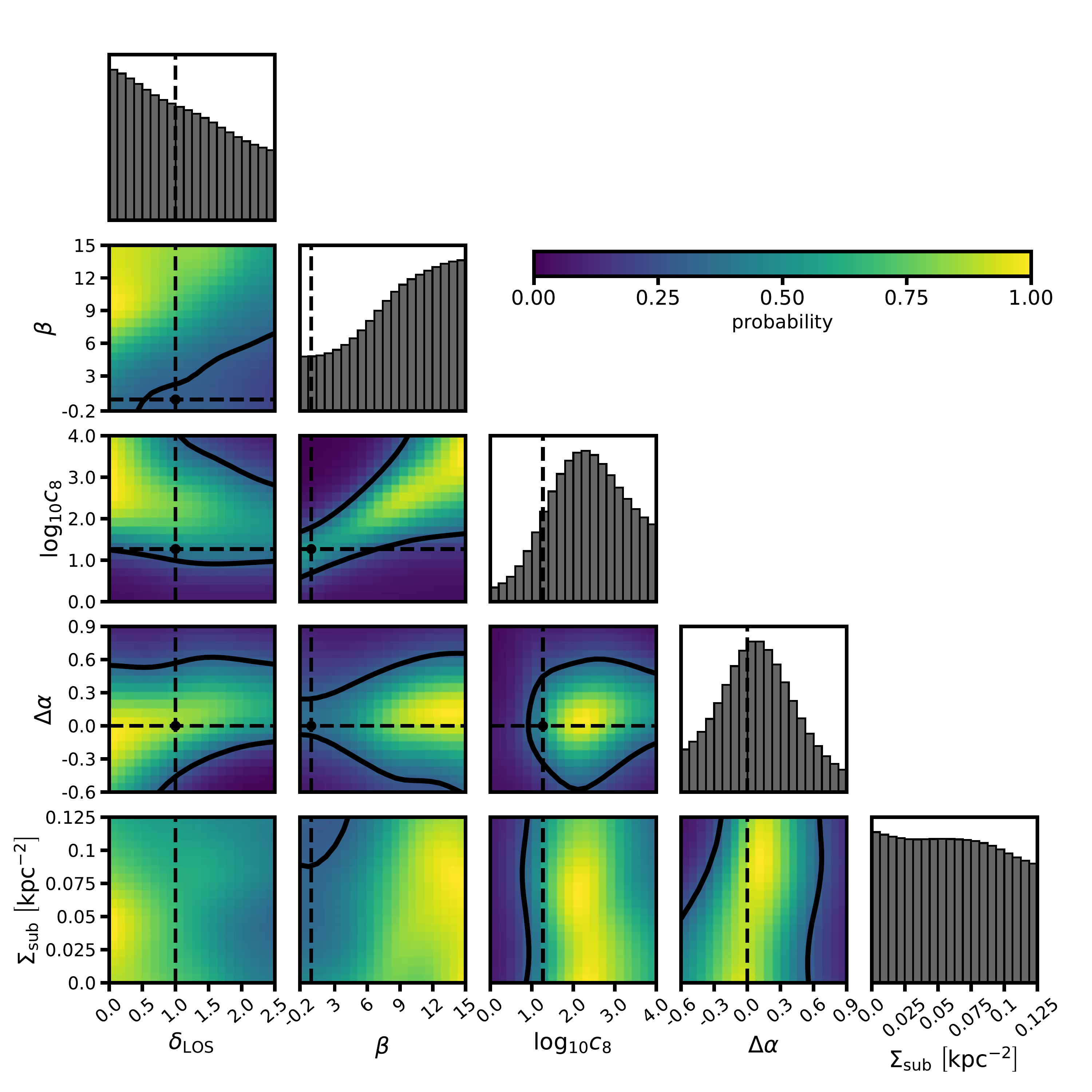}
		\caption{\label{fig:lensinginferencefull} The joint probability distribution $p\left(\qsub | \boldsymbol{D} \right)$ obtained by applying the inference method described in Section \ref{sec:inference} to a sample of eleven quads. The parameters in $\qsub$, which we summarize in Table \ref{tab:params}, include $\Sigma_{\rm{sub}}$, the amplitude of the subhalo mass function, and $\delta_{\rm{LOS}}$, the amplitude of the field halo mass function relative to the prediction of the Sheth-Tormen model in $\Lambda$CDM. We vary the logarithmic slopes of the mass functions through the parameter $\Delta \alpha$ around the $\Lambda$CDM prediction $\Delta \alpha = 0$. The parameters $\beta$ and $c_8$ set the logarithmic slope and amplitude of the concentration-mass relation, respectively. Black contours in the joint distributions show the $68 \%$ confidence region, and a black point identifies the $\Lambda$CDM prediction for each parameter, with the exception of $\Sigma_{\rm{sub}}$ (see main text).}
	\end{figure*}
	\begin{figure*}
		\includegraphics[clip,trim=0cm 0cm 1cm
		1cm,width=.95\textwidth,keepaspectratio,angle=0]{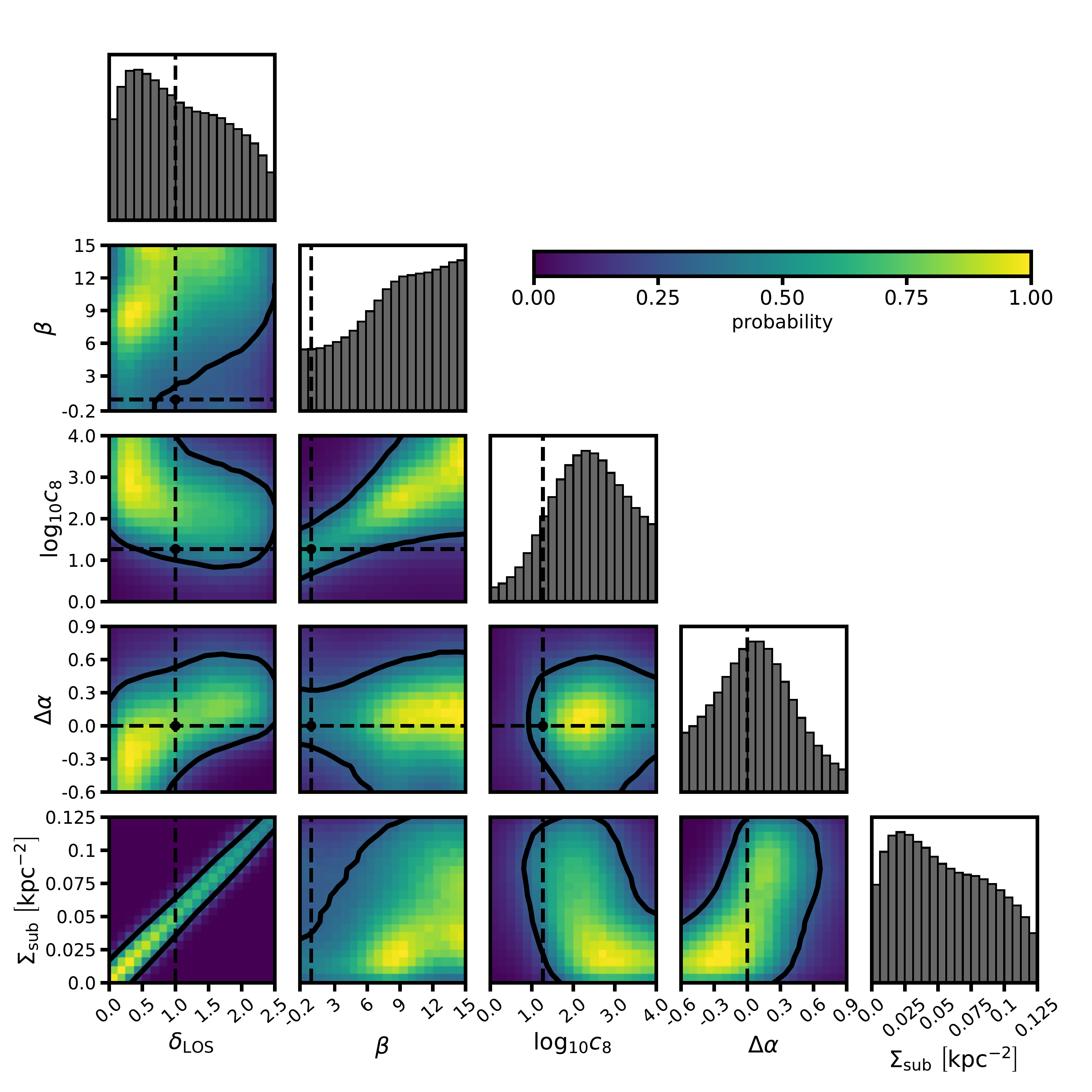}
		\caption{\label{fig:lensinginferenceprior} The joint probability distribution $p\left(\Sigma_{\rm{sub}}, \delta_{\rm{LOS}}\right) \times p\left(\qsub | \boldsymbol{D} \right)$, where the first term is an informative prior on the combination of the normalization of the subhalo mass function and the field halo mass function. The prior couples the amplitudes of the field halo and subhalo mass functions such that they vary proportionally, reflecting the assumption that subhalos are accreted from the field. We assume a subhalo mass function amplitude $\Sigma_{\rm{sub(predicted)}} = 0.05 \  \rm{kpc^{-2}}$, which corresponds to twice as efficient tidal disruption of halos in the Milky Way relative to massive elliptical galaxies. We repeat the inference assuming $\Sigma_{\rm{sub(predicted)}} = 0.025 \ \rm{kpc^{-2}}$ in Appendix \ref{app:sigmasubprior}. Black contours in the joint distributions show the $68 \%$ confidence region, and a black point identifies the $\Lambda$CDM prediction for each parameter.}
	\end{figure*}
	\begin{figure}
		\includegraphics[clip,trim=0cm 0cm 0cm
		0cm,width=.45\textwidth,keepaspectratio,angle=0]{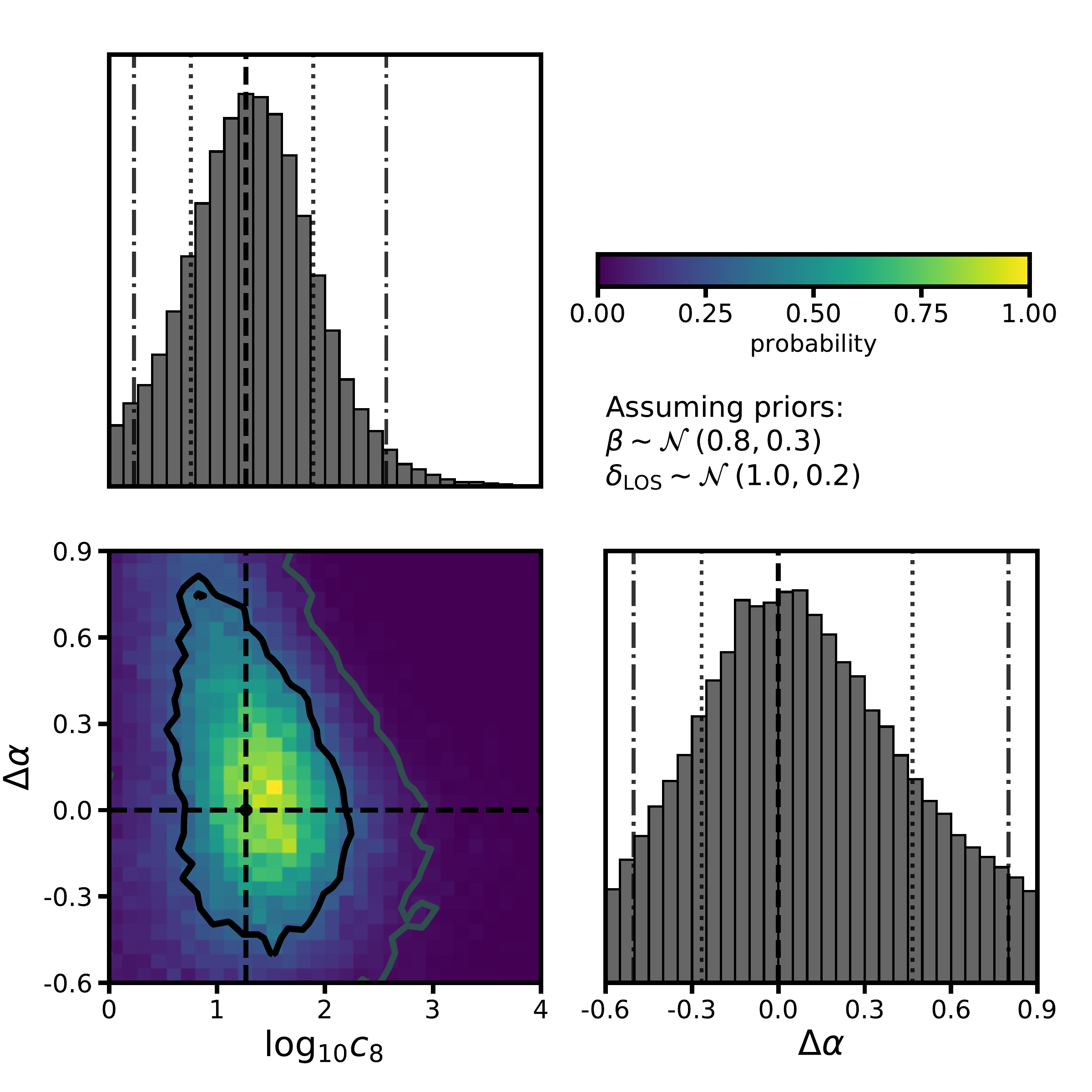}
		{\caption{\label{fig:lensinginferenceCDM}} Joint inference of the amplitude of the concentration-mass relation at $10^8 M_{\odot}$, $c_8$, and deviations of the logarithmic slope of the halo mass function $\Delta \alpha$, with the $\Lambda$CDM predictions highlighted with the black cross hairs and vertical dashed lines. Other vertical lines and contours denote $68\%$ and $95\%$ confidence intervals. We obtain this inference by multiplying the joint distribution shown in Figure \ref{fig:lensinginferencefull} by a prior that enforces the $\Lambda$CDM prediction for the amplitude of the halo mass function $\delta_{\rm{LOS}} = 1 \pm 0.2$ and $\beta = 0.8 \pm 0.3$, while marginalizing over $\Sigma_{\rm{sub}}$.}
	\end{figure}

	\begin{figure*}
	\includegraphics[clip,trim=0cm 0cm 0cm
	0cm,width=.45\textwidth,keepaspectratio,angle=0]{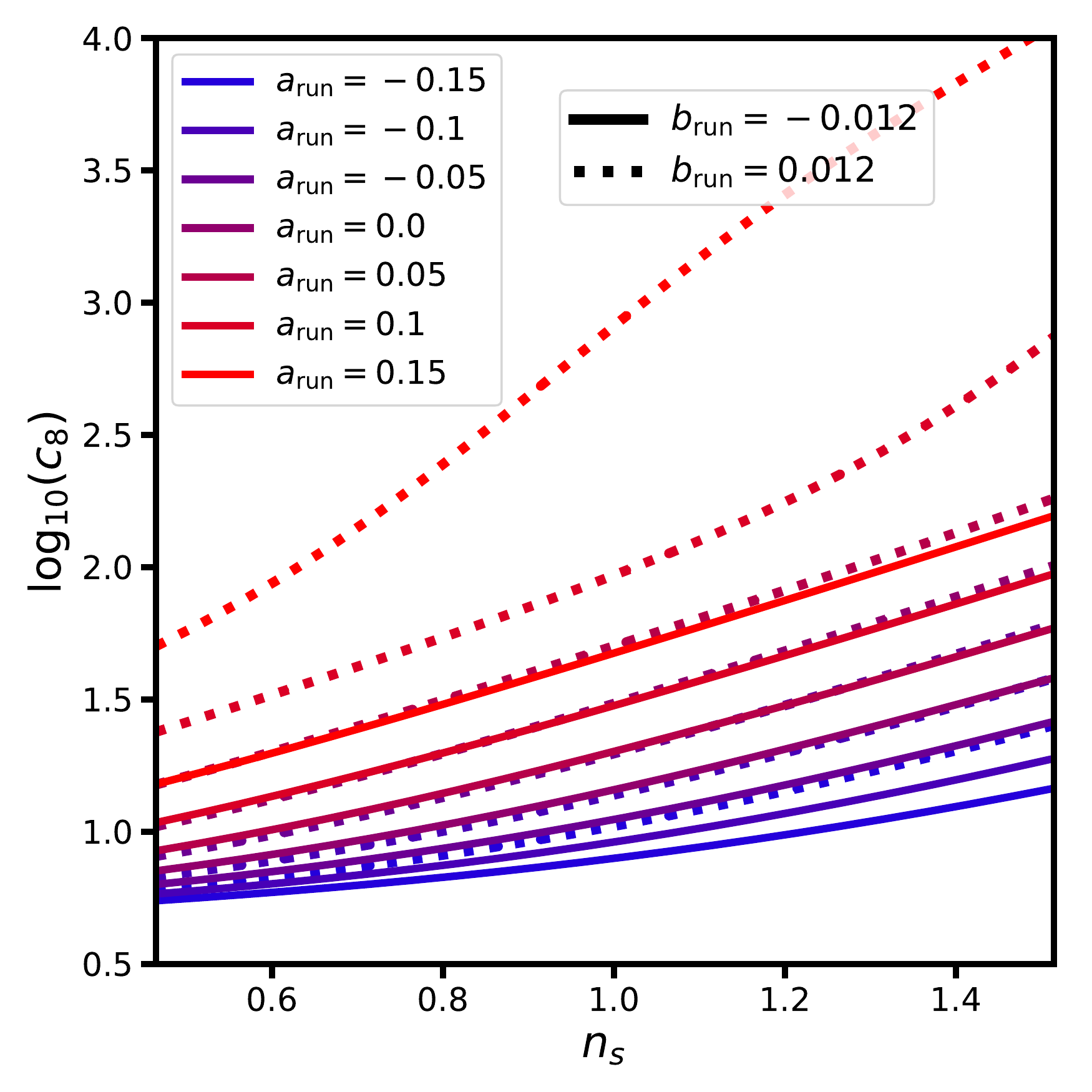}
	\includegraphics[clip,trim=0cm 0cm 0cm
	0cm,width=.45\textwidth,keepaspectratio,angle=0]{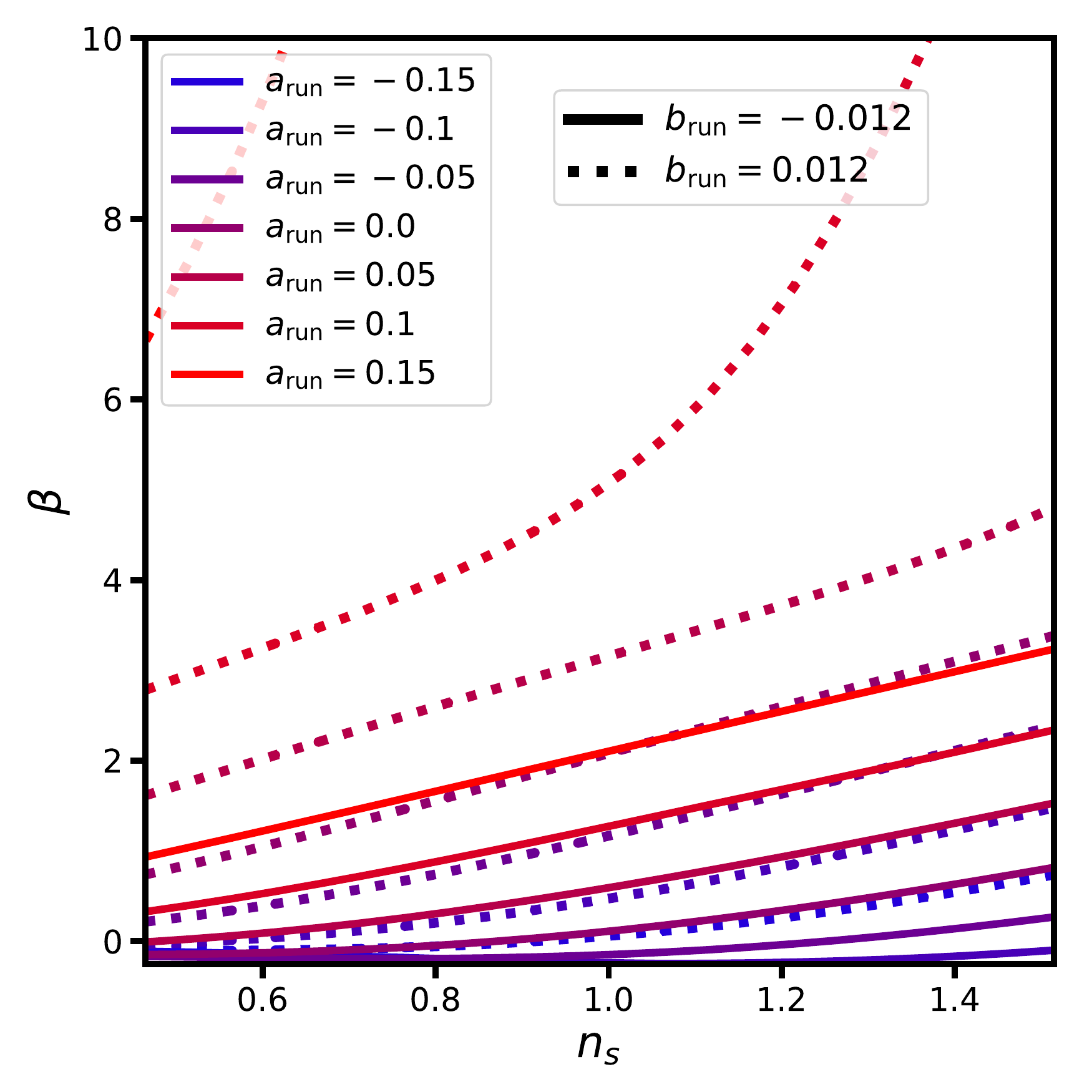}
	\includegraphics[clip,trim=0cm 0cm 0cm
	0cm,width=.45\textwidth,keepaspectratio,angle=0]{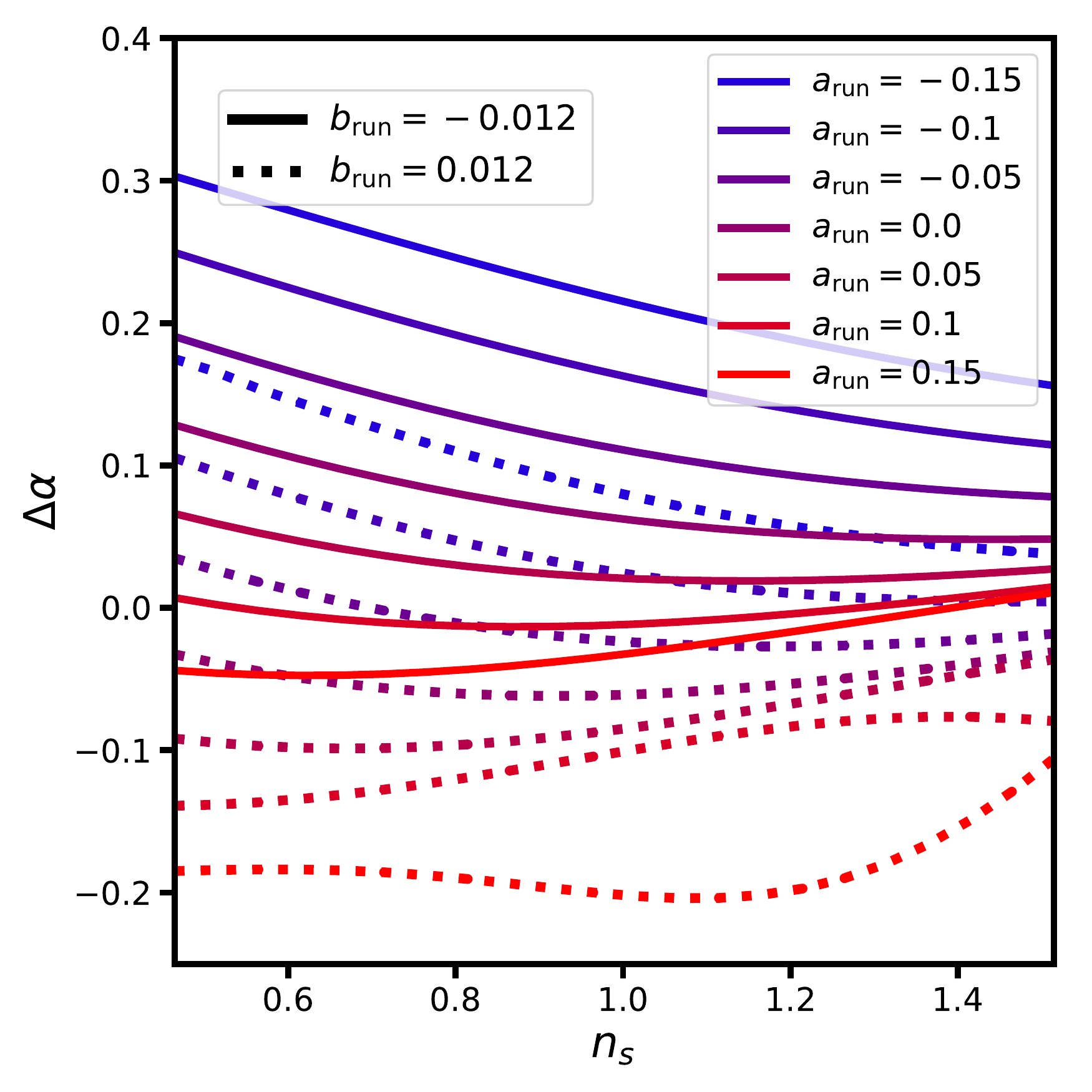}
	\includegraphics[clip,trim=0cm 0cm 0cm
	0cm,width=.45\textwidth,keepaspectratio,angle=0]{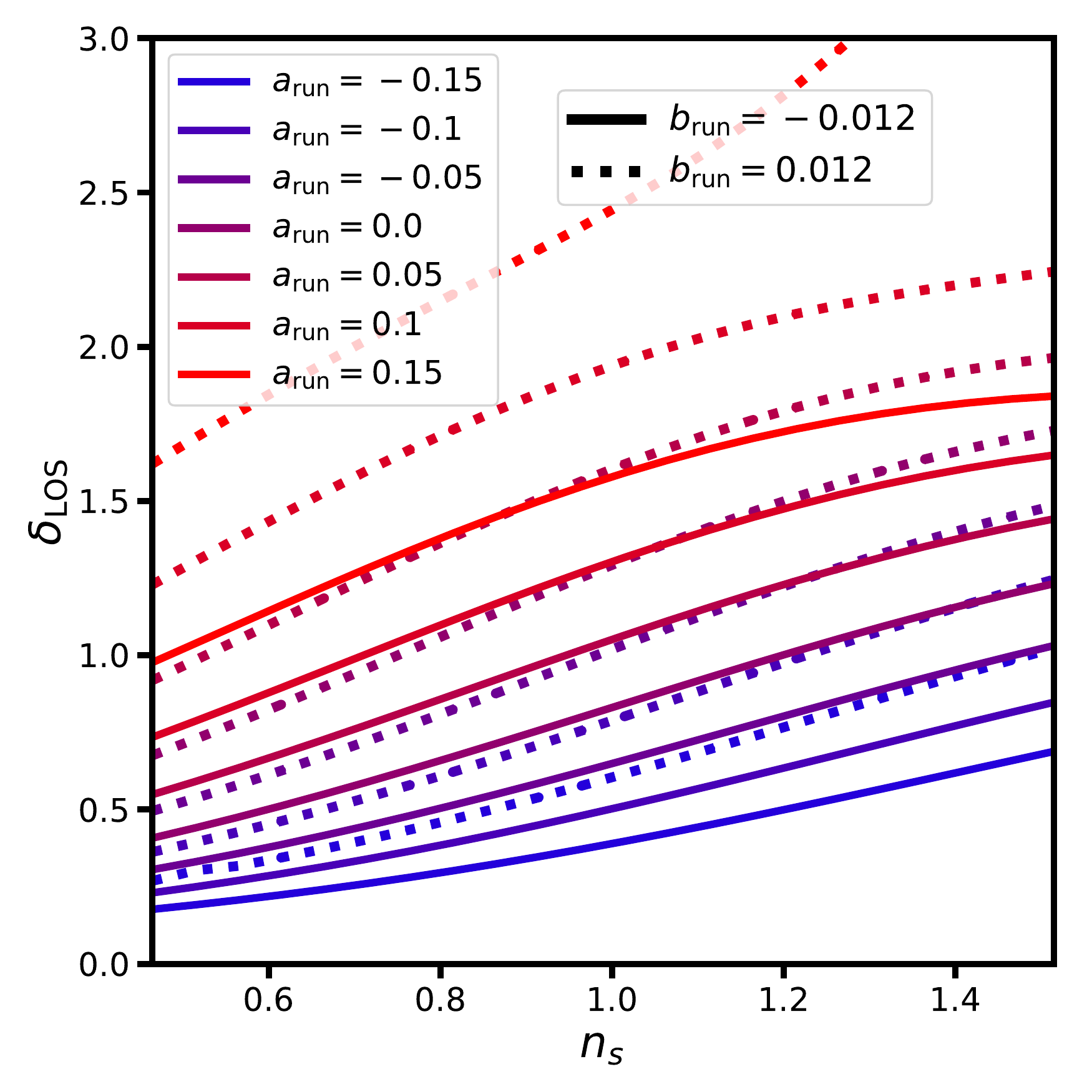}
	\caption{\label{fig:mappings} The correspondence between the hyper-parameters sampled in the lensing analysis $\qsub$, and the parameters in $\qp$, $n_s$, $a_{\rm{run}}$, and $b_{\rm{run}}$, that determine the form of $P\left(k\right)$. The dashed and solid lines show the mapping $\qp \rightarrow \qsub$ for values of $b_{\rm{run}} = -0.012$ and $b_{\rm{run}} = 0.012$, respectively. The color of each curve corresponds to the value of $a_{\rm{run}}$, and the x-axis shows the value of $n_s$. The y-axis indicates the value of parameters in $\qsub$ that corresponds to each form of the power spectrum. }
\end{figure*}

\begin{figure*}
	\includegraphics[clip,trim=0cm 0cm 0cm
	0cm,width=.33\textwidth,keepaspectratio,angle=0]{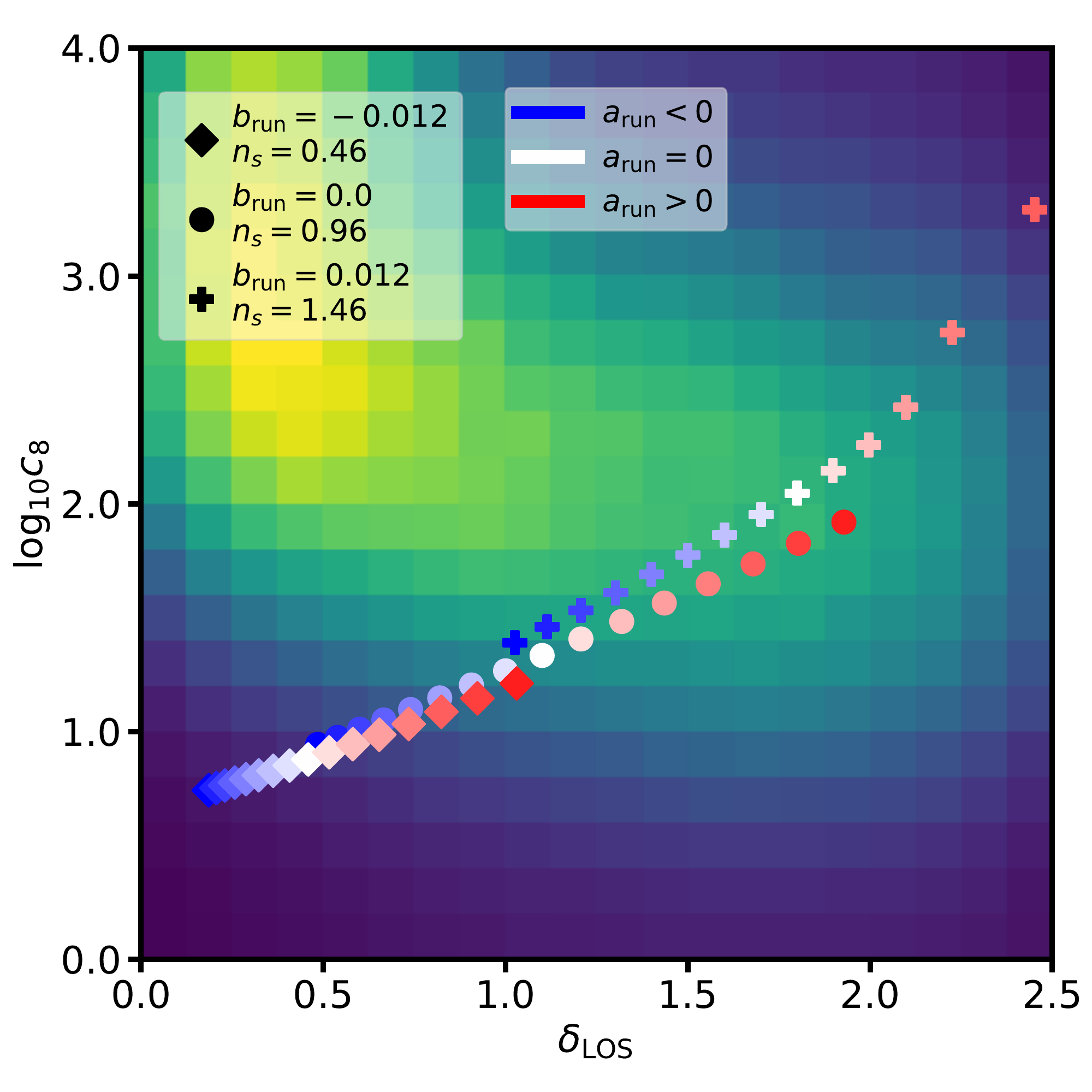}
	\includegraphics[clip,trim=0cm 0cm 0cm
	0cm,width=.33\textwidth,keepaspectratio,angle=0]{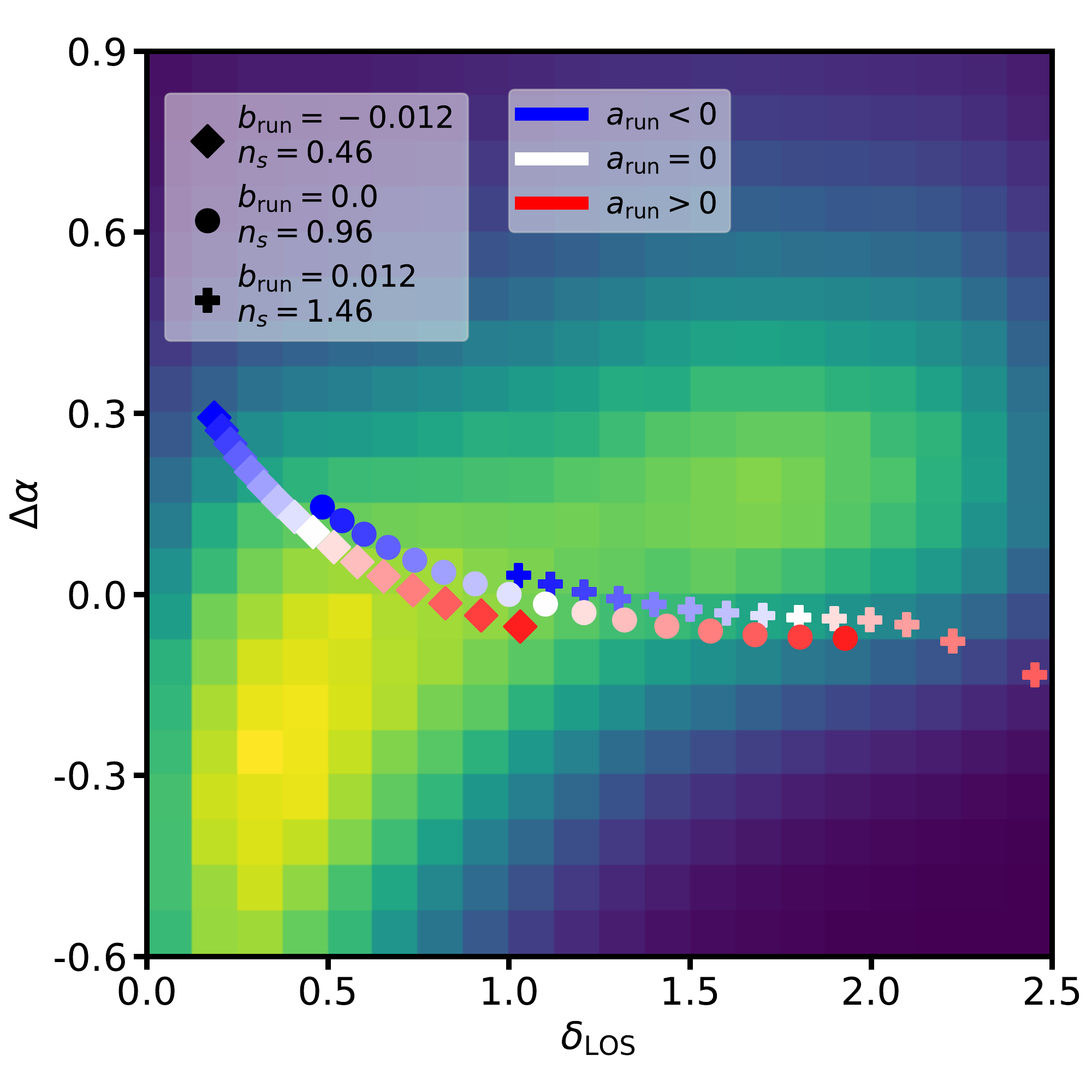}
	\includegraphics[clip,trim=0cm 0cm 0cm
	0cm,width=.33\textwidth,keepaspectratio,angle=0]{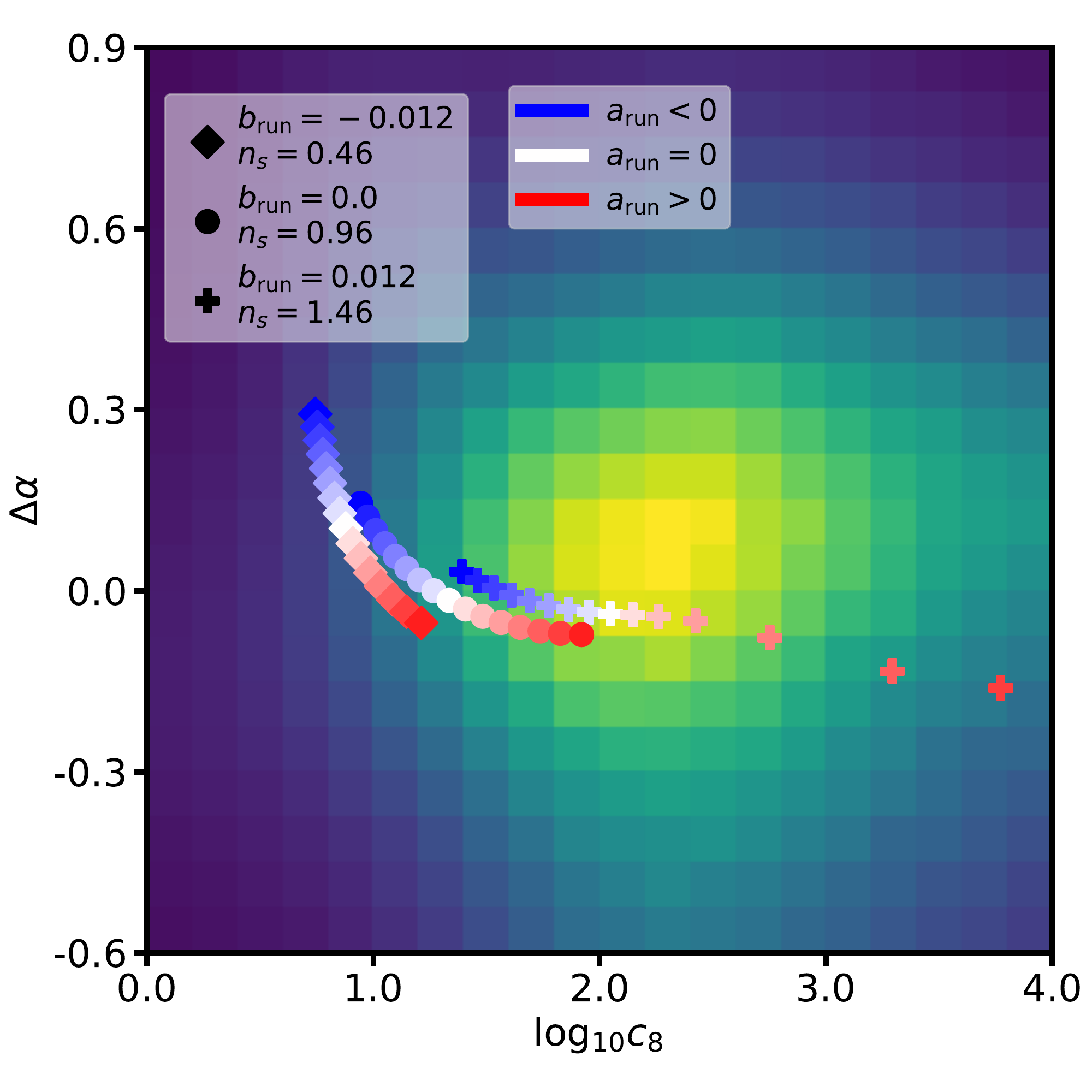}
	\caption{\label{fig:jointpdfmapping} This figure shows points corresponding to different power spectra overlaid on the joint likelihood of several parameters sampled in the lensing analysis. The color of the points corresponds to the amplitude of $a_{\rm{run}}$ sampled between $-0.1$ and $0.1$, while the marker style indicates values of $b_{\rm{run}}$ and $n_s$. First, we note that the likelihood contours inferred from the lenses, particularly in the left and center panels, run orthogonal to the relationship between model parameters that results from adding or reducing small-scale power. For this reason, we can constrain the form of the power spectrum without a statistically significant marginal constrain on any single parameter analyzed in the lensing analysis. Second, we note that models with reduced small-scale power introduced through $b_{\rm{run}} = -0.012$ and $n_s = 0.46$, but enchanced small scale power through $a_{\rm{run}}$, map to similar regions of parameter space as models with enchanced small scale power through $b_{\rm{run}} = 0.012$ and $n_s = 1.46$, but negative values of $a_{\rm{run}}$. Thus, we expect the joint probability distribution $p\left(\qp | \boldsymbol{D}\right)$ to exhibit an anti-correlation between each pair of $\qp$ parameters.}
\end{figure*}
	\label{ssec:lensingmeasurement}
	Applying the inference framework described in Section \ref{sec:inference} to the structure formation model described in Section \ref{sec:models}, we compute the joint probability distribution of the hyper-parameters $\qsub$ that describe the subhalo and halo mass functions, and the concentration mass relation. Figure \ref{fig:lensinginferencefull} shows the posterior distribution $p\left(\qsub | \boldsymbol{D}\right)$. The $\Lambda$CDM prediction for each parameter is marked in black. The amplitude of the subhalo mass function $\Sigma_{\rm{sub}}$ depends, among other factors, on the tidal stripping efficiency of the massive elliptical galaxies that act as strong lenses. For this reason, we do not include a $\Lambda$CDM prediction for $\Sigma_{\rm{sub}}$ in the figure. We note that the binning and the kernel density estimator applied to the individual lens likelihoods slightly alters the probability density, particularly at the edge of the prior, but not by an amount that significantly affects our results\footnote{In particular, the bin closest to $\Sigma_{\rm{sub}} = 0$ and $\delta_{\rm{LOS}} = 0$ does not actually contain zero halos, as it includes models with low amplitudes for both $\Sigma_{\rm{sub}}$ and $\delta_{\rm{LOS}}$. These models require extremely high concentration parameters $c_8$ to explain the data.}.
	
	While we cannot constrain $\Sigma_{\rm{sub}}$ independently from the other parameters in $\qsub$, we can introduce a theoretically motivated prior that couples the amplitude of the subhalo mass function, $\Sigma_{\rm{sub}}$, to the amplitude of the line of sight halo mass function, $\delta_{\rm{LOS}}$, at the pivot scale of $10^8 M_{\odot}$. As subhalos are accreted from the field, we expect that a universe with more (fewer) $10^8 M_{\odot}$ halos in the field should have more (fewer) subhalos with an infall mass of $10^8 M_{\odot}$. Given an expected subhalo mass function amplitude\footnote{We factor the redshift evolution and host halo mass dependence out of the normalization $\Sigma_{\rm{sub}}$, so we expect a common value for all of the lenses deflectors in our sample.} $\Sigma_{\rm{sub(predicted)}}$ in $\Lambda$CDM, we define the relative excess or deficit of subhalos $\delta \Sigma \equiv \Sigma_{\rm{sub}} / \Sigma_{\rm{sub(predicted)}}$, and demand that this relative excess or deficit vary proportionally with the relative excess or deficit of line-of-sight halos around the Sheth-Tormen prediction $\delta_{\rm{LOS}}=1$. We enforce this prior by adding importance sampling weights $w \equiv p\left(\Sigma_{\rm{sub}}, \delta_{\rm{LOS}}\right)$ given by
		\begin{ceqn}
	\begin{equation}
	w = \exp \left(\frac{-\left(\delta \Sigma - \delta_{\rm{LOS}}\right)^2}{2{\delta_{\Sigma}}^2}\right).
	\end{equation}
	\end{ceqn}  
	To be clear, we do not anchor the amplitudes of $\Sigma_{\rm{sub}}$ and $\delta_{\rm{LOS}}$ on small scales, which would trivially bring the inferred $P\left(k\right)$ into agreement with theoretical expectations. The total number of subhalos and line of sight halos can still vary freely within the bounds of the chosen priors (see Table \ref{tab:params}). 
	
	Assuming that massive elliptical galaxies and the Milky Way tidally disrupt subhalos with equal efficiency, we would expect an amplitude $\Sigma_{\rm{sub(predicted)}} \sim 0.025 \  \rm{kpc^{-2}}$ for a universe with $n_s = 0.9645$, $a_{\rm{run}} = 0$, and $b_{\rm{run}} = 0$, as shown by \citet{Nadler++21}. We expect, however, that the Milky Way's disk destroys subhalos more efficiently than an elliptical galaxy. Observational support for this assumption comes from \citet{Nadler++21}, who find that a differential tidal disruption efficiency of $\sim 2$ appears more consistent with the number of satellite galaxies in the Milky Way. 
	
	Taking these considerations into account, in the following sections we assume a value of $\Sigma_{\rm{sub(predicted)}} = 0.05 \ \rm{kpc^{-2}}$, and couple the normalizations with $\delta_{\Sigma} = 0.2$, or a $20 \%$ intrinsic scatter between the amplitudes of the field halo and subhalo mass functions that could arise from halo to halo variance \citep{Jiang++17}. Figure \ref{fig:lensinginferenceprior} shows the resulting joint distribution of $w \times p\left(\delta_{\rm{LOS}}, \beta, \log_{10}c_8, \Delta \alpha | \boldsymbol{D}\right)$.  In Appendix \ref{app:sigmasubprior}, we perform our analysis assuming $\Sigma_{\rm{sub(predicted)}} = 0.025 \ \rm{kpc^{-2}}$, and show that the effect of assuming a different value for $\Sigma_{\rm{sub(predicted)}}$ has a smaller affect on our results than the statistical uncertainty of our measurement. 
	
	As Figures \ref{fig:lensinginferencefull} and \ref{fig:lensinginferenceprior} clearly demonstrate, the $\Lambda$CDM prediction lies within the $68 \%$ confidence intervals, so our results agree with the predictions of $\Lambda$CDM. However, we can make the agreement more striking by assigning an informative prior to the logarithmic slope of the concentration-mass relation $\beta$, and the amplitude of the line of sight halo mass function $\delta_{\rm{LOS}}$. Doing so, we obtain the joint distribution $p\left(\beta, \delta_{\rm{LOS}} | \Lambda\rm{CDM} \right) \times p\left(\qsub | \boldsymbol{D}\right)$, where the first term corresponds to Gaussian priors on $\beta$ and $\delta_{\rm{LOS}}$ of $\mathcal{N}\left(0.8, 0.3\right)$ and $\mathcal{N}\left(1.0, 0.2\right)$, respectively. Figure \ref{fig:lensinginferenceCDM} shows the resulting joint distribution of $\Delta \alpha$ and $\log_{10} c_8$, marginalized over the uniform prior on $\Sigma_{\rm{sub}}$ (without the importance weights $w$). We infer $\log_{10} c_8 =$ $\mcconstraintonesigma$ and $\mcconstrainttwosigma$ at $68 \%$ and $95 \%$ confidence, respectively. We infer $\Delta \alpha = $ $\deltalphainferenceonesigma$ at $68 \%$ confidence. Both results are in excellent agreement with the predictions of $\Lambda$CDM. 
	
		\begin{figure*}
		\includegraphics[clip,trim=0cm 0cm 0cm
		0cm,width=.95\textwidth,keepaspectratio,angle=0]{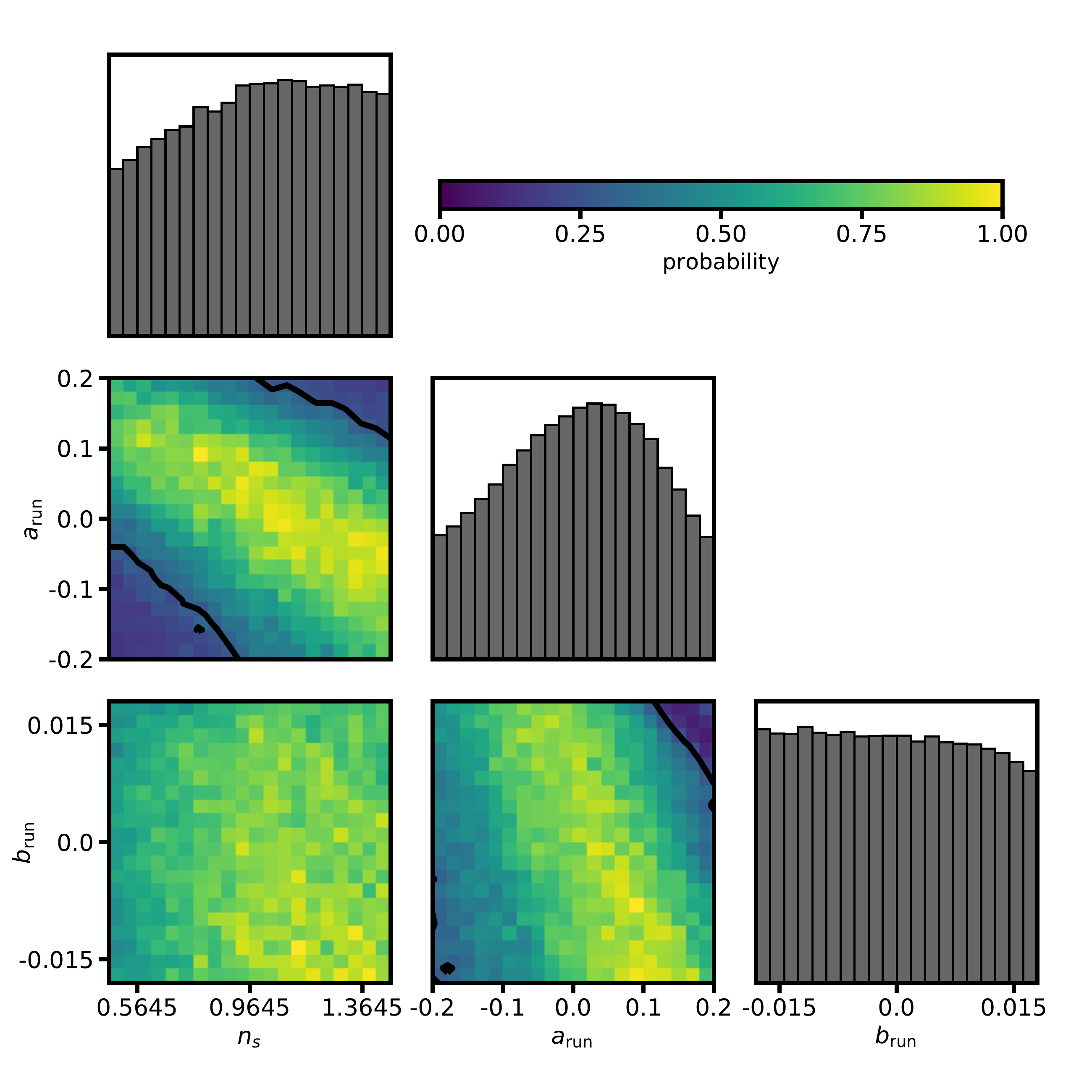}
		\caption{\label{fig:likelihood_PK} The posterior probability distribution $p\left(\qp | \boldsymbol{D} \right)$, obtained by mapping each point in the $\qp$ parameter space to the corresponding set of $\qsub$ parameters, and evaluating the likelihood shown in Figure \ref{fig:lensinginferenceprior}. Contours show $68 \%$ confidence intervals.}
	\end{figure*}
	\begin{figure}
		\includegraphics[clip,trim=0cm 0cm 0cm
		0cm,width=.45\textwidth,keepaspectratio,angle=0]{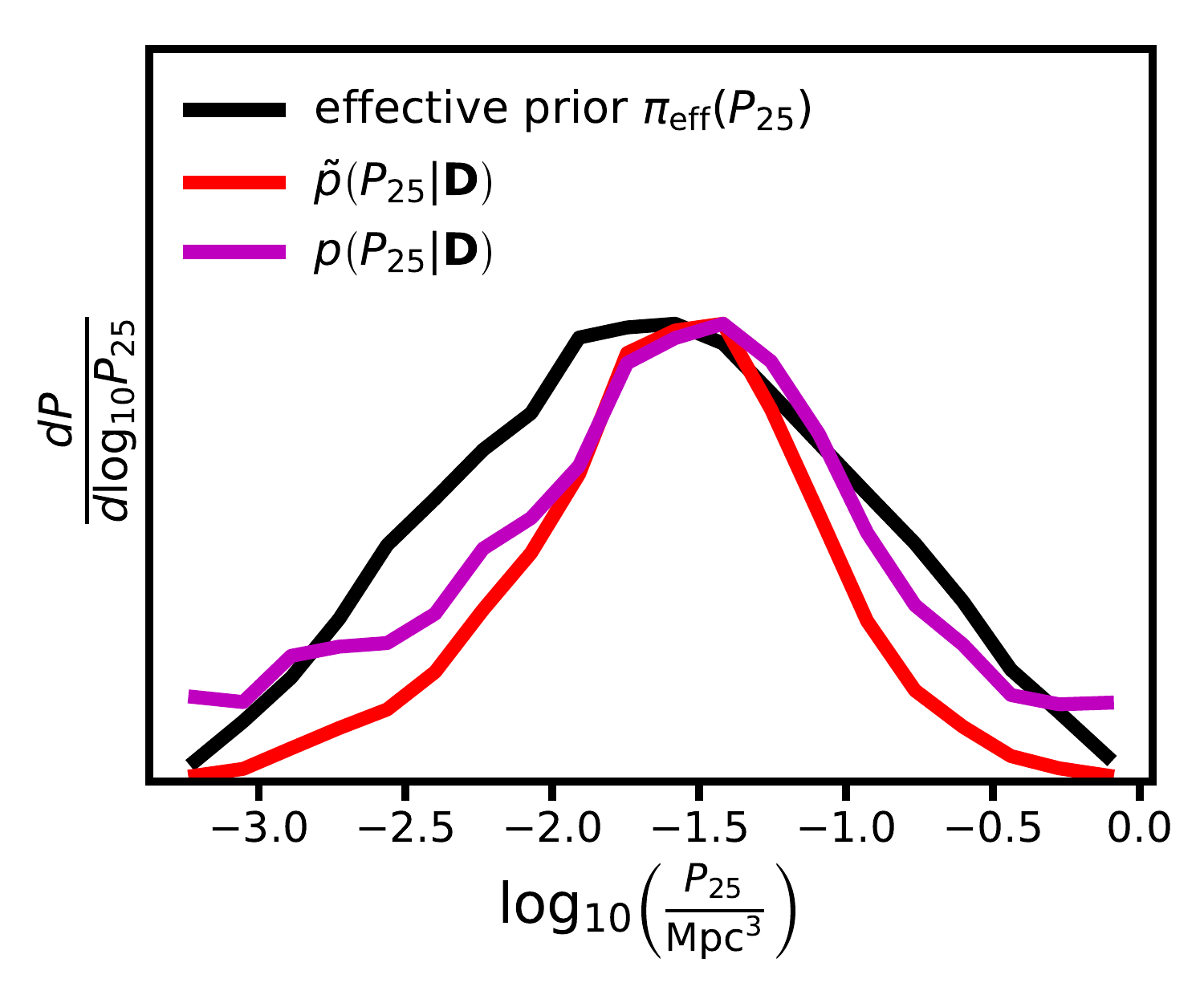}
		\caption{\label{fig:likepriorpost} The effective prior $\pi_{\rm{eff}} \left(P_{25}\right)$ on $P_{25}$, the power spectrum amplitude evaluated at $25 \ \rm{Mpc^{-1}}$, that corresponding to a uniform prior on $\qp$ parameters (black), the posterior distribution $\tilde{p}\left(P_{25} | \boldsymbol{D}\right)$ obtained from sampling the probability density shown in Figure \ref{fig:likelihood_PK} and evaluating Equation \ref{eqn:Pkequation} (red), and the probability distribution $p\left(P_{25} | \boldsymbol{D}\right) \propto \frac{\tilde{p}\left(P_{25} | \boldsymbol{D}\right)}{\pi_{\rm{eff}} \left(P_{25}\right)}$ (magenta). The magenta curve has the interpretation of the posterior distribution on $P_{25}$ given the data, assuming a log-uniform prior on $P_{25}$. We use a value of $k=25 \ \rm{Mpc^{-1}}$ to create the figure, but this choice is arbitrary. When stating a constraint on the power spectrum amplitude at any scale (labeled $P_k$), we follow the procedure illustrated in the figure to compute $p\left(P_k | \boldsymbol{D}\right)$, and use this distribution to compute the median and confidence intervals. The $\Lambda$CDM prediction for the power spectrum amplitude at this scale is $10^{-1.6} \rm{Mpc^{3}}$.}
	\end{figure}
	
	\subsection{Recasting the lensing measurement in terms of $n_s$, $a_{\rm{run}}$, and $b_{\rm{run}}$}
	\label{ssec:recasting}
	To recast our lensing measurement in terms of the primordial power spectrum, we compute $p\left(\qp | \boldsymbol{D}\right)$, the probability distribution of the parameters describing the power spectrum model in Equation \ref{eqn:Pkequation}, given the data. We can express this probability distribution as the prior on the $\qp = \left(n_s, a_{\rm{run}}, b_{\rm{run}}\right)$ parameters, $\pi\left(\qp\right)$, times the likelihood of the data given the parameters
	\begin{eqnarray}
	\label{eqn:pdfqp}
	\nonumber p\left(\qp | \boldsymbol{D} \right) & \propto & \pi\left(\qp\right) \mathcal{L}\left(\boldsymbol{D} | \qp \right) \\
	& \propto & \pi\left(\qp\right) \int \mathcal{L}\left(\boldsymbol{D} | \qsub \right) p\left(\qsub | \qp \right) d \qsub. 
	\end{eqnarray}
	
	In the second line, we have expressed the likelihood of $\qp$ in terms of $\qsub$ through the conditional probability $p\left(\qsub | \qp \right)$. To evaluate this term, we use {\tt{galacticus}} to compute predictions for the mass function and the concentration-mass relation using the theoretical frameworks discussed in Section \ref{sec:pkandobservables}. For each predicted mass function and concentration-mass relation, we determine the set of $\qsub$ parameters that minimize the residuals between the theoretical prediction, and the model in terms of $\qsub$. Figure \ref{fig:mcrel} show an example of this procedure. Solid lines show the theoretically-predicted mass functions and concentration-mass relations, and dashed lines show the best-fit model for the mass function in terms of $\delta_{\rm{LOS}}$ and $\Delta \alpha$ using Equation \ref{eqn:massfunction}, and the best-fit concentration-mass relation parameterized in terms of $c_8$ and $\beta$ with Equation \ref{eqn:mcrelation}. To compute a corresponding set of $\qsub$ parameters for any $\qp$, we repeat the fit of $\qsub$ parameters to the $\qp$ predictions for values of $n_s$ between $0.3645$ and $1.5645$, $a_{\rm{run}}$ between $-0.2$ and $0.2$, and $b_{\rm{run}}$ between $-0.018$ and $0.018$. Using $5,000$ unique combinations of $\qp$ parameters, we interpolate the mapping to establish a continuous transformation between the parameters describing the power spectrum, and parameters sampled in the lensing analysis. 
	
	Figure \ref{fig:mappings} shows the correspondence between $\qp$ and $\qsub$ as a function of $n_s$, $a_{\rm{run}}$, and two values of $b_{\rm{run}}$. The line style indicates the value of $b_{\rm{run}}$, while the color of each curve indicates the value of $a_{\rm{run}}$. The y-axis shows the value of each $\qsub$ parameter as a function of $n_s$. The information conveyed by the curves in the different panels of Figure \ref{fig:mappings} complements the discussion in the opening paragraph of Section \ref{sec:pkandobservables}. Increasing the amount of small-scale power leads to covariant changes in the mass function and concentration-mass relation. Increasing power raises the amplitude of the mass function at $10^{8} M_{\odot}$, and makes halos more concentrated. Decreasing power makes halos less concentrated, lowers the amplitude of the mass function at $10^{8} M_{\odot}$, and leads to a shallower halo mass function slope (positive $\Delta \alpha$). The other mass function models we consider (see Appendix \ref{app:massfunctions}) express similar trends between the $\qp$ and $\qsub$ parameters, although they predict slightly higher mass function amplitudes $\delta_{\rm{los}}$\footnote{In Appendix \ref{app:massfunctions}, we repeat the analysis presented in this section using the other models for the halo mass function, and show that the statistical measurement uncertainties are larger than the effect of using a different model for the halo mass function.}. 
	
	Using the continuous transformation $\qp \rightarrow \qsub$ illustrated in Figure \ref{fig:mappings}, we evaluate Equation \ref{eqn:pdfqp} by, first, sampling uniform priors in $\qp$ space. If we had no data, then each $\qp$ sample would correspond to an equally probable point in $\qsub$ space, as we assigned equal prior probability to points in $\qsub$ space. However, we do have informative data, and we can assign each sample drawn from the $\qp$ prior the likelihood $\mathcal{L}\left(\boldsymbol{D} | \qsub\left(\qp\right) \right)$, where we have now expressed the $\qsub$ parameters as functions of $\qp$ reflecting the mapping $\qp \rightarrow \qsub$. The prior for the parameters used in the lensing analysis, $\pi\left(\qsub\right)$, plays no role in this computation, because we only make use of the likelihood $\mathcal{L}\left(\boldsymbol{D} | \qsub \right)$. As we have used uniform priors to derive the posterior distributions shown in Figures \ref{fig:lensinginferencefull} and \ref{fig:lensinginferenceprior}, the posterior distributions shown in the figures vary proportionally with the likelihood, and we can sample them directly to evaluate $\mathcal{L}\left(\boldsymbol{D} | \qsub\right)$. 
	
	In practice, the models implemented for the mass function and the concentration-mass relation will not perfectly describe the theoretically-predicted relations. This outcome can result from a breakdown of the theoretical models themselves, or because the analytic formulas implemented in the lensing analysis cannot fit the predicted form of the mass function and concentration-mass relation perfectly. As shown in Figure \ref{fig:mcrel}, the second outcome occurs for some combinations of $\qp$ parameters that predict enhanced small-scale power, causing the concentration-mass relation to deviate from a power law in peak height (Equation \ref{eqn:mcrelation}). To address these sources of systematic uncertainty, we can re-compute our main results by changing the mass function and concentration-mass models used to perform the mapping $\qp \rightarrow \qsub$. This test is done in Appendix \ref{app:massfunctions} with two other models for the halo mass function. If a future investigation presents a theoretical framework specifically tailored to predict mass functions and concentrations from the initial power spectrum model we implemented, it would be straightforward to re-interpret the inference shown in Figures \ref{fig:lensinginferencefull} and \ref{fig:lensinginferenceprior} in terms of the new model, when it becomes available. 
	
	To address the second source of systematic uncertainty associated with the mapping $\qp \rightarrow \qsub$, we estimate systematic uncertainties associated with this transformation, and propagate them through our model. For each point in $\qp$ space, we compute a set of systematic errors $\Delta\qsub \left(\qp\right)$, and propagate them through the model by altering the mapping: $\qp \rightarrow \qsub + \Delta \qsub \left(\qp\right)$. We have written the change in $\qsub$ parameters as a function of $\qp$ because we compute these systematic uncertainties for each point in $\qp$ space. Appendix \ref{app:systematics} gives additional details about how we estimate these systematic uncertainties $\Delta \qsub \left(\qp\right)$, how we propagate them through our model, and by how much they increase the uncertainties.

To gain physical intuition for how the data can constrain the power spectrum, Figure \ref{fig:jointpdfmapping} illustrates how the $\qp$ parameters map to $\qsub$. The physics relevant for structure formation constrains the mapping, and predicts certain correlations between halo abundance and concentration that cuts through the parameter space spanned by the prior $\pi\left(\qsub\right)$. Each panel in the figure shows a joint likelihood between $\qsub$ parameters overlaid with three curves differentiated by the marker style. Diamonds indicate power spectrum models with suppressed small scale power, with $b_{\rm{run}} = -0.012$ and $n_s = 0.46$, crosses have enhanced small-scale power with $b_{\rm{run}} = 0.012$ and $n_s = 1.46$, and circles lie in between, with $b_{\rm{run}} = 0$ and $n_s = 0.96$. The color of each point indicates the value of $a_{\rm{run}}$, with negative (positive) values of $a_{\rm{run}}$ corresponding to blue (red) points. The figure clearly shows that an enhancement of small-scale power introduce through one parameter can offset a suppression of small scale power introduced through another. Models with positive $a_{\rm{run}}$, but suppressed small scale power through low $n_s$ and negative $b_{\rm{run}}$, map to similar regions of $\qsub$ parameter space\footnote{For example, the reddest diamond, which has $b_{\rm{run}} = -0.012$, $n_s = 046$, and $a_{\rm{run}} = 0.15$, has a similar likelihood as the bluest cross, which has $b_{\rm{run}} = 0.12$, $n_s = 1.46$, and $a_{\rm{run}} = -0.15$.}, and hence have roughly equal likelihoods, as models with negative $a_{\rm{run}}$, and enhanced small-scale power through large $n_s$ and positive $b_{\rm{run}}$. We therefore expect anti-correlations between pairs of $\qp$ parameters. 

Figure \ref{fig:jointpdfmapping} also demonstrates why we can constrain $P\left(k\right)$ using strong lensing data, even though the lensing inference itself exhibits significant covariance between model parameters that precludes statistically significant marginalized constraints on their values. Intuitively, our lensing inference identifies a volume of parameters space in $\qsub$ that predicts a constant amount of magnification perturbation in the data. Simultaneously increasing the number of halos and their concentrations increases the amount of perturbation to image magnifications. Conversely, simultaneously lowering the amplitude of the mass function, and making halos less concentrated, reduces the amount of perturbation. To reproduce the amount of perturbation as in the data, strong lensing inferences therefore exhibit an anti-correlation between the amplitude of the halo mass function and the concentration-mass relation. Recalling the discussion in opening paragraph of Section \ref{sec:pkandobservables} related to how altering the power spectrum at a scale $\tilde{k}$ affects dark matter structure, increasing the amount of power at some scale $\tilde{k}$ simultaneously increases the concentration and abundance of halos with mass $\tilde{m}$. As Figure \ref{fig:jointpdfmapping} illustrates, the positive correlation between halo abundance and concentration that results from altering $P\left(k\right)$ runs orthogonal to the anti-correlation between halo abundance and concentration allowed by the data, leading to constraints on the amplitude of $P\left(k\right)$ from above and below. 
		
	\subsection{Reconstructing the primordial matter power spectrum}
	\label{ssec:inferencepk}
	
	We use the mapping between $\qp$ and $\qsub$ to compute the joint distribution $p\left(\qp | \boldsymbol{D} \right)$. First, we generate 20,000 samples of $\qp$ from uniform priors on $n_s$, $a_{\rm{run}}$, and $b_{\rm{run}}$ in the ranges indicated in Table \ref{tab:params}. Then, we assign each sample a likelihood by mapping each $\qp$ point to the corresponding point in the space of $\qsub$, and evaluate the probability of the $\qsub$ point using the likelihood in Figure \ref{fig:lensinginferenceprior}. Figure \ref{fig:likelihood_PK} shows the resulting joint distribution $p\left(\qp| \boldsymbol{D}\right)$. We recover the expected anti-correlation between $\qp$ parameters predicted in the previous section. Due to the correlated posterior distributions, the marginal likelihoods appear somewhat unconstrained. However, the constraining power on the power spectrum comes from the full three dimensional joint distribution, where our data rules out values of $n_s$, $a_{\rm{run}}$, and $b_{\rm{run}}$ that produce an large excess or deficit of small-scale power relative to the $\Lambda$CDM prediction, which we associate with $n_s = 0.965$, $a_{\rm{run}} = 0$, and $b_{\rm{run}} = 0$. 
	
	\begin{figure*}
		\includegraphics[clip,trim=0cm 0.5cm 0cm
		0cm,width=.95\textwidth,keepaspectratio,angle=0]{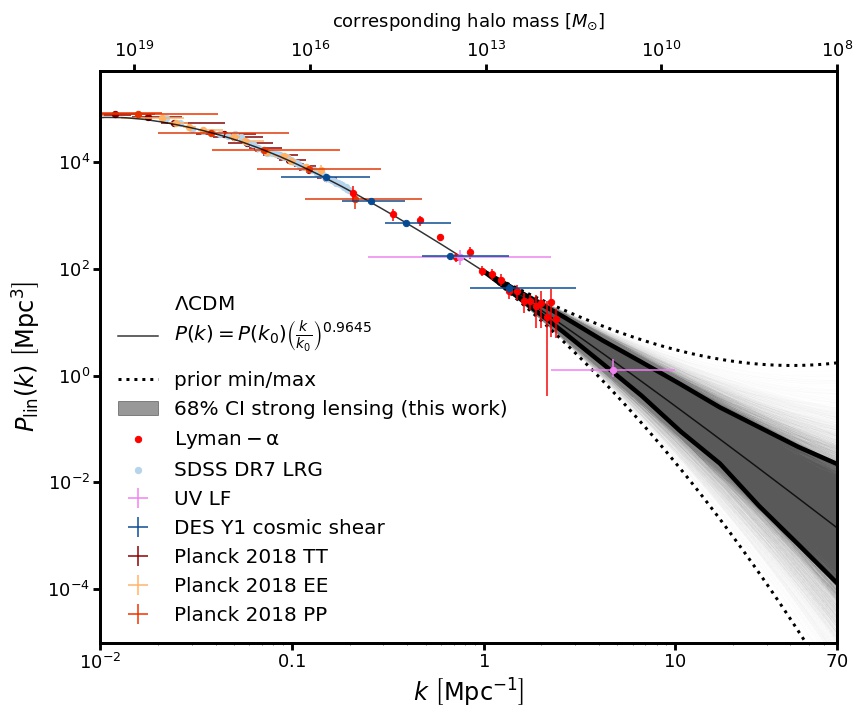}
		\caption{\label{fig:inference} Measurements (colored points) and the $\Lambda$CDM prediction (black curve) for the matter power spectrum $P_{\rm{lin}}\equiv P\left(k\right) T^2\left(k\right)$, where $P\left(k\right)$ is the primordial matter power spectrum, and $T\left(k\right)$ is the linear theory transfer function \citep{EisensteinHu98}. Measurements shown in the figure come from analyses of CMB from the Planck satellite \citep{Planck2020a}, galaxy clustering measured by the Dark Energy Survey (DES) \citep{Troxel++2018} and the Sloan Digital Sky Survey (SDSS) \citep{Reid++10}, the Lyman-$\alpha$ forest \citep{Viel++04,Chabanier++19,Chabanier++2019b}, and the ultra-violet luminosity function of distant galaxies \citep[][with label UV LF]{Sabti++21}. The lower x-axis shows the wavenumber $k$, and the upper axis shows a corresponding halo mass scale $m = (4 \pi/3) \rho_m \left(2 \pi / k\right)^3$ computed with respect to the contribution of matter to the critical density of the Universe $\rho_m$. The light gray curves show 20,000 individual realizations of the power spectrum obtained by sampling the likelihood shown in Figure \ref{fig:likelihood_PK}, and evaluating Equation \ref{eqn:Pkequation}. The dark gray band enclosed by the thick black curves shows the $68 \%$ confidence intervals of the power spectrum amplitude at each value of $k$ sampled from $p\left(P_k | \boldsymbol{D}\right)$ (see the discussion in the caption of Figure \ref{fig:likepriorpost}). The upper and lower dashed black curves show the maximum and minimum power spectrum amplitude allowed by our model at each $k$ scale, assuming the uniform priors on the parameters describing $P\left(k\right)$ summarized in Table \ref{tab:params}. We adapted this figure from \citet{Chabanier++19}, using data products presented by \citet{Planck2020a}.}
	\end{figure*}

	By sampling $\qp$ parameters from the joint distribution $p\left(\qp | \boldsymbol{D} \right)$ and evaluating Equation \ref{eqn:Pkequation}, we can compute the posterior distribution $\tilde{p}\left(P_k | \boldsymbol{D}\right)$ of the power spectrum amplitude at any scale $k$, which we label $P_k$, given a uniform prior on $n_s$, $a_{\rm{run}}$, and $b_{\rm{run}}$. However, a uniform prior on $n_s$, $a_{\rm{run}}$, and $b_{\rm{run}}$ does not correspond to a uniform prior on $P_k$. Rather, a uniform prior on $\qp$ parameters corresponds to an effective prior $\pi_{\rm{eff}}\left(P_k\right)$ for the power spectrum amplitude evaluated at $k$. When presenting our inference on the power spectrum at any scale throughout the remainder of this section, we divide the posterior distribution $\tilde{p}\left(P_k | \boldsymbol{D}\right)$ by this effective prior to obtain a new probability distribution $p\left(P_k | \boldsymbol{D} \right)$, which has the interpretation of the posterior distribution of $P_k$ given the data, assuming a log-uniform prior on $P_k$. Figure \ref{fig:likepriorpost} shows the three distributions discussed in this paragraph - the posterior $\tilde{p}\left(P_{25} | \boldsymbol{D}\right)$, the implicit prior $\pi_{\rm{eff}}\left(P_{25}\right)$, and the posterior distribution $p\left(P_{25} | \boldsymbol{D}\right)$, evaluating the power spectrum model at $k = 25 \ \rm{Mpc^{-1}}$. 
	
	Using the likelihood $p\left(\qp | \boldsymbol{D}\right)$, we can reconstruct the shape of $P\left(k\right)$, assuming the model for the power spectrum given by Equation \ref{eqn:Pkequation}. The light gray curves in Figure \ref{fig:inference} depict 20,000 individual power spectra that correspond to 20,000 samples of $\qp$ parameters drawn from the joint distribution $p\left(\qp | \boldsymbol{D} \right)$. The dark gray shaded region bounded by the black curves shows the $68 \%$ confidence interval of the distribution $p\left(P_k | \boldsymbol{D} \right)$, or the inference of the power spectrum amplitude given a log-uniform prior on the power spectrum amplitude, at each value of $k$. The inference obtained from the eleven lenses in our sample is consistent with the predictions of $\Lambda$CDM in the range of $k$ values shown in the figure, between $1 - 70 \ \rm{Mpc^{-1}}$. 
	
	\subsection{The scales probed by our data}
	\label{ssec:pivotscales}
	We can discuss in order-of-magnitude terms what $k$ modes drive the inference shown in Figure \ref{fig:inference} based on the structure and abundance of halos that we infer from from the data. A useful proxy for halo mass in this context is the wavenumber that corresponds to the Lagrangian radius $R_l$ of a halo of mass $m$, defined in the introduction as $m\equiv \frac{4 \pi}{3} \Omega_m \rho_{\rm{crit}} R_l^3$. Rearranging this equation in terms of a wavenumber $k_l \equiv \frac{2 \pi}{R_l}$, we find $k_l = 50 \left(\frac{2\times 10^8 }{m \left[M_{\odot}\right]}\right)^{\frac{1}{3}} \rm{Mpc^{-1}}$. For galaxies inhabiting halos with masses between $10^{11} - 10^{13} M_{\odot}$, the corresponding modes are between $1 - 10 \  \rm{Mpc^{-1}}$. \citet{Sabti++21} recently presented constraints on the matter power spectrum on these scales by analyzing the luminosity function of galaxies, casting this measurement in terms of the halo mass function and therefore the linear matter power spectrum. We can extend this reasoning down to the mass ranges we probe with strong lensing, keeping in mind that we have access to additional information regarding $P\left(k\right)$ because we also constrain the concentration-mass relation. The largest halos we rendered in our simulations have masses of $10^{10} M_{\odot}$, correspond to $k \sim 10 \ \rm{Mpc^{-1}}$, while we estimate sensitivity\footnote{See Section \ref{ssec:mfuncmcrelmodels} for a discussion of the halo mass scales probed by the data.} to halos down to roughly $10^7 M_{\odot}$, or $k \sim 100 \ \rm{Mpc^{-1}}$. The constraints from a sample of eight lensed quasars presented \citet{Gilman++20a} ruled out warm dark matter models with turnovers in the mass function above $6 \times 10^{7} M_{\odot}$ at $95 \%$ confidence, supporting the argument that we can probe structure below $10^{8} M_{\odot}$ with existing data. 
			
	The line of reasoning discussed in the previous paragraph provides general intuition for the scales we can constrain with our data, although it almost certainly oversimplifies the problem for three reasons. First, halos with masses of $10^7 \ M_{\odot}$ halos do not necessarily contribute the same amount of signal as halos with masses of $10^8 \ M_{\odot}$ or $10^{9} \ M_{\odot}$. Thus, the signal we measure has an integrated contribution from many different scales, with each scale not necessarily contributing in equal measure. Second, the theoretical models linking the linear matter power spectrum to the halo mass function and concentration-mass relation depend on integrals and derivatives of $P\left(k\right)$, further blending the contribution from different $k$ modes to the structure and abundance of halos of any given mass. Third, the analytic model we implemented for $P\left(k\right)$ enforces certain correlations, or relative contributions, from different $k$ scales to the signal we extract from the data. Taking these complications into account, it is likely that large-scale modes with $k < 10 \ \rm{Mpc^{-1}}$ filter into the signal we measure, while the analytic form of the power spectrum we used will couple constraints on $P\left(k\right)$ from large scales with those from small scales, and vice versa. 
	
	Despite these complications, we can make some headway towards understanding what scales drive our constraints by experimenting with the pivot scale $k_0$. As the form of the power spectrum is fixed to the $\Lambda$CDM prediction on scales $k < k_0$, we can assess where our data begins to probe the power spectrum by shifting the pivot to larger scales, and repeating our analysis. More specifically, by shifting the pivot scale towards smaller values of $k$ (larger scales), we can assess to what degree modes with $k < 1 \ \rm{Mpc^{-1}}$ contribute the measured signal, relative to modes with $k > 1 \  \rm{Mpc^{-1}}$. If shifting the pivot scale to, for example, $0.1 \ \rm{Mpc^{-1}}$, results in a large increase in the uncertainty of the power spectrum amplitude inferred at $30 \ \rm{Mpc^{-1}}$ or $50 \ \rm{Mpc^{-1}}$, then we would conclude that the power spectrum on scales $0.1 - 1 \ \rm{Mpc^{-1}}$ contributes more signal than the power spectrum on scales $k > 1 \  \rm{Mpc^{-1}}$. On the other hand, if we find a similar degree of constraining power at $k > 1 \ \rm{Mpc^{-1}}$ when shifting the pivot to smaller $k$, then we would conclude that scales $k > 1 \ \rm{Mpc^{-1}}$ contributes more signal than $k < 1 \ \rm{Mpc^{-1}}$. 
	
\begin{figure}
		\includegraphics[clip,trim=0cm 0.5cm 0.3cm
		0.5cm,width=.48\textwidth,keepaspectratio,angle=0]{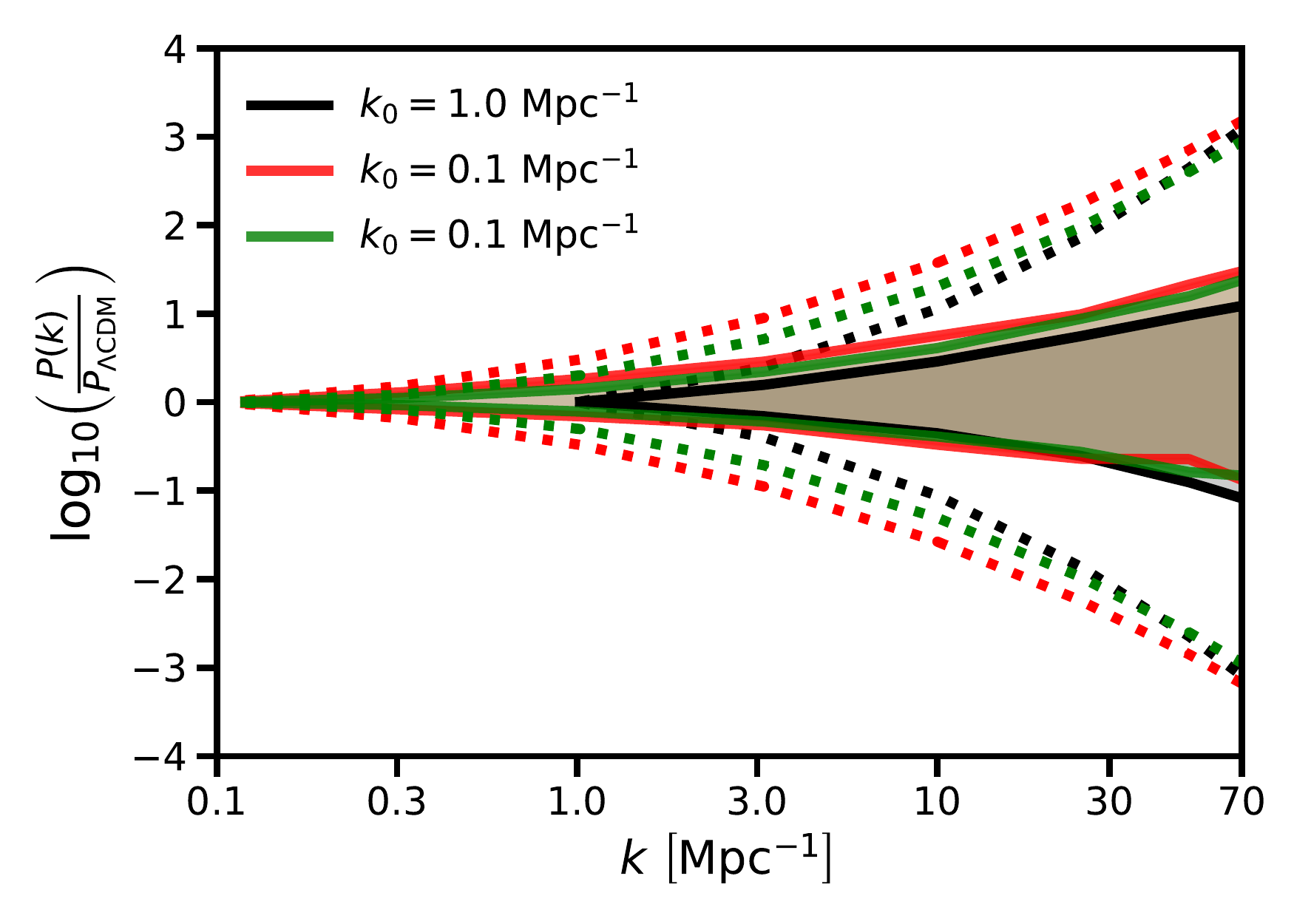}
		\caption{\label{fig:pivoteffect} The inference on the power spectrum amplitude using a model with a pivot at $1 \ \rm{Mpc^{-1}}$ (black), and two models with pivots at $k_0 = 0.1 \  \rm{Mpc^{-1}}$ (red and green). Shaded regions show the $68 \%$ confidence interval of the inference on the power spectrum amplitude at each scale. The dotted lines show the minimum and maximum values the power spectrum amplitude could have taken, given the priors on the $\qp$ parameters. The red and green models share the same pivot scale, but have different priors on the $\qp$ parameters that alter the uncertainty on the power spectrum amplitude relative to the black model with the pivot at $1 \ \rm{Mpc^{-1}}$.}
	\end{figure}

	Figure \ref{fig:pivoteffect} shows the result of repeating our analysis with a pivot at $k_0 = 0.1 \ \rm{Mpc^{-1}}$, relative to the power spectrum inference shown in Figure \ref{fig:inference} (black). The red and green curves correspond to models each with a pivot at $k_0 = 0.1 \  \rm{Mpc^{-1}}$, but with different priors on $\qp$ parameters that alter the uncertainty in the power spectrum amplitude as a function of $k$. The y-axis shows the power spectrum amplitude relative to the $\Lambda$CDM prediction, and the shaded region represents the $68 \%$ confidence interval as a function of $k$. As in Figure \ref{fig:inference}, the dotted lines correspond to the implicit prior on the power spectrum amplitude at each value of $k$, or the maximum and minimum value $P\left(k\right)$ could have at each scale, given the priors on $\qp$. We have chosen the priors on $\qp$ parameters for the red and green models to increase the uncertainties on the power spectrum amplitude on scales $1 - 10 \ \rm{Mpc^{-1}}$ relative to the black model with the pivot at $1 \ \rm{Mpc^{-1}}$, but set the priors on the green model to approximately match the uncertainty in the power spectrum amplitude of the black model on scales $k > 25 \ \rm{Mpc^{-1}}$. 
	
	By comparing the inference on $P\left(k\right)$ obtained with the three models, we can assess to what degree uncertainty in the power spectrum amplitude on relatively large scales, $k < 1 \ \rm{Mpc^{-1}}$ propagates onto the constraints on modes $k > 10 \ \rm{Mpc^{-1}}$. Evaluating the model for $P\left(k\right)$ at $k = 10$, $25$, and $50 \ \rm{Mpc^{-1}}$, and placing the pivot at $1 \ \rm{Mpc^{-1}}$, we infer $\log_{10}\left(P / P_{\Lambda \rm{CDM}}\right)$, the power spectrum amplitude relative to the $\Lambda$CDM prediction, of $\pkone$, $\pktwo$, and $\pkthree$, respectively. Shifting the pivot to $k_0 = 0.1 \ \rm{Mpc^{-1}}$, we infer $\log_{10}\left(P / P_{\Lambda \rm{CDM}}\right) = 0.2_{-0.4}^{+0.6}$, $0.2_{-0.6}^{+0.9}$, and $0.3_{-0.8}^{+1.2}$ for the green model shown in Figure \ref{fig:pivoteffect}. Using the red model, we infer $\log_{10}\left(P / P_{\Lambda \rm{CDM}}\right) = 0.2_{-0.5}^{+0.8}$, $0.3_{-0.6}^{+1.0}$, and $0.4_{-0.6}^{+1.3}$. Moving the pivot to larger scales leads to the outcome we would expect if the majority of the constraining power comes from small scales, with $k > 10 \ \rm{Mpc^{-1}}$; extrapolating the model over a large range of $k$ modes naturally increases the uncertainty on the power spectrum amplitude on small scales, but by an amount much smaller than the statistical measurement uncertainties.
	
	Each of the three inferences depicted in Figure \ref{fig:pivoteffect} is consistent with the predictions of $\Lambda$CDM and single-field slow-roll inflation over the $k$ scales shown in the figure. However, as discussed in the previous paragraphs, the use of an analytic model for $P\left(k\right)$ couples the power spectrum amplitude at different scales, and the inferences on $P\left(k\right)$ at different scales are therefore not independent. The covariance matrix $\Sigma$ between the inferences for the model with the pivot at $1 \ \rm{Mpc^{-1}}$ is 
	\begin{ceqn}
	\[
	\Sigma=
	\begin{pmatrix}
	0.10 & 0.16 & 0.21 \\
	0.16 & 0.27 & 0.37 \\
	0.21 & 0.37 & 0.53 \\
	\end{pmatrix}.
	\]
	\end{ceqn}
	where the first, second, and third entries correspond to 
	samples from the posterior distribution $\tilde{p}\left(P_k | \boldsymbol{D}\right)$, or $\log_{10} P\left(k\right)$ evaluated at $k = 10$, $25$, and $50 \ \rm{Mpc^{-1}}$.

	\section{Discussion}
	\label{sec:conclusions}
	Using a sample of eleven quadruply-imaged quasars, we have performed a simultaneous inference of the halo mass function and the concentration-mass relation in the mass range $10^7 - 10^{10} M_{\odot}$, and interpret the results in terms of the primordial matter power spectrum. Our analysis shows that the reach of strong lensing as a probe of fundamental physics extends beyond the confines of $\Lambda$CDM and warm dark matter, the two theoretical frameworks that typically underpin a galaxy-scale strong lensing analysis and the interpretation of results. Strong lensing of compact, unresolved sources can also constrain the initial conditions for structure formation quantified by the primordial matter power spectrum, which itself encodes properties of the inflaton. We summarize our main results as follows:
	
	\begin{itemize}
		\item Assuming an analytic model for $P\left(k\right)$ with a varying spectral index, running of the spectral index, and running-of-the-running beyond a pivot scale at $1 \  \rm{Mpc^{-1}}$, we constrain the form of the primordial matter power. Relative to $P_{\Lambda\rm{CDM}}$, a model with no scale dependence of the spectral index, and with an amplitude anchored on large scales, we infer power spectrum amplitudes $\log_{10}\left(P / P_{\Lambda \rm{CDM}}\right)$ at $k = 10$, $25$, and $50$ $\rm{Mpc^{-1}}$ of $\pkone$, $\pktwo$, and $\pkthree$ at $68 \%$ confidence, in agreement with the predictions of $\Lambda$CDM and slow-roll inflation. These constraints correspond to power spectrum amplitudes in physical units $\rm{Mpc^{3}}$ of $\log_{10} \left(\frac{P\left(k\right)}{\rm{Mpc^{3}}}\right)$ of  $-0.5_{-0.4}^{+0.5}$, $-1.5_{-0.6}^{+0.7}$, and $-2.4_{-0.9}^{+1.0}$, respectively. Since we assume an analytic model for the power spectrum, the inferences on these scales are not independent, and we provide the covariance matrix in Section \ref{ssec:pivotscales}. 
		\item Assuming the $\Lambda$CDM prediction for the logarithmic slope of the concentration-mass relation and the amplitude of the halo mass function, we infer an amplitude $c_8$ of the concentration-mass relation at $10^{8} M_{\odot}$ of $\log_{10} c_8 =$ $\mcconstraintonesigma$ and $\mcconstrainttwosigma$ at $68 \%$ confidence and $95 \%$ confidence, respectively, and constrain deviations of the logarithmic slope of the halo mass function around the CDM prediction $\Delta \alpha = 0$ to $\Delta \alpha = \deltalphainferenceonesigma$ at $68 \%$ confidence. These results are marginalized over the amplitude of the subhalo mass function, and are in excellent agreement with the predictions of $\Lambda$CDM. 
	\end{itemize}

	Our results should be interpreted within the context of the analytic form for the primordial power spectrum we have assumed throughout this work. As the models we used for the halo mass function and concentration-mass relation are not arbitrarily flexible, we could only consider models for the power spectrum with certain properties, namely, a variable spectral index, and scale-dependent terms. When interpreting Figure \ref{fig:inference}, one should take into consideration that the constraints we present have meaning specifically within the context of this model. The constraints at different scales are not independent, and thus the confidence interval band shown in Figure \ref{fig:inference} represents a series of correlated inferences at different scales, unlike the independent measurements shown in the figure. A different way of phrasing the discussion surrounding the analytic model is that we have built into our inferences certain assumptions about the relative contribution from different $k$-scales to the signal we measure. Determining how to de-couple the information from different $k$-scales could form the basis for future investigation.
	
	 These complications aside, our findings suggest that the small-scale behavior of the power spectrum does not significantly deviate from the power spectrum predicted by single-field slow-roll inflation, characterized by $a_{\rm{run}}$ and $b_{\rm{run}}$ small in magnitude compared to $|n_s - 1|$ \citep{LiddleLyth2000}. Future analyses that build on the framework we present, utilizing a larger sample size of lenses and possibly refined calibrations of the halo mass function and concentration-mass relation, will enable tighter constraints on the abundance and concentration of low-mass dark matter halos, opening new avenues to constrain the properties of the early Universe.
	
	As a necessary intermediate step in our analysis, we measured the abundance and concentrations of dark matter halos. In particular, assuming the $\Lambda$CDM prediction for the logarithmic slope of the concentration-mass relation and the amplitude of the field halo mass function, we obtained the tightest constraints on the amplitude of the concentration-mass relation and the slope of the halo mass function presented to date. The constraint on the concentration-mass relation is consistent with our previous work \citep{Gilman++20b}. Our measurement of the logarithmic slope of the halo mass function is consistent with the measurement presented by \citet{Vegetti++14}, although our measurement reaches higher precision.
	
	To date, inferences with strong lensing of alternative theories to $\Lambda$CDM have focused mainly on models of warm dark matter, characterized by a cutoff in the linear matter power spectrum and a suppression of small scale structure \citep{Inoue++15,Birrer++17,Vegetti++18,Gilman++20a,Hsueh++20}. The cutoff in the power spectrum in these models comes from dark matter physics that alters the transfer function, namely, free-streaming of dark matter particles, and is not of primordial origin. Other analyses have sought to detect individual dark matter halos \citep{Vegetti++14,Nierenberg++14,Hezaveh++16,Nierenberg++17,Sengul++21}, or characterize general properties of the population of perturbing halos without attempting to distinguish between dark matter models \citep{Dalal++02,Gilman++20b}. As we discuss in Section \ref{ssec:recasting}, the constraining power of strong lensing over the primordial spectrum $P\left(k\right)$ comes from the joint distribution of mass function and concentration-mass relation amplitudes and logarithmic slopes. While some lensing analyses have constrained the mass function or concentration-mass relation independently of one another, no previous work has simultaneously measured these relations to account for covariance between them. For this reason, we cannot compare our results with previous lensing measurements in the context of the power spectrum. 
	
	We have performed our analysis with three different models of the halo mass function (see Appendix \ref{app:massfunctions}), and find that using a different model to connect the mass function to the power spectrum leads to shifts in our results by less than one standard deviation. However, as the sample size of lenses suitable for this analysis grows with forthcoming surveys and the James Webb Space telescope, the constraining power of strong lensing over the primordial matter power spectrum will grow. As the information content of the dataset expands, systematic uncertainties associated with the halo mass function and concentration-mass relation will become comparable with the statistical measurement uncertainties, and we will likely require a refined model for the halo mass function \citep[e.g.][]{Stafford++20,Brown++20,Ondaro-Mallea++21} that is calibrated in the mass and redshift range relevant for a strong lensing analysis, $10^6 - 10^{10} M_{\odot}$ and $z = 0 - 3$, respectively. In addition to the mass function, other potential systematic effects could warrant further study, including the presence of filamentary structure along the line of sight, in addition to halos \citep[e.g.][]{Richardson++21}.
	
	We conclude by revisiting the topic of assuming an analytic model for the power spectrum. Two practical considerations motivated this choice. First, in order to compute the concentration-mass relation and halo mass function, we must be able to integrate and differentiate the power spectrum. A free-form model, in which the amplitude varies freely in separate $k$ bins, would cause problems in the computation of the mass function and concentration-mass relation, unless the $k$ bins become extremely narrow and approach a continuous representation of the power spectrum at different scales. Second, our strategy for constraining $P\left(k\right)$ involves an intermediate measurement of several hyper-parameters $\qsub$, which we must then connect with a separate set of parameters that specify the form of the power spectrum. One advantage of this approach is that we can use different halo mass function and concentration-mass relation models to map $\qsub \rightarrow \qp$. The price we pay for this flexibility is that we must restrict our analysis to models of $P\left(k\right)$ that predict halo mass functions and concentration-mass relations that we can fit using our parameterization in terms of $\qsub$. In other words, we must restrict our analysis to models of $P\left(k\right)$ where the concentration-mass relation is approximately given by a power law in peak height, and the where the halo mass function resembles a re-scaled version of Sheth-Tormen with a different logarithmic slope. An emulator that rapidly maps an arbitrary form for the power spectrum into a halo mass function and concentration-mass relation, effectively performing the forward modeling of observables directly on the level of the power spectrum, would allow us to explore a wider variety of forms for $P\left(k\right)$, facilitating more direct contact between strong lensing data and the properties of the early Universe. 
	
	\section*{Acknowledgments}
	We thank Krishna Choudhary, David Gilman, and Prateek Puri for comments on a draft version of this paper. We also thank an anonymous referee for constructive feedback. 
	
	DG was partially supported by a HQP grant from the McDonald Institute (reference number HQP 2019-4-2). DG and JB acknowledge financial support from NSERC (funding reference number RGPIN-2020-04712). AJB was supported in part by the NASA Astrophysics Theory Program, under grant 80NSSC18K1014. Support for this work was also provided by the National Science Foundation (NSF) through NSF AST-1716527. TT acknowledges support by NSF through grant AST-1714953, and grant AST-1836016, by NASA through and HST-GO-15652, by the Packard Foundation through a Packard Research Fellowship, and by the Moore Foundation through grant 8548. This research is based on observations made with the NASA/ESA Hubble Space Telescope obtained from the Space Telescope Science Institute, which is operated by the Association of Universities for Research in Astronomy, Inc., under NASA contract NAS 5–26555. These observations are associated with programs GO-15177 and GO-13732.
	
	We used three computing clusters to perform the ray-tracing simulations discussed in this paper. First, we used the Niagara supercomputer at the SciNet HPC Consortium with support provided by: the Canada Foundation for Innovation; the Government of Ontario; Ontario Research Fund - Research Excellence; and the University of Toronto. Second, we used computational and storage services associated with the Hoffman2 Shared Cluster provided by the UCLA Institute for Digital Research and Education's Research Technology Group. Third, we used the {\tt memex} compute cluster, a resource provided by the Carnegie Institution for Science.  
	
	\section*{Data Availability}
	Data used in this article came from observing programs HST-GO-13732 and HST-GO-15177. The data and analysis scripts used to prepare this article can be accessed at the following github repository https://github.com/dangilman/lenslikelihood.
	
	\bibliographystyle{mnras}
	\bibliography{bibliography}

\begin{thebibliography}{}
\makeatletter
\relax
\def\mn@urlcharsother{\let\do\@makeother \do\$\do\&\do\#\do\^\do\_\do\%\do\~}
\def\mn@doi{\begingroup\mn@urlcharsother \@ifnextchar [ {\mn@doi@}
  {\mn@doi@[]}}
\def\mn@doi@[#1]#2{\def\@tempa{#1}\ifx\@tempa\@empty \href
  {http://dx.doi.org/#2} {doi:#2}\else \href {http://dx.doi.org/#2} {#1}\fi
  \endgroup}
\def\mn@eprint#1#2{\mn@eprint@#1:#2::\@nil}
\def\mn@eprint@arXiv#1{\href {http://arxiv.org/abs/#1} {{\tt arXiv:#1}}}
\def\mn@eprint@dblp#1{\href {http://dblp.uni-trier.de/rec/bibtex/#1.xml}
  {dblp:#1}}
\def\mn@eprint@#1:#2:#3:#4\@nil{\def\@tempa {#1}\def\@tempb {#2}\def\@tempc
  {#3}\ifx \@tempc \@empty \let \@tempc \@tempb \let \@tempb \@tempa \fi \ifx
  \@tempb \@empty \def\@tempb {arXiv}\fi \@ifundefined
  {mn@eprint@\@tempb}{\@tempb:\@tempc}{\expandafter \expandafter \csname
  mn@eprint@\@tempb\endcsname \expandafter{\@tempc}}}

\bibitem[\protect\citeauthoryear{{Albrecht} \& {Steinhardt}}{{Albrecht} \&
  {Steinhardt}}{1982}]{AlbrechtSteinhardt82}
{Albrecht} A.,  {Steinhardt} P.~J.,  1982, \mn@doi [\prl]
  {10.1103/PhysRevLett.48.1220}, \href
  {https://ui.adsabs.harvard.edu/abs/1982PhRvL..48.1220A} {48, 1220}

\bibitem[\protect\citeauthoryear{{Amorisco} et~al.,}{{Amorisco}
  et~al.}{2021}]{Amorisco++21}
{Amorisco} N.~C.,  et~al., 2021, arXiv e-prints, \href
  {https://ui.adsabs.harvard.edu/abs/2021arXiv210900018A} {p. arXiv:2109.00018}

\bibitem[\protect\citeauthoryear{{Anderson}}{{Anderson}}{2016}]{Anderson++16}
{Anderson} J.,  2016, Technical report, {Empirical Models for the WFC3/IR PSF}

\bibitem[\protect\citeauthoryear{{Astropy Collaboration} et~al.,}{{Astropy
  Collaboration} et~al.}{2018}]{Astropy}
{Astropy Collaboration} et~al., 2018, \mn@doi [\aj] {10.3847/1538-3881/aabc4f},
  \href {https://ui.adsabs.harvard.edu/abs/2018AJ....156..123A} {156, 123}

\bibitem[\protect\citeauthoryear{{Auger}, {Treu}, {Bolton}, {Gavazzi},
  {Koopmans}, {Marshall}, {Moustakas}  \& {Burles}}{{Auger}
  et~al.}{2010}]{Auger++10}
{Auger} M.~W.,  {Treu} T.,  {Bolton} A.~S.,  {Gavazzi} R.,  {Koopmans}
  L.~V.~E.,  {Marshall} P.~J.,  {Moustakas} L.~A.,   {Burles} S.,  2010,
  \mn@doi [\apj] {10.1088/0004-637X/724/1/511}, \href
  {https://ui.adsabs.harvard.edu/abs/2010ApJ...724..511A} {724, 511}

\bibitem[\protect\citeauthoryear{{Baltz}, {Marshall}  \& {Oguri}}{{Baltz}
  et~al.}{2009}]{Baltz++09}
{Baltz} E.~A.,  {Marshall} P.,   {Oguri} M.,  2009, \mn@doi [\jcap]
  {10.1088/1475-7516/2009/01/015}, \href
  {https://ui.adsabs.harvard.edu/abs/2009JCAP...01..015B} {2009, 015}

\bibitem[\protect\citeauthoryear{{Bardeen}, {Steinhardt}  \&
  {Turner}}{{Bardeen} et~al.}{1983}]{Bardeen++83}
{Bardeen} J.~M.,  {Steinhardt} P.~J.,   {Turner} M.~S.,  1983, \mn@doi [\prd]
  {10.1103/PhysRevD.28.679}, \href
  {https://ui.adsabs.harvard.edu/abs/1983PhRvD..28..679B} {28, 679}

\bibitem[\protect\citeauthoryear{{Bender}, {Surma}, {Doebereiner},
  {Moellenhoff}  \& {Madejsky}}{{Bender} et~al.}{1989}]{Bender++89}
{Bender} R.,  {Surma} P.,  {Doebereiner} S.,  {Moellenhoff} C.,   {Madejsky}
  R.,  1989, \aap, \href
  {https://ui.adsabs.harvard.edu/abs/1989A&A...217...35B} {217, 35}

\bibitem[\protect\citeauthoryear{{Benson}}{{Benson}}{2012}]{Benson++12}
{Benson} A.~J.,  2012, \mn@doi [\na] {10.1016/j.newast.2011.07.004}, \href
  {https://ui.adsabs.harvard.edu/abs/2012NewA...17..175B} {17, 175}

\bibitem[\protect\citeauthoryear{{Bhattacharya}, {Heitmann}, {White},
  {Luki{\'c}}, {Wagner}  \& {Habib}}{{Bhattacharya}
  et~al.}{2011}]{Bhattacharya++11}
{Bhattacharya} S.,  {Heitmann} K.,  {White} M.,  {Luki{\'c}} Z.,  {Wagner} C.,
   {Habib} S.,  2011, \mn@doi [\apj] {10.1088/0004-637X/732/2/122}, \href
  {https://ui.adsabs.harvard.edu/abs/2011ApJ...732..122B} {732, 122}

\bibitem[\protect\citeauthoryear{{Birrer} \& {Amara}}{{Birrer} \&
  {Amara}}{2018}]{BirrerAmara++18}
{Birrer} S.,  {Amara} A.,  2018, \mn@doi [Physics of the Dark Universe]
  {10.1016/j.dark.2018.11.002}, \href
  {https://ui.adsabs.harvard.edu/abs/2018PDU....22..189B} {22, 189}

\bibitem[\protect\citeauthoryear{{Birrer}, {Amara}  \& {Refregier}}{{Birrer}
  et~al.}{2017}]{Birrer++17}
{Birrer} S.,  {Amara} A.,   {Refregier} A.,  2017, \mn@doi [\jcap]
  {10.1088/1475-7516/2017/05/037}, \href
  {https://ui.adsabs.harvard.edu/abs/2017JCAP...05..037B} {2017, 037}

\bibitem[\protect\citeauthoryear{{Birrer} et~al.,}{{Birrer}
  et~al.}{2021}]{Birrer++21}
{Birrer} S.,  et~al., 2021, \mn@doi [The Journal of Open Source Software]
  {10.21105/joss.03283}, \href
  {https://ui.adsabs.harvard.edu/abs/2021JOSS....6.3283B} {6, 3283}

\bibitem[\protect\citeauthoryear{{Blandford} \& {Narayan}}{{Blandford} \&
  {Narayan}}{1986}]{BlanfordNarayan86}
{Blandford} R.,  {Narayan} R.,  1986, \mn@doi [\apj] {10.1086/164709}, \href
  {https://ui.adsabs.harvard.edu/abs/1986ApJ...310..568B} {310, 568}

\bibitem[\protect\citeauthoryear{{Bode}, {Ostriker}  \& {Turok}}{{Bode}
  et~al.}{2001}]{Bode++01}
{Bode} P.,  {Ostriker} J.~P.,   {Turok} N.,  2001, \mn@doi [\apj]
  {10.1086/321541}, \href {http://adsabs.harvard.edu/abs/2001ApJ...556...93B}
  {556, 93}

\bibitem[\protect\citeauthoryear{{Bohr}, {Zavala}, {Cyr-Racine}  \&
  {Vogelsberger}}{{Bohr} et~al.}{2021}]{Bohr++21}
{Bohr} S.,  {Zavala} J.,  {Cyr-Racine} F.-Y.,   {Vogelsberger} M.,  2021,
  \mn@doi [\mnras] {10.1093/mnras/stab1758}, \href
  {https://ui.adsabs.harvard.edu/abs/2021MNRAS.506..128B} {506, 128}

\bibitem[\protect\citeauthoryear{{Brown}, {McCarthy}, {Diemer}, {Font},
  {Stafford}  \& {Pfeifer}}{{Brown} et~al.}{2020}]{Brown++20}
{Brown} S.~T.,  {McCarthy} I.~G.,  {Diemer} B.,  {Font} A.~S.,  {Stafford}
  S.~G.,   {Pfeifer} S.,  2020, \mn@doi [\mnras] {10.1093/mnras/staa1491},
  \href {https://ui.adsabs.harvard.edu/abs/2020MNRAS.495.4994B} {495, 4994}

\bibitem[\protect\citeauthoryear{{Bullock}, {Kolatt}, {Sigad}, {Somerville},
  {Kravtsov}, {Klypin}, {Primack}  \& {Dekel}}{{Bullock}
  et~al.}{2001}]{Bullock++01}
{Bullock} J.~S.,  {Kolatt} T.~S.,  {Sigad} Y.,  {Somerville} R.~S.,  {Kravtsov}
  A.~V.,  {Klypin} A.~A.,  {Primack} J.~R.,   {Dekel} A.,  2001, \mn@doi
  [\mnras] {10.1046/j.1365-8711.2001.04068.x}, \href
  {https://ui.adsabs.harvard.edu/abs/2001MNRAS.321..559B} {321, 559}

\bibitem[\protect\citeauthoryear{{Chabanier}, {Millea}  \&
  {Palanque-Delabrouille}}{{Chabanier} et~al.}{2019a}]{Chabanier++19}
{Chabanier} S.,  {Millea} M.,   {Palanque-Delabrouille} N.,  2019a, \mn@doi
  [\mnras] {10.1093/mnras/stz2310}, \href
  {https://ui.adsabs.harvard.edu/abs/2019MNRAS.489.2247C} {489, 2247}

\bibitem[\protect\citeauthoryear{{Chabanier} et~al.,}{{Chabanier}
  et~al.}{2019b}]{Chabanier++2019b}
{Chabanier} S.,  et~al., 2019b, \mn@doi [\jcap]
  {10.1088/1475-7516/2019/07/017}, \href
  {https://ui.adsabs.harvard.edu/abs/2019JCAP...07..017C} {2019, 017}

\bibitem[\protect\citeauthoryear{{Chiba}, {Minezaki}, {Kashikawa}, {Kataza}  \&
  {Inoue}}{{Chiba} et~al.}{2005}]{Chiba++05}
{Chiba} M.,  {Minezaki} T.,  {Kashikawa} N.,  {Kataza} H.,   {Inoue} K.~T.,
  2005, \mn@doi [\apj] {10.1086/430403}, \href
  {https://ui.adsabs.harvard.edu/abs/2005ApJ...627...53C} {627, 53}

\bibitem[\protect\citeauthoryear{{Chluba}, {Erickcek}  \& {Ben-Dayan}}{{Chluba}
  et~al.}{2012}]{Chluba++12}
{Chluba} J.,  {Erickcek} A.~L.,   {Ben-Dayan} I.,  2012, \mn@doi [\apj]
  {10.1088/0004-637X/758/2/76}, \href
  {https://ui.adsabs.harvard.edu/abs/2012ApJ...758...76C} {758, 76}

\bibitem[\protect\citeauthoryear{{Dalal} \& {Kochanek}}{{Dalal} \&
  {Kochanek}}{2002}]{Dalal++02}
{Dalal} N.,  {Kochanek} C.~S.,  2002, \mn@doi [\apj] {10.1086/340303}, \href
  {https://ui.adsabs.harvard.edu/abs/2002ApJ...572...25D} {572, 25}

\bibitem[\protect\citeauthoryear{{Despali}, {Giocoli}, {Angulo}, {Tormen},
  {Sheth}, {Baso}  \& {Moscardini}}{{Despali} et~al.}{2016}]{Despali++16}
{Despali} G.,  {Giocoli} C.,  {Angulo} R.~E.,  {Tormen} G.,  {Sheth} R.~K.,
  {Baso} G.,   {Moscardini} L.,  2016, \mn@doi [\mnras]
  {10.1093/mnras/stv2842}, \href
  {https://ui.adsabs.harvard.edu/abs/2016MNRAS.456.2486D} {456, 2486}

\bibitem[\protect\citeauthoryear{{Diemer}}{{Diemer}}{2018}]{Diemer18}
{Diemer} B.,  2018, \mn@doi [\apjs] {10.3847/1538-4365/aaee8c}, \href
  {https://ui.adsabs.harvard.edu/abs/2018ApJS..239...35D} {239, 35}

\bibitem[\protect\citeauthoryear{{Diemer} \& {Joyce}}{{Diemer} \&
  {Joyce}}{2019}]{DiemerJoyce19}
{Diemer} B.,  {Joyce} M.,  2019, \mn@doi [\apj] {10.3847/1538-4357/aafad6},
  \href {https://ui.adsabs.harvard.edu/abs/2019ApJ...871..168D} {871, 168}

\bibitem[\protect\citeauthoryear{{Diemer}, {More}  \& {Kravtsov}}{{Diemer}
  et~al.}{2013}]{Diemer++13}
{Diemer} B.,  {More} S.,   {Kravtsov} A.~V.,  2013, \mn@doi [\apj]
  {10.1088/0004-637X/766/1/25}, \href
  {https://ui.adsabs.harvard.edu/abs/2013ApJ...766...25D} {766, 25}

\bibitem[\protect\citeauthoryear{{Dobler} \& {Keeton}}{{Dobler} \&
  {Keeton}}{2006}]{Dobler++06}
{Dobler} G.,  {Keeton} C.~R.,  2006, \mn@doi [\mnras]
  {10.1111/j.1365-2966.2005.09809.x}, \href
  {https://ui.adsabs.harvard.edu/abs/2006MNRAS.365.1243D} {365, 1243}

\bibitem[\protect\citeauthoryear{{Eisenstein} \& {Hu}}{{Eisenstein} \&
  {Hu}}{1998}]{EisensteinHu98}
{Eisenstein} D.~J.,  {Hu} W.,  1998, \mn@doi [\apj] {10.1086/305424}, \href
  {https://ui.adsabs.harvard.edu/abs/1998ApJ...496..605E} {496, 605}

\bibitem[\protect\citeauthoryear{{Eke}, {Navarro}  \& {Steinmetz}}{{Eke}
  et~al.}{2001}]{Eke++01}
{Eke} V.~R.,  {Navarro} J.~F.,   {Steinmetz} M.,  2001, \mn@doi [\apj]
  {10.1086/321345}, \href
  {https://ui.adsabs.harvard.edu/abs/2001ApJ...554..114E} {554, 114}

\bibitem[\protect\citeauthoryear{{Falco} et~al.,}{{Falco}
  et~al.}{1999}]{Fal++99}
{Falco} E.~E.,  et~al., 1999, \mn@doi [\apj] {10.1086/307758}, \href
  {http://adsabs.harvard.edu/cgi-bin/nph-bib_query?bibcode=1999ApJ...523..617F&db_key=AST}
  {523, 617}

\bibitem[\protect\citeauthoryear{{Fedeli}, {Finelli}  \& {Moscardini}}{{Fedeli}
  et~al.}{2010}]{Fedeli++10}
{Fedeli} C.,  {Finelli} F.,   {Moscardini} L.,  2010, \mn@doi [\mnras]
  {10.1111/j.1365-2966.2010.17026.x}, \href
  {https://ui.adsabs.harvard.edu/abs/2010MNRAS.407.1842F} {407, 1842}

\bibitem[\protect\citeauthoryear{{Ferrari}, {Pastoriza}, {Macchetto}  \&
  {Caon}}{{Ferrari} et~al.}{1999}]{Ferrari++99}
{Ferrari} F.,  {Pastoriza} M.~G.,  {Macchetto} F.,   {Caon} N.,  1999, \mn@doi
  [\aaps] {10.1051/aas:1999465}, \href
  {https://ui.adsabs.harvard.edu/abs/1999A&AS..136..269F} {136, 269}

\bibitem[\protect\citeauthoryear{{Gelman}, {Carlin}, {Stern}, {Dunson}, {Aki}
  \& {Rubin}}{{Gelman} et~al.}{2013}]{Gelman++13}
{Gelman} A.~R.,  {Carlin} J.~B.,  {Stern} H.~S.,  {Dunson} D.~B.,  {Aki} V.,
  {Rubin} D.~B.,  2013, {Bayesian Data Analysis}.
CRC Press

\bibitem[\protect\citeauthoryear{{Gilman}, {Agnello}, {Treu}, {Keeton}  \&
  {Nierenberg}}{{Gilman} et~al.}{2017}]{Gilman++17}
{Gilman} D.,  {Agnello} A.,  {Treu} T.,  {Keeton} C.~R.,   {Nierenberg} A.~M.,
  2017, \mn@doi [\mnras] {10.1093/mnras/stx158}, \href
  {https://ui.adsabs.harvard.edu/abs/2017MNRAS.467.3970G} {467, 3970}

\bibitem[\protect\citeauthoryear{{Gilman}, {Birrer}, {Treu}, {Keeton}  \&
  {Nierenberg}}{{Gilman} et~al.}{2018}]{Gilman++18}
{Gilman} D.,  {Birrer} S.,  {Treu} T.,  {Keeton} C.~R.,   {Nierenberg} A.,
  2018, \mn@doi [\mnras] {10.1093/mnras/sty2261}, \href
  {https://ui.adsabs.harvard.edu/abs/2018MNRAS.481..819G} {481, 819}

\bibitem[\protect\citeauthoryear{{Gilman}, {Birrer}, {Treu}, {Nierenberg}  \&
  {Benson}}{{Gilman} et~al.}{2019}]{Gilman++19}
{Gilman} D.,  {Birrer} S.,  {Treu} T.,  {Nierenberg} A.,   {Benson} A.,  2019,
  \mn@doi [\mnras] {10.1093/mnras/stz1593}, \href
  {https://ui.adsabs.harvard.edu/abs/2019MNRAS.487.5721G} {487, 5721}

\bibitem[\protect\citeauthoryear{{Gilman}, {Birrer}, {Nierenberg}, {Treu}, {Du}
   \& {Benson}}{{Gilman} et~al.}{2020a}]{Gilman++20a}
{Gilman} D.,  {Birrer} S.,  {Nierenberg} A.,  {Treu} T.,  {Du} X.,   {Benson}
  A.,  2020a, \mn@doi [\mnras] {10.1093/mnras/stz3480}, \href
  {https://ui.adsabs.harvard.edu/abs/2020MNRAS.491.6077G} {491, 6077}

\bibitem[\protect\citeauthoryear{{Gilman}, {Du}, {Benson}, {Birrer},
  {Nierenberg}  \& {Treu}}{{Gilman} et~al.}{2020b}]{Gilman++20b}
{Gilman} D.,  {Du} X.,  {Benson} A.,  {Birrer} S.,  {Nierenberg} A.,   {Treu}
  T.,  2020b, \mn@doi [\mnras] {10.1093/mnrasl/slz173}, \href
  {https://ui.adsabs.harvard.edu/abs/2020MNRAS.492L..12G} {492, L12}

\bibitem[\protect\citeauthoryear{{Gilman}, {Bovy}, {Treu}, {Nierenberg},
  {Birrer}, {Benson}  \& {Sameie}}{{Gilman} et~al.}{2021}]{Gilman++21}
{Gilman} D.,  {Bovy} J.,  {Treu} T.,  {Nierenberg} A.,  {Birrer} S.,  {Benson}
  A.,   {Sameie} O.,  2021, \mn@doi [\mnras] {10.1093/mnras/stab2335}, \href
  {https://ui.adsabs.harvard.edu/abs/2021MNRAS.507.2432G} {507, 2432}

\bibitem[\protect\citeauthoryear{{Green} \& {van den Bosch}}{{Green} \& {van
  den Bosch}}{2019}]{GreenvandenBosch19}
{Green} S.~B.,  {van den Bosch} F.~C.,  2019, \mn@doi [\mnras]
  {10.1093/mnras/stz2767}, \href
  {https://ui.adsabs.harvard.edu/abs/2019MNRAS.490.2091G} {490, 2091}

\bibitem[\protect\citeauthoryear{{Guth}}{{Guth}}{1981}]{Guth81}
{Guth} A.~H.,  1981, \mn@doi [\prd] {10.1103/PhysRevD.23.347}, \href
  {https://ui.adsabs.harvard.edu/abs/1981PhRvD..23..347G} {23, 347}

\bibitem[\protect\citeauthoryear{{Guth} \& {Pi}}{{Guth} \&
  {Pi}}{1982}]{GuthPi82}
{Guth} A.~H.,  {Pi} S.~Y.,  1982, \mn@doi [\prl] {10.1103/PhysRevLett.49.1110},
  \href {https://ui.adsabs.harvard.edu/abs/1982PhRvL..49.1110G} {49, 1110}

\bibitem[\protect\citeauthoryear{{Hezaveh} et~al.,}{{Hezaveh}
  et~al.}{2016}]{Hezaveh++16}
{Hezaveh} Y.~D.,  et~al., 2016, \mn@doi [\apj] {10.3847/0004-637X/823/1/37},
  \href {https://ui.adsabs.harvard.edu/abs/2016ApJ...823...37H} {823, 37}

\bibitem[\protect\citeauthoryear{{Hinshaw} et~al.,}{{Hinshaw}
  et~al.}{2013}]{Hinshaw++13}
{Hinshaw} G.,  et~al., 2013, \mn@doi [\apjs] {10.1088/0067-0049/208/2/19},
  \href {https://ui.adsabs.harvard.edu/abs/2013ApJS..208...19H} {208, 19}

\bibitem[\protect\citeauthoryear{{Hsueh}, {Fassnacht}, {Vegetti}, {McKean},
  {Spingola}, {Auger}, {Koopmans}  \& {Lagattuta}}{{Hsueh}
  et~al.}{2016}]{Hsueh++16}
{Hsueh} J.~W.,  {Fassnacht} C.~D.,  {Vegetti} S.,  {McKean} J.~P.,  {Spingola}
  C.,  {Auger} M.~W.,  {Koopmans} L.~V.~E.,   {Lagattuta} D.~J.,  2016, \mn@doi
  [\mnras] {10.1093/mnrasl/slw146}, \href
  {https://ui.adsabs.harvard.edu/abs/2016MNRAS.463L..51H} {463, L51}

\bibitem[\protect\citeauthoryear{{Hsueh} et~al.,}{{Hsueh}
  et~al.}{2017}]{Hsueh++17}
{Hsueh} J.~W.,  et~al., 2017, \mn@doi [\mnras] {10.1093/mnras/stx1082}, \href
  {https://ui.adsabs.harvard.edu/abs/2017MNRAS.469.3713H} {469, 3713}

\bibitem[\protect\citeauthoryear{{Hsueh}, {Despali}, {Vegetti}, {Xu},
  {Fassnacht}  \& {Metcalf}}{{Hsueh} et~al.}{2018}]{Hsueh++18}
{Hsueh} J.-W.,  {Despali} G.,  {Vegetti} S.,  {Xu} D.,  {Fassnacht} C.~D.,
  {Metcalf} R.~B.,  2018, \mn@doi [\mnras] {10.1093/mnras/stx3320}, \href
  {https://ui.adsabs.harvard.edu/abs/2018MNRAS.475.2438H} {475, 2438}

\bibitem[\protect\citeauthoryear{{Hsueh}, {Enzi}, {Vegetti}, {Auger},
  {Fassnacht}, {Despali}, {Koopmans}  \& {McKean}}{{Hsueh}
  et~al.}{2020}]{Hsueh++20}
{Hsueh} J.~W.,  {Enzi} W.,  {Vegetti} S.,  {Auger} M.~W.,  {Fassnacht} C.~D.,
  {Despali} G.,  {Koopmans} L.~V.~E.,   {McKean} J.~P.,  2020, \mn@doi [\mnras]
  {10.1093/mnras/stz3177}, \href
  {https://ui.adsabs.harvard.edu/abs/2020MNRAS.492.3047H} {492, 3047}

\bibitem[\protect\citeauthoryear{{Inoue} \& {Chiba}}{{Inoue} \&
  {Chiba}}{2005}]{Inoue++05}
{Inoue} K.~T.,  {Chiba} M.,  2005, \mn@doi [\apj] {10.1086/496870}, \href
  {https://ui.adsabs.harvard.edu/abs/2005ApJ...634...77I} {634, 77}

\bibitem[\protect\citeauthoryear{{Inoue}, {Takahashi}, {Takahashi}  \&
  {Ishiyama}}{{Inoue} et~al.}{2015}]{Inoue++15}
{Inoue} K.~T.,  {Takahashi} R.,  {Takahashi} T.,   {Ishiyama} T.,  2015,
  \mn@doi [\mnras] {10.1093/mnras/stv194}, \href
  {https://ui.adsabs.harvard.edu/abs/2015MNRAS.448.2704I} {448, 2704}

\bibitem[\protect\citeauthoryear{{Ir{\v{s}}i{\v{c}}}
  et~al.,}{{Ir{\v{s}}i{\v{c}}} et~al.}{2017}]{Irsic++17}
{Ir{\v{s}}i{\v{c}}} V.,  et~al., 2017, \mn@doi [\prd]
  {10.1103/PhysRevD.96.023522}, \href
  {https://ui.adsabs.harvard.edu/abs/2017PhRvD..96b3522I} {96, 023522}

\bibitem[\protect\citeauthoryear{{Jiang} \& {van den Bosch}}{{Jiang} \& {van
  den Bosch}}{2017}]{Jiang++17}
{Jiang} F.,  {van den Bosch} F.~C.,  2017, \mn@doi [\mnras]
  {10.1093/mnras/stx1979}, \href
  {https://ui.adsabs.harvard.edu/abs/2017MNRAS.472..657J} {472, 657}

\bibitem[\protect\citeauthoryear{{Lazar}, {Bullock}, {Boylan-Kolchin},
  {Feldmann}, {{\c{C}}atmabacak}  \& {Moustakas}}{{Lazar}
  et~al.}{2021}]{Lazar++21}
{Lazar} A.,  {Bullock} J.~S.,  {Boylan-Kolchin} M.,  {Feldmann} R.,
  {{\c{C}}atmabacak} O.,   {Moustakas} L.,  2021, \mn@doi [\mnras]
  {10.1093/mnras/stab448}, \href
  {https://ui.adsabs.harvard.edu/abs/2021MNRAS.502.6064L} {502, 6064}

\bibitem[\protect\citeauthoryear{{Liddle} \& {Lyth}}{{Liddle} \&
  {Lyth}}{2000}]{LiddleLyth2000}
{Liddle} A.~R.,  {Lyth} D.~H.,  2000, {Cosmological Inflation and Large-Scale
  Structure}

\bibitem[\protect\citeauthoryear{{Linde}}{{Linde}}{1982}]{Linde82}
{Linde} A.~D.,  1982, \mn@doi [Physics Letters B]
  {10.1016/0370-2693(82)91219-9}, \href
  {https://ui.adsabs.harvard.edu/abs/1982PhLB..108..389L} {108, 389}

\bibitem[\protect\citeauthoryear{{Minor}, {Kaplinghat}, {Chan}  \&
  {Simon}}{{Minor} et~al.}{2021}]{Minor++21}
{Minor} Q.,  {Kaplinghat} M.,  {Chan} T.~H.,   {Simon} E.,  2021, \mn@doi
  [\mnras] {10.1093/mnras/stab2209}, \href
  {https://ui.adsabs.harvard.edu/abs/2021MNRAS.507.1202M} {507, 1202}

\bibitem[\protect\citeauthoryear{{More}, {Diemer}  \& {Kravtsov}}{{More}
  et~al.}{2015}]{More++15}
{More} S.,  {Diemer} B.,   {Kravtsov} A.~V.,  2015, \mn@doi [\apj]
  {10.1088/0004-637X/810/1/36}, \href
  {https://ui.adsabs.harvard.edu/abs/2015ApJ...810...36M} {810, 36}

\bibitem[\protect\citeauthoryear{{M{\"u}ller-S{\'a}nchez}, {Prieto}, {Hicks},
  {Vives-Arias}, {Davies}, {Malkan}, {Tacconi}  \&
  {Genzel}}{{M{\"u}ller-S{\'a}nchez} et~al.}{2011}]{MullerSanchez++11}
{M{\"u}ller-S{\'a}nchez} F.,  {Prieto} M.~A.,  {Hicks} E.~K.~S.,  {Vives-Arias}
  H.,  {Davies} R.~I.,  {Malkan} M.,  {Tacconi} L.~J.,   {Genzel} R.,  2011,
  \mn@doi [\apj] {10.1088/0004-637X/739/2/69}, \href
  {https://ui.adsabs.harvard.edu/abs/2011ApJ...739...69M} {739, 69}

\bibitem[\protect\citeauthoryear{{Nadler}, {Birrer}, {Gilman}, {Wechsler},
  {Du}, {Benson}, {Nierenberg}  \& {Treu}}{{Nadler} et~al.}{2021}]{Nadler++21}
{Nadler} E.~O.,  {Birrer} S.,  {Gilman} D.,  {Wechsler} R.~H.,  {Du} X.,
  {Benson} A.,  {Nierenberg} A.~M.,   {Treu} T.,  2021, \mn@doi [\apj]
  {10.3847/1538-4357/abf9a3}, \href
  {https://ui.adsabs.harvard.edu/abs/2021ApJ...917....7N} {917, 7}

\bibitem[\protect\citeauthoryear{{Navarro}, {Frenk}  \& {White}}{{Navarro}
  et~al.}{1997}]{Navarro++97}
{Navarro} J.~F.,  {Frenk} C.~S.,   {White} S. D.~M.,  1997, \mn@doi [\apj]
  {10.1086/304888}, \href
  {https://ui.adsabs.harvard.edu/abs/1997ApJ...490..493N} {490, 493}

\bibitem[\protect\citeauthoryear{{Nierenberg}, {Treu}, {Wright}, {Fassnacht}
  \& {Auger}}{{Nierenberg} et~al.}{2014}]{Nierenberg++14}
{Nierenberg} A.~M.,  {Treu} T.,  {Wright} S.~A.,  {Fassnacht} C.~D.,   {Auger}
  M.~W.,  2014, \mn@doi [\mnras] {10.1093/mnras/stu862}, \href
  {https://ui.adsabs.harvard.edu/abs/2014MNRAS.442.2434N} {442, 2434}

\bibitem[\protect\citeauthoryear{{Nierenberg} et~al.,}{{Nierenberg}
  et~al.}{2017}]{Nierenberg++17}
{Nierenberg} A.~M.,  et~al., 2017, \mn@doi [\mnras] {10.1093/mnras/stx1400},
  \href {https://ui.adsabs.harvard.edu/abs/2017MNRAS.471.2224N} {471, 2224}

\bibitem[\protect\citeauthoryear{{Nierenberg} et~al.,}{{Nierenberg}
  et~al.}{2020}]{Nierenberg++21}
{Nierenberg} A.~M.,  et~al., 2020, \mn@doi [\mnras] {10.1093/mnras/stz3588},
  \href {https://ui.adsabs.harvard.edu/abs/2020MNRAS.492.5314N} {492, 5314}

\bibitem[\protect\citeauthoryear{{Ondaro-Mallea}, {Angulo}, {Zennaro},
  {Contreras}  \& {Aric{\`o}}}{{Ondaro-Mallea}
  et~al.}{2021}]{Ondaro-Mallea++21}
{Ondaro-Mallea} L.,  {Angulo} R.~E.,  {Zennaro} M.,  {Contreras} S.,
  {Aric{\`o}} G.,  2021, arXiv e-prints, \href
  {https://ui.adsabs.harvard.edu/abs/2021arXiv210208958O} {p. arXiv:2102.08958}

\bibitem[\protect\citeauthoryear{{Planck Collaboration} et~al.,}{{Planck
  Collaboration} et~al.}{2014}]{Planck2014}
{Planck Collaboration} et~al., 2014, \mn@doi [\aap]
  {10.1051/0004-6361/201321591}, \href
  {https://ui.adsabs.harvard.edu/abs/2014A&A...571A..16P} {571, A16}

\bibitem[\protect\citeauthoryear{{Planck Collaboration} et~al.,}{{Planck
  Collaboration} et~al.}{2020a}]{Planck2020a}
{Planck Collaboration} et~al., 2020a, \mn@doi [\aap]
  {10.1051/0004-6361/201833880}, \href
  {https://ui.adsabs.harvard.edu/abs/2020A&A...641A...1P} {641, A1}

\bibitem[\protect\citeauthoryear{{Planck Collaboration} et~al.,}{{Planck
  Collaboration} et~al.}{2020b}]{Planck2020params}
{Planck Collaboration} et~al., 2020b, \mn@doi [\aap]
  {10.1051/0004-6361/201833910}, \href
  {https://ui.adsabs.harvard.edu/abs/2020A&A...641A...6P} {641, A6}

\bibitem[\protect\citeauthoryear{{Planck Collaboration} et~al.,}{{Planck
  Collaboration} et~al.}{2020c}]{Planck2020b}
{Planck Collaboration} et~al., 2020c, \mn@doi [\aap]
  {10.1051/0004-6361/201833886}, \href
  {https://ui.adsabs.harvard.edu/abs/2020A&A...641A...8P} {641, A8}

\bibitem[\protect\citeauthoryear{{Planck Collaboration} et~al.,}{{Planck
  Collaboration} et~al.}{2020d}]{Planck2020}
{Planck Collaboration} et~al., 2020d, \mn@doi [\aap]
  {10.1051/0004-6361/201833887}, \href
  {https://ui.adsabs.harvard.edu/abs/2020A&A...641A..10P} {641, A10}

\bibitem[\protect\citeauthoryear{{Prada}, {Klypin}, {Cuesta}, {Betancort-Rijo}
  \& {Primack}}{{Prada} et~al.}{2012}]{Prada++12}
{Prada} F.,  {Klypin} A.~A.,  {Cuesta} A.~J.,  {Betancort-Rijo} J.~E.,
  {Primack} J.,  2012, \mn@doi [\mnras] {10.1111/j.1365-2966.2012.21007.x},
  \href {https://ui.adsabs.harvard.edu/abs/2012MNRAS.423.3018P} {423, 3018}

\bibitem[\protect\citeauthoryear{{Press} \& {Schechter}}{{Press} \&
  {Schechter}}{1974}]{PressSchechter++74}
{Press} W.~H.,  {Schechter} P.,  1974, \mn@doi [\apj] {10.1086/152650}, \href
  {https://ui.adsabs.harvard.edu/abs/1974ApJ...187..425P} {187, 425}

\bibitem[\protect\citeauthoryear{{Reid} et~al.,}{{Reid}
  et~al.}{2010}]{Reid++10}
{Reid} B.~A.,  et~al., 2010, \mn@doi [\mnras]
  {10.1111/j.1365-2966.2010.16276.x}, \href
  {https://ui.adsabs.harvard.edu/abs/2010MNRAS.404...60R} {404, 60}

\bibitem[\protect\citeauthoryear{{Richardson}, {St{\"u}cker}, {Angulo}  \&
  {Hahn}}{{Richardson} et~al.}{2021}]{Richardson++21}
{Richardson} T.~R.~G.,  {St{\"u}cker} J.,  {Angulo} R.~E.,   {Hahn} O.,  2021,
  arXiv e-prints, \href {https://ui.adsabs.harvard.edu/abs/2021arXiv210107806R}
  {p. arXiv:2101.07806}

\bibitem[\protect\citeauthoryear{{Rodr{\'\i}guez-Puebla}, {Behroozi},
  {Primack}, {Klypin}, {Lee}  \& {Hellinger}}{{Rodr{\'\i}guez-Puebla}
  et~al.}{2016}]{Rodriguez-Puebla++16}
{Rodr{\'\i}guez-Puebla} A.,  {Behroozi} P.,  {Primack} J.,  {Klypin} A.,  {Lee}
  C.,   {Hellinger} D.,  2016, \mn@doi [\mnras] {10.1093/mnras/stw1705}, \href
  {https://ui.adsabs.harvard.edu/abs/2016MNRAS.462..893R} {462, 893}

\bibitem[\protect\citeauthoryear{{Rogers} \& {Peiris}}{{Rogers} \&
  {Peiris}}{2021}]{Rogers++21}
{Rogers} K.~K.,  {Peiris} H.~V.,  2021, \mn@doi [\prl]
  {10.1103/PhysRevLett.126.071302}, \href
  {https://ui.adsabs.harvard.edu/abs/2021PhRvL.126g1302R} {126, 071302}

\bibitem[\protect\citeauthoryear{Rubin}{Rubin}{1984}]{Rubin1984}
Rubin D.~B.,  1984, The Annals of Statistics, 12, 1151

\bibitem[\protect\citeauthoryear{{Sabti}, {Mu{\~n}oz}  \& {Blas}}{{Sabti}
  et~al.}{2021}]{Sabti++21}
{Sabti} N.,  {Mu{\~n}oz} J.~B.,   {Blas} D.,  2021, arXiv e-prints, \href
  {https://ui.adsabs.harvard.edu/abs/2021arXiv211013161S} {p. arXiv:2110.13161}

\bibitem[\protect\citeauthoryear{{Schneider}, {Smith}, {Macci{\`o}}  \&
  {Moore}}{{Schneider} et~al.}{2012}]{Schneider++12}
{Schneider} A.,  {Smith} R.~E.,  {Macci{\`o}} A.~V.,   {Moore} B.,  2012,
  \mn@doi [\mnras] {10.1111/j.1365-2966.2012.21252.x}, \href
  {http://adsabs.harvard.edu/abs/2012MNRAS.424..684S} {424, 684}

\bibitem[\protect\citeauthoryear{{Sheth}, {Mo}  \& {Tormen}}{{Sheth}
  et~al.}{2001}]{ST99}
{Sheth} R.~K.,  {Mo} H.~J.,   {Tormen} G.,  2001, \mn@doi [\mnras]
  {10.1046/j.1365-8711.2001.04006.x}, \href
  {http://adsabs.harvard.edu/abs/2001MNRAS.323....1S} {323, 1}

\bibitem[\protect\citeauthoryear{{Springel} et~al.,}{{Springel}
  et~al.}{2008}]{Springel++08}
{Springel} V.,  et~al., 2008, \mn@doi [\mnras]
  {10.1111/j.1365-2966.2008.14066.x}, \href
  {https://ui.adsabs.harvard.edu/abs/2008MNRAS.391.1685S} {391, 1685}

\bibitem[\protect\citeauthoryear{{Stacey} \& {McKean}}{{Stacey} \&
  {McKean}}{2018}]{Stacey++18}
{Stacey} H.~R.,  {McKean} J.~P.,  2018, \mn@doi [\mnras]
  {10.1093/mnrasl/sly153}, \href
  {https://ui.adsabs.harvard.edu/abs/2018MNRAS.481L..40S} {481, L40}

\bibitem[\protect\citeauthoryear{{Stacey}, {Lafontaine}  \& {McKean}}{{Stacey}
  et~al.}{2020}]{Stacey++20}
{Stacey} H.~R.,  {Lafontaine} A.,   {McKean} J.~P.,  2020, \mn@doi [\mnras]
  {10.1093/mnras/staa494}, \href
  {https://ui.adsabs.harvard.edu/abs/2020MNRAS.493.5290S} {493, 5290}

\bibitem[\protect\citeauthoryear{{Stafford}, {Brown}, {McCarthy}, {Font},
  {Robertson}  \& {Poole-McKenzie}}{{Stafford} et~al.}{2020}]{Stafford++20}
{Stafford} S.~G.,  {Brown} S.~T.,  {McCarthy} I.~G.,  {Font} A.~S.,
  {Robertson} A.,   {Poole-McKenzie} R.,  2020, \mn@doi [\mnras]
  {10.1093/mnras/staa2059}, \href
  {https://ui.adsabs.harvard.edu/abs/2020MNRAS.497.3809S} {497, 3809}

\bibitem[\protect\citeauthoryear{{Starobinsky}}{{Starobinsky}}{1982}]{Starobinsky82}
{Starobinsky} A.~A.,  1982, \mn@doi [Physics Letters B]
  {10.1016/0370-2693(82)90541-X}, \href
  {https://ui.adsabs.harvard.edu/abs/1982PhLB..117..175S} {117, 175}

\bibitem[\protect\citeauthoryear{{Steinhardt} \& {Turner}}{{Steinhardt} \&
  {Turner}}{1984}]{SteinhardtTurner84}
{Steinhardt} P.~J.,  {Turner} M.~S.,  1984, \mn@doi [\prd]
  {10.1103/PhysRevD.29.2162}, \href
  {https://ui.adsabs.harvard.edu/abs/1984PhRvD..29.2162S} {29, 2162}

\bibitem[\protect\citeauthoryear{{Sugai}, {Kawai}, {Shimono}, {Hattori},
  {Kosugi}, {Kashikawa}, {Inoue}  \& {Chiba}}{{Sugai} et~al.}{2007}]{Sugai++07}
{Sugai} H.,  {Kawai} A.,  {Shimono} A.,  {Hattori} T.,  {Kosugi} G.,
  {Kashikawa} N.,  {Inoue} K.~T.,   {Chiba} M.,  2007, \mn@doi [\apj]
  {10.1086/513731}, \href
  {https://ui.adsabs.harvard.edu/abs/2007ApJ...660.1016S} {660, 1016}

\bibitem[\protect\citeauthoryear{{Tinker}, {Kravtsov}, {Klypin}, {Abazajian},
  {Warren}, {Yepes}, {Gottl{\"o}ber}  \& {Holz}}{{Tinker}
  et~al.}{2008}]{Tinker++08}
{Tinker} J.,  {Kravtsov} A.~V.,  {Klypin} A.,  {Abazajian} K.,  {Warren} M.,
  {Yepes} G.,  {Gottl{\"o}ber} S.,   {Holz} D.~E.,  2008, \mn@doi [\apj]
  {10.1086/591439}, \href
  {https://ui.adsabs.harvard.edu/abs/2008ApJ...688..709T} {688, 709}

\bibitem[\protect\citeauthoryear{{Troxel} et~al.,}{{Troxel}
  et~al.}{2018}]{Troxel++2018}
{Troxel} M.~A.,  et~al., 2018, \mn@doi [\prd] {10.1103/PhysRevD.98.043528},
  \href {https://ui.adsabs.harvard.edu/abs/2018PhRvD..98d3528T} {98, 043528}

\bibitem[\protect\citeauthoryear{{Vegetti}, {Koopmans}, {Auger}, {Treu}  \&
  {Bolton}}{{Vegetti} et~al.}{2014}]{Vegetti++14}
{Vegetti} S.,  {Koopmans} L.~V.~E.,  {Auger} M.~W.,  {Treu} T.,   {Bolton}
  A.~S.,  2014, \mn@doi [\mnras] {10.1093/mnras/stu943}, \href
  {https://ui.adsabs.harvard.edu/abs/2014MNRAS.442.2017V} {442, 2017}

\bibitem[\protect\citeauthoryear{{Vegetti}, {Despali}, {Lovell}  \&
  {Enzi}}{{Vegetti} et~al.}{2018}]{Vegetti++18}
{Vegetti} S.,  {Despali} G.,  {Lovell} M.~R.,   {Enzi} W.,  2018, \mn@doi
  [\mnras] {10.1093/mnras/sty2393}, \href
  {https://ui.adsabs.harvard.edu/abs/2018MNRAS.481.3661V} {481, 3661}

\bibitem[\protect\citeauthoryear{{Viel}, {Weller}  \& {Haehnelt}}{{Viel}
  et~al.}{2004}]{Viel++04}
{Viel} M.,  {Weller} J.,   {Haehnelt} M.~G.,  2004, \mn@doi [\mnras]
  {10.1111/j.1365-2966.2004.08498.x}, \href
  {https://ui.adsabs.harvard.edu/abs/2004MNRAS.355L..23V} {355, L23}

\bibitem[\protect\citeauthoryear{{Viel}, {Becker}, {Bolton}  \&
  {Haehnelt}}{{Viel} et~al.}{2013}]{Viel++13}
{Viel} M.,  {Becker} G.~D.,  {Bolton} J.~S.,   {Haehnelt} M.~G.,  2013, \mn@doi
  [\prd] {10.1103/PhysRevD.88.043502}, \href
  {https://ui.adsabs.harvard.edu/abs/2013PhRvD..88d3502V} {88, 043502}

\bibitem[\protect\citeauthoryear{{Vogelsberger}, {Zavala}, {Cyr-Racine},
  {Pfrommer}, {Bringmann}  \& {Sigurdson}}{{Vogelsberger}
  et~al.}{2016}]{Vogelsberger++16}
{Vogelsberger} M.,  {Zavala} J.,  {Cyr-Racine} F.-Y.,  {Pfrommer} C.,
  {Bringmann} T.,   {Sigurdson} K.,  2016, \mn@doi [\mnras]
  {10.1093/mnras/stw1076}, \href
  {http://adsabs.harvard.edu/abs/2016MNRAS.460.1399V} {460, 1399}

\bibitem[\protect\citeauthoryear{{Wechsler}, {Bullock}, {Primack}, {Kravtsov}
  \& {Dekel}}{{Wechsler} et~al.}{2002}]{Wechsler++02}
{Wechsler} R.~H.,  {Bullock} J.~S.,  {Primack} J.~R.,  {Kravtsov} A.~V.,
  {Dekel} A.,  2002, \mn@doi [\apj] {10.1086/338765}, \href
  {https://ui.adsabs.harvard.edu/abs/2002ApJ...568...52W} {568, 52}

\bibitem[\protect\citeauthoryear{{{\c{C}}a{\u{g}}an {\c{S}}eng{\"u}l},
  {Dvorkin}, {Ostdiek}  \& {Tsang}}{{{\c{C}}a{\u{g}}an {\c{S}}eng{\"u}l}
  et~al.}{2021}]{Sengul++21}
{{\c{C}}a{\u{g}}an {\c{S}}eng{\"u}l} A.,  {Dvorkin} C.,  {Ostdiek} B.,
  {Tsang} A.,  2021, arXiv e-prints, \href
  {https://ui.adsabs.harvard.edu/abs/2021arXiv211200749C} {p. arXiv:2112.00749}

\makeatother
\end{thebibliography}
	
	\appendix
	
	\section{Quantifying the effects of systematic modeling uncertainties}
	\label{app:systematics}
	In this appendix, we discuss three sources of potential systematic errors in our analysis related to the theoretical models we implement for the mass function and concentration-mass relation, the process of mapping the theoretical prediction onto the parameters we inferred from the lenses $\qsub$, and finally, the overall normalization of the power spectrum determined by anchoring the power spectrum to large-scale measurements.
	
	\subsection{Systematic errors in the theoretical predictions}
	 If the models we use for the mass function and concentration-mass relation break down for certain combinations of $n_s$, $a_{\rm{run}}$ and $b_{\rm{run}}$, then the predicted mass functions and concentration-mass relations will not have the correct dependence on the power spectrum parameters. To quantify the extent to which this occurs, if it occurs at all, will require targeted N-body simulations, and possibility a refinement of the model for halo mass function and the concentration-mass relation. Further investigation of how changes to the power spectrum affect dark matter structure, similar to the analyses carried out by \citep{Brown++20,Ondaro-Mallea++21}, can help address this issue, and potentially improve on the results we present.  
	 
	\subsection{Systematic errors in the mapping $\qp \rightarrow \qsub$}
	The second source of systematic uncertainty relates to our strategy for connecting the parameters $\delta_{\rm{LOS}}$, $\beta$, $c_8$, and $\Delta \alpha$ to the power spectrum. If, for example, the concentration-mass relation deviates from a power-law in the peak height, or if the halo mass function deviates from a re-scaled version of the Sheth-Tormen mass function with a modified logartihmic slope, then our parameterization in terms of $\beta$ and $\Delta \alpha$ would not exactly describe the populations of dark matter halos that would exist for a given $n_s$, $a_{\rm{run}}$ and $b_{\rm{run}}$. The lower right panel of Figure \ref{fig:mcrel} shows an example of this, as the theoretical prediction for the concentration-mass relation with $b_{\rm{run}} = 0.01$ and $a_{\rm{run}} = 0.14$ curves away from the model (solid curve) with a positive second derivative. 
	
	To account for uncertainties related to the logarithmic slopes in different mass ranges, we re-calibrate the mapping between $\qsub$ and $\qp$ by fitting the mass function and concentration-mass relation in different mass ranges around the baseline range of $10^7 - 10^9 M_{\odot}$\footnote{We identify this mass range as the most relevant for lensing analyses with flux ratios given the sensitivity of the data determined by the background source size, and the masses and abundance of halos.}. By computing best fit values of $\beta$ and $\Delta \alpha$ in the range $10^7 - 10^8 M_{\odot}$ and $10^8 - 10^{9} M_{\odot}$, we can estimate how much our results depend on the deviation of the logarithmic slope of the mass function and concentration-mass relation around the functional form assumed in our model. We define the systematic errors in $\beta$ and $\Delta \alpha$, $\delta F_{\beta}$ and $\delta F_{\Delta \alpha}$, respectively. We compute these quantities as a function of $n_s$, $a_{\rm{run}}$, and $b_{\rm{run}}$, so that we can estimate the systematic error for any point in the space of $\qp$ parameters. 
	
	By construction, the normalization parameters $c_8$ and $\delta_{\rm{LOS}}$ are defined at a particular pivot scale of $10^8 M_{\odot}$, and therefore do not depend on the mass range where we establish the mapping from $\qp$ to $\qsub$. However, as we have modeled the mass function in the lensing analysis relative to Sheth-Tormen with $n_s = 0.9645$, $a_{\rm{run}} = 0$, and $b_{\rm{run}} = 0$, we do not explicitly account for different redshift evolution of the mass function with the $\qp$ parameters. Similarly, the redshift evolution of the concentration-mass relation may also depend on $\qp$ in a manner that we do explicitly include in the lensing measurement. 
	
	To assess the importance of these effects, we can estimate the systematic error associated with the redshift evolution of the halo mass function $n\left(m, z, n_s, a_{\rm{run}}, b_{\rm{run}}\right)$ by computing the absolute error in the redshift evolution as a function of the power spectrum parameters
	
	\begin{ceqn}
	\begin{equation} 
	\delta F_{\rm{\delta_{\rm{LOS}}}} \equiv \frac{n\left(10^8, z, n_s, a_{\rm{run}}, b_{\rm{run}}\right)}{n\left(10^8, 0, n_s, a_{\rm{run}}, b_{\rm{run}}\right)} - \frac{n\left(10^8, z, 0.9645, 0, 0\right)}{n\left(10^8, 0, 0.9645, 0, 0\right)}.
	\end{equation}
	\end{ceqn}
	We use an analogous expression for the redshift evolution of the concentration-mass relation $c\left(m, z, n_s, a_{\rm{run}}, b_{\rm{run}}\right)$ to compute $\delta F_{c_8}$. Given that redshift range relevant for strong lensing is approximately $z = 0 - 2$, we evaluate the expressions for $\delta F_{c_8}$ and $\delta F_{\delta_{\rm{LOS}}}$ at $z = 1$. 
	
\begin{figure}
		\includegraphics[clip,trim=0cm 0cm 0cm
		0cm,width=.45\textwidth,keepaspectratio,angle=0]{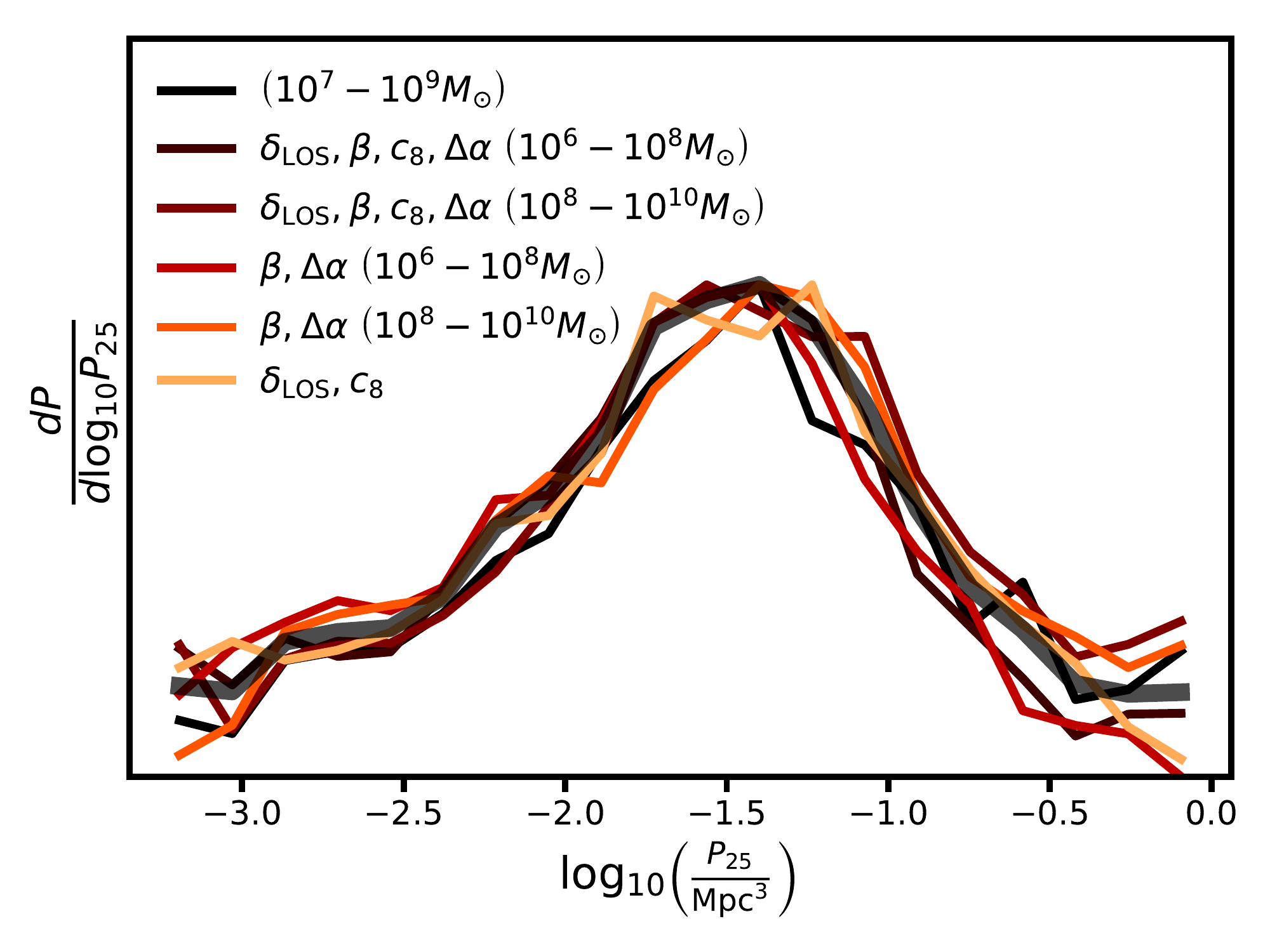}
		{\caption{\label{fig:systematiclikelihood}} An illustration of how systematic modeling uncertainties associated with the mapping between $\qp$ and $\qsub$ parameters affect our main results. Each colored curve depicts an inference of $P\left(k\right)$ at $25 \ \rm{Mpc^{-1}}$ using an altered mapping $\qsub \rightarrow \qp + \delta \qp$, where $\bf{{\delta}} \qp$ includes errors associated with the amplitudes and logarithmic slopes of the mass function and concentration-mass relation. The mass range in parenthesis indicates where we compute the mapping $\qp \rightarrow \qsub$, which can impact the logarithmic slopes predicted by a particular set of $n_s$, $a_{\rm{run}}$, and $b_{\rm{run}}$. We estimate the errors in the mass function amplitudes by performing the mapping to $c_8$ and $\delta_{\rm{LOS}}$ at $z=1$. The parameters listed in front of the mass range in parenthesis indicate which systematic errors are included in each curve. The medians and confidence intervals we present in Section \ref{sec:conclusions} are computed by averaging the inferences together, shown in the figure as a thick black curve.}
	\end{figure}

	Figure \ref{fig:systematiclikelihood} illustrates the effect of the aforementioned systematic errors in our analysis. The figure shows the inference of the power spectrum amplitude evaluated at $25 \ \rm{Mpc^{-1}}$ (the exact $k$ value where we compute the power is irrelevant for this discussion) for different perturbed mappings $\qp \rightarrow \qsub + \boldsymbol{\Delta \qsub}$, where $\boldsymbol{\Delta \qsub} \equiv \left(\delta F_{\delta \rm{LOS}}, \delta F_{c_8}, \delta F_{\Delta \alpha}, \delta F_{\beta}\right)$. Each color denotes a different combination of systematics. Our results do not change significantly for two reasons: First, changes to the power spectrum result in order of magnitude changes in $c_8$, and much larger changes in $\beta$ and $\Delta \alpha$ than those associated with the systematic uncertainties in our model. Second, due to the relatively small sample size of eleven lenses, our statistical measurement uncertainties are large enough that we cannot constrain the model parameters at the level of the systematic errors. To obtain our main results, and when quoting an inference of $P\left(k\right)$ at any scale, we have taken the average of the six probability distributions shown in Figure \ref{fig:systematiclikelihood}.
	
	\subsection{Effect of the amplitude of the power spectrum on large scales} 
	
\begin{figure}
	\includegraphics[clip,trim=0cm 0cm 0cm
	0cm,width=.45\textwidth,keepaspectratio,angle=0]{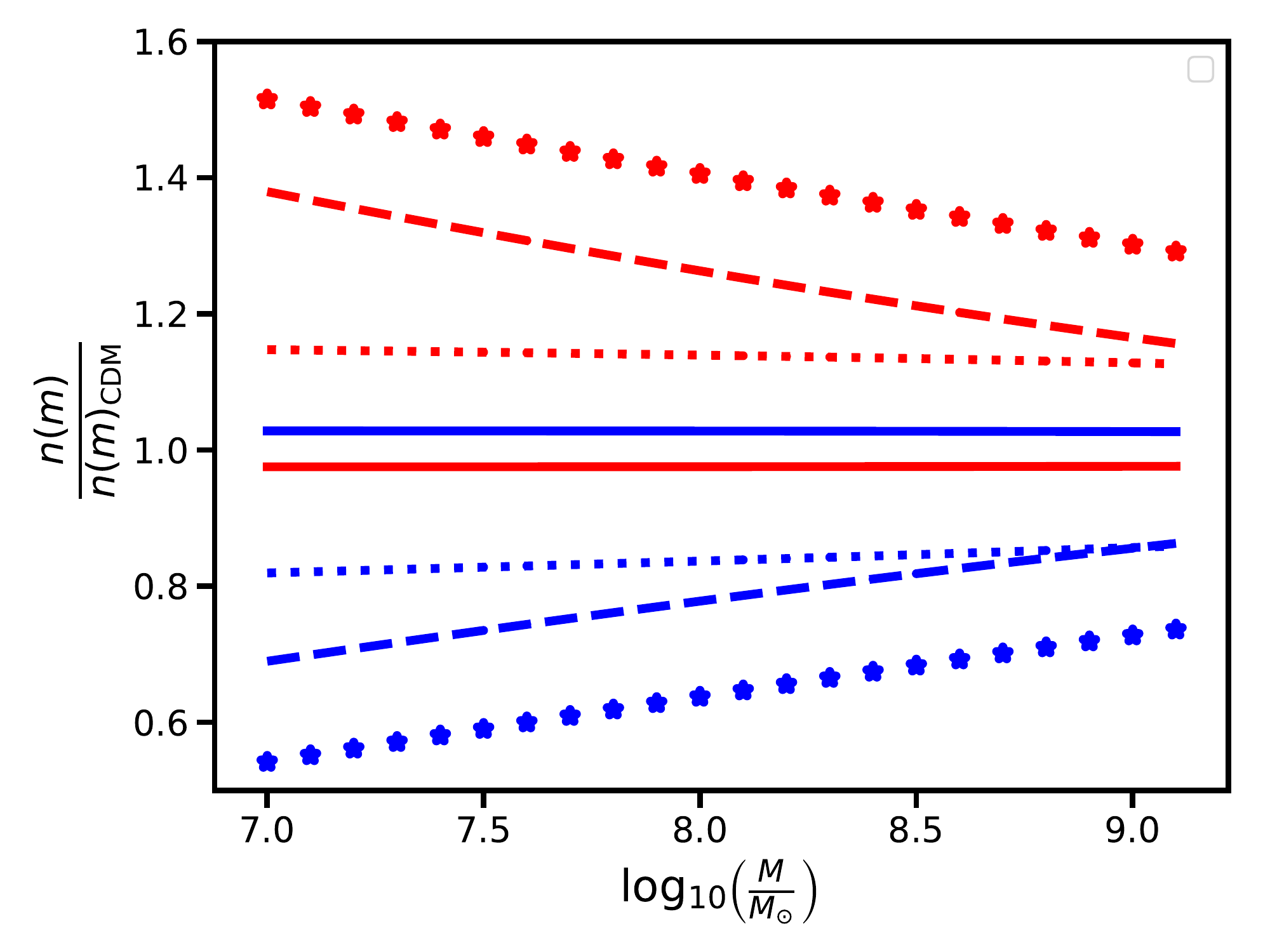}
	\includegraphics[clip,trim=0cm 0cm 0cm
	0cm,width=.45\textwidth,keepaspectratio,angle=0]{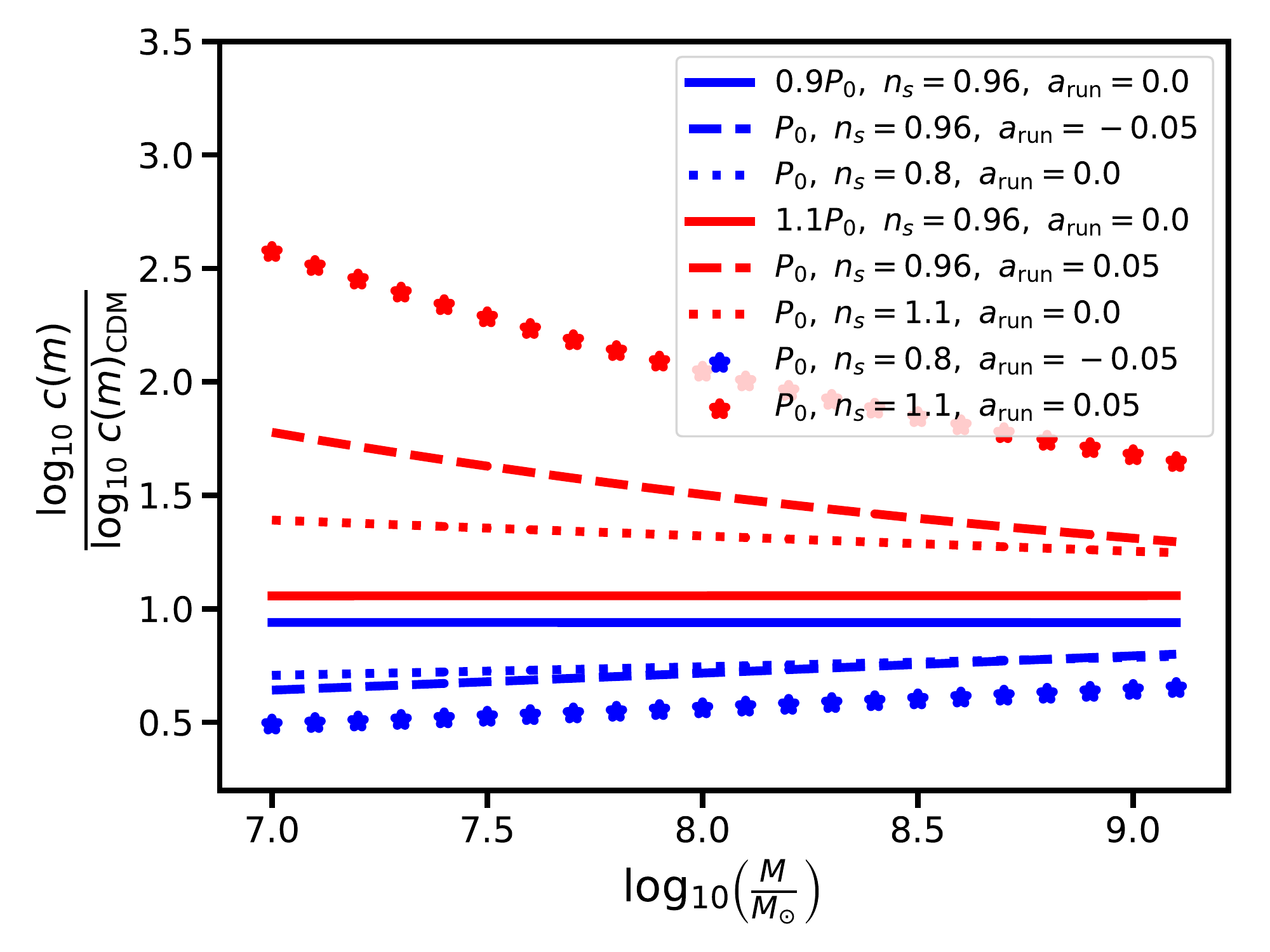}
	{\caption{\label{fig:rescalepk}} The halo mass function (upper panel) and concentration-mass relation (lower panel) relative the $\Lambda$CDM prediction, for models with increased power (red) and decreased power (blue).  The solid lines have increased and decreased power by re-scaling the amplitude of the entire power spectrum by $10\%$. The different line styles show the effect of changing the spectral index $n_s$, and the running parameters $a_{\rm{run}}$.}
\end{figure}

	We anchor the amplitude of the power spectrum on large scales to the amplitude inferred from the CMB \citep{Planck2020params}. We can estimate the uncertainties on the amplitude of the power spectrum inferred from the CMB through the uncertainty on $\sigma_8$, which is approximately $1\%$. Thus, we can investigate to what degree uncertainties in the amplitude of the power spectrum of a few-percent might play in our analysis. We do this by computing the mass function and concentration-mass relation with a power spectrum re-scaled by $10 \%$, and compare the effect with changing and $n_s$ and $a_{\rm{run}}$. For this test, we place the pivot scale at $1 \ \rm{Mpc^{-1}}$. Figure \ref{fig:rescalepk} shows the resulting mass function and concentration-mass relation, plotted relative to the $\Lambda$CDM prediction. 
			
	The effect on the mass function and concentration-mass relation from rescaling the amplitude of the large-scale power spectrum by $\pm 10 \%$ is shown in the figure as a solid line. Scaling the amplitude of the entire power spectrum causes a corresponding scaling of the mass function and concentration-mass relation. For comparison, we also show curves representing the amplitude of the relations for different values of $n_s$ and $a_{\rm{run}}$. Changing the spectral indices has a significantly larger impact on the mass function and concentration-mass relation than a (likely exaggerated) uncertainty of $10 \%$ in the overall amplitude. This makes intuitive sense, as the parameters in our model appear in the exponent of the scale $\frac{k}{k_0}$, and therefore act on long lever arms for the small scales relevant for our analysis. Finally we comment on the perhaps un-intuitive feature of Figure \ref{fig:rescalepk} that increasing the large-scale power $P_0$ slightly lowers the amplitude of the mass function on small scales. This is because increasing power on large scales effectively shifts the mass function laterally to higher masses, leaving less mass on small-scales that can collapse into halos. 
	
	\section{Results with different mass function models}
	\label{app:massfunctions}
	We consider two mass function models in addition to Sheth-Tormen that we use to connect the hyper-parameters sampled in the lensing analysis, $\qsub$, to the parameters describing the power spectrum, $\qp$. First, we consider is the mass function model presented \citet[hereafter Rodriguez-Puebla]{Rodriguez-Puebla++16}. The Rodriguez-Puebla model used the same model for $f\left(\sigma\right)$ as \citet{Tinker++08} 
		\begin{ceqn}
	\begin{equation}
	f\left(\sigma\right)= A \left[\left(\frac{\sigma}{b}\right)^{-a} + 1\right] \exp\left(\frac{-b}{\sigma^2}\right),
	\end{equation}
	\end{ceqn}
	but the parameters $a$ and $b$ are re-calibrated to the Planck 2013 cosmology \citep{Planck2014} down to halo masses of $\sim 10^{10} M_{\odot}$, and redshifts $z = 0-9$. 
	
	In addition to Sheth-Tormen and Rodriguez-Puebla, we repeat our analysis using the model presented by \citet{Bhattacharya++11} (hereafter referred to as the Bhattacharya model). \citet{Bhattacharya++11} added an additional free parameter to make an empirical adjustment to the model for $f\left(\sigma, z\right)$ in the Sheth-Tormen model. The additional term has the form $\left(\frac{\delta_c \sqrt{a}}{\sigma}\right)^{q}$, with the additional parameter $q$ fit simultaneously with $a$ and $p$. \citet{Bhattacharya++11} included the redshift evolution of the mass function is explicitly in their model in order to match simulations on mass scales above $6 \times 10^{11} M_{\odot}$, and $z=0-2$.  
	
\begin{figure}
		\includegraphics[clip,trim=0cm 0.5cm 0cm
		0cm,width=.48\textwidth,keepaspectratio,angle=0]{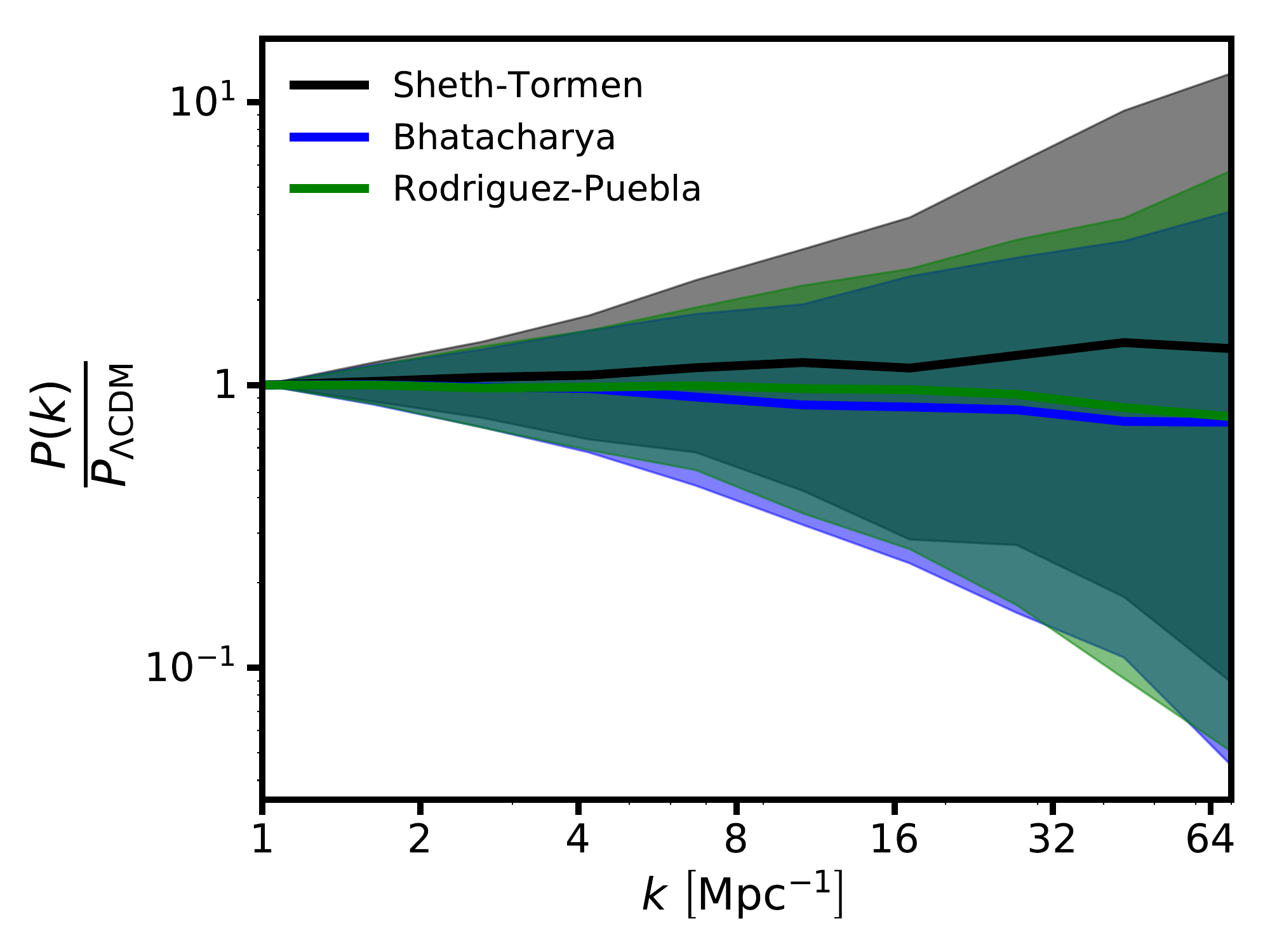}
		{\caption{\label{fig:mfunccomparison}} The reconstructed primordial matter power spectrum using the mass function model presented by \citet{Bhattacharya++11} (blue) and \citet{Rodriguez-Puebla++16} (green) to connect the primordial matter power spectrum to abundance of dark matter halos, in comparison with the Sheth-Tormen model (black).}
	\end{figure}

	Both the Bhattacharya and Rodriguez-Puebla models predict a $\sim 25\%$ higher mass function amplitude $\delta_{\rm{LOS}}$ than the Sheth-Tormen model, and very similar logarithmic slopes $\Delta \alpha$. The higher mass function amplitude introduces more halos into the lens models, which in turn enables in tighter constraints on the concentration-mass relation and the logarithmic slope. For this reason, our constraints on $P\left(k\right)$ become stronger when using the Bhattacharya and Rodriguez-Puebla models. Figure \ref{fig:mfunccomparison} shows the power spectrum inference relative to the $\Lambda$CDM prediction obtained with the three mass function models we implemented. 
	
	\section{Results with different coupling between the subhalo and field halo mass functions}
	\label{app:sigmasubprior}
	The subhalo mass function grows by accretion of field halos, so we expect correlation between the logarithmic slopes of the field halo and subhalo mass function, and their amplitudes. We have explicitly included the coupling between the logarithmic slopes of the mass functions through the parameter $\Delta \alpha$. The amplitude of the subhalo mass function, however, depends on how efficiently elliptical galaxies disrupt halos after accretion, and it is therefore subject to additional theoretical uncertainty. 
	
	We assume that that the amplitude of the subhalo mass function varies proportionally with the amplitude of the field halo mass function. In terms of the parameters implemented in our analysis, an amplitude of the subhalo mass function $\Sigma_{\rm{sub(predicted)}}$ would correspond to the $\Lambda$CDM prediction for the amplitude of the field halo mass function $\delta_{\rm{LOS}} = 1.0$. If the amplitude of the field halo mass function were $50 \%$ higher, we would therefore expect to measure $\delta_{\rm{LOS}} = 1.5$ and $\sigma_{\rm{sub}} = 1.5 \times \Sigma_{\rm{sub(predicted)}}$. As discussed in Section \ref{sec:constrainpk}, we implement this assumption by adding importance weights $w$ to the samples in the posterior distribution $p\left(\qsub | \boldsymbol{D}\right)$. 
	
	Motivated by the results presented by \citet{Nadler++21}, who compared strong lensing inferences of the subhalo mass function with measurements of Milky Way satellites, we assume a value $\Sigma_{\rm{sub(predicted)}} = 0.05 \  \rm{Mpc^{-1}}$ in the analysis presented in Section \ref{sec:constrainpk}. The value for $\Sigma_{\rm{sub(predicted)}}$ used in the main analysis corresponds to a scenario in which disruption by the disk in our galaxy tidally disrupts halos twice as efficiently as the baryonic potential of a massive elliptical galaxy. However, our anlaysis does not depend on this choice, and we can repeat it with other values of $\Sigma_{\rm{sub(predicted)}}$. 
	
\begin{figure}
		\includegraphics[clip,trim=0cm 0.5cm 0cm
		0cm,width=.48\textwidth,keepaspectratio,angle=0]{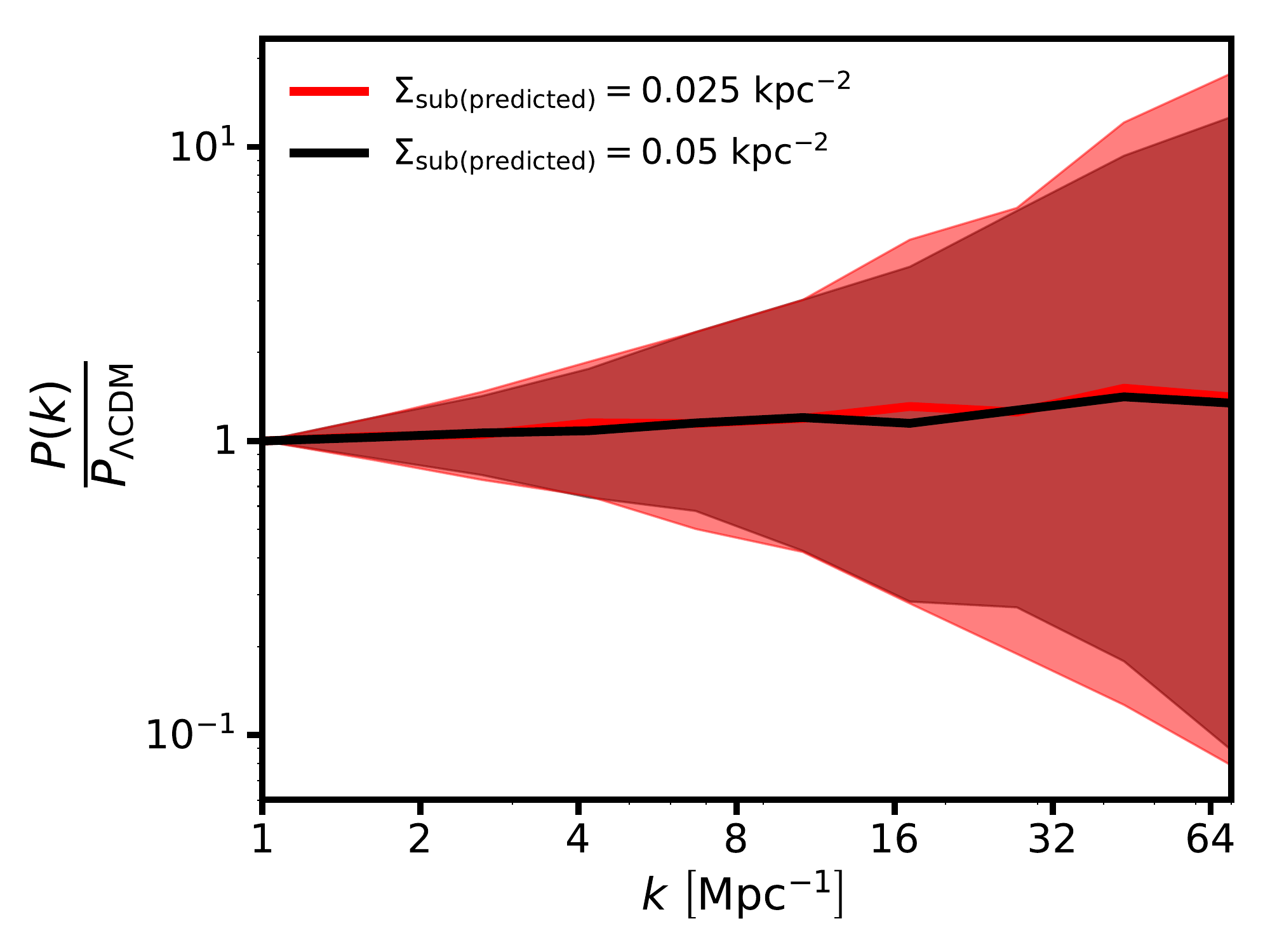}
		{\caption{\label{fig:sigmasubcomparison}} The inference of the primordial matter power spectrum relative to the $\Lambda$CDM prediction assuming a subhalo mass function amplitude of $\Sigma_{\rm{sub(predicted)}} = 0.05 \ \rm{kpc^{-1}}$ (black) and $\Sigma_{\rm{sub(predicted)}} = 0.025 \ \rm{kpc^{-1}}$ (red).}
	\end{figure}

	Figure \ref{fig:sigmasubcomparison} shows the inference of the power spectrum assuming $\Sigma_{\rm{sub(predicted)}} = 0.025 \ \rm{Mpc^{-1}}$. As there are fewer total halos, the data places less stringent constraints on the mass-concentration relation, which in turn leads to slightly weaker constraints on the power spectrum. 

	\section{Tests on simulated data}
	\label{app:simdatatest}
    We have validated the inference method we use to infer hyper-parameters from a sample of quads with simulated datasets \citep{Gilman++18,Gilman++19,Gilman++21}. However, the models we implement for halo mass function and concentration-mass relation in this work differ significantly from those we have previously analyzed. In addition, we have incorporated an additional step in our analysis, wherein we interpret the lensing inference in terms of the primordial matter power spectrum. To verify that the inference method to provides unbiased results at each stage in our analysis, we perform tests with simulated data to test how well we can recover the form of an input power spectrum given set of simulated $\qsub$ parameters.  
	
	To generate the mock data, we first choose a set of power spectrum parameters $\qp$, and use the mapping from $\qp$ to $\qsub$ to obtain a set of parameters $\delta_{\rm{LOS}}$, $\beta$, $\log_{\rm{10}}c_8$, and $\Delta \alpha$. We assume an amplitude for the subhalo mass function $\Sigma_{\rm{sub(predicted)}} = 0.05 \ \rm{kpc^{-2}}$. Then, we generate a simulated set of flux ratios for each lens by creating a realization of dark matter halos from a model specified by these hyper-parameters, and replaced the measured flux ratios with the simulated ones. We then apply the inference method to the mock datasets as described in Section 2 to obtain the posterior $p\left(\qsub | \boldsymbol{D^{\prime}}\right)$, where $\boldsymbol{D^{\prime}}$ represents the simulated data. Then, we apply the methodology described in Section \ref{sec:constrainpk} to infer $\qp$, and recover the power spectrum. 
	
	First, we generate a data from a model specified by $n_s = 0.965$, $a_{\rm{run}} = 0.0$, and $b_{\rm{run}} = 0$. This set of parameters (which corresponds to $\Lambda$CDM) predicts $\left(\delta_{\rm{LOS}}, \beta, \log_{\rm{10}} c_8, \Delta \alpha, \Sigma_{\rm{sub}}\right) = \left(1.0, 0.8, 1.3, 0.0, 0.05 \rm{kpc^{-2}}\right)$. Second, we generate a data from a model with suppressed small-scale power relative to $\Lambda$CDM, with $n_s = 0.6$, $a_{\rm{run}} = -0.1$, and $b_{\rm{run}} = -0.008$, and $\left(\delta_{\rm{LOS}}, \beta, \log_{\rm{10}} c_8, \Delta \alpha, \Sigma_{\rm{sub}}\right) = \left(0.3, 0.5, 0.8, 0.2, 0.015 \rm{kpc^{-2}}\right)$. Third, we consider a model with enhanced small-scale power relative to $\Lambda$CDM, with  $n_s = 1.3$, $a_{\rm{run}} = 0.06$, and $b_{\rm{run}} = 0.016$, and $\left(\delta_{\rm{LOS}}, \beta, \log_{\rm{10}} c_8, \Delta \alpha, \Sigma_{\rm{sub}}\right) = \left(2.0, 8.1, 2.2, -0.1, 0.1 \rm{kpc^{-2}}\right)$. 
	
\begin{figure}
		\includegraphics[clip,trim=0cm 0.5cm 0cm
		0cm,width=.4\textwidth,keepaspectratio,angle=0]{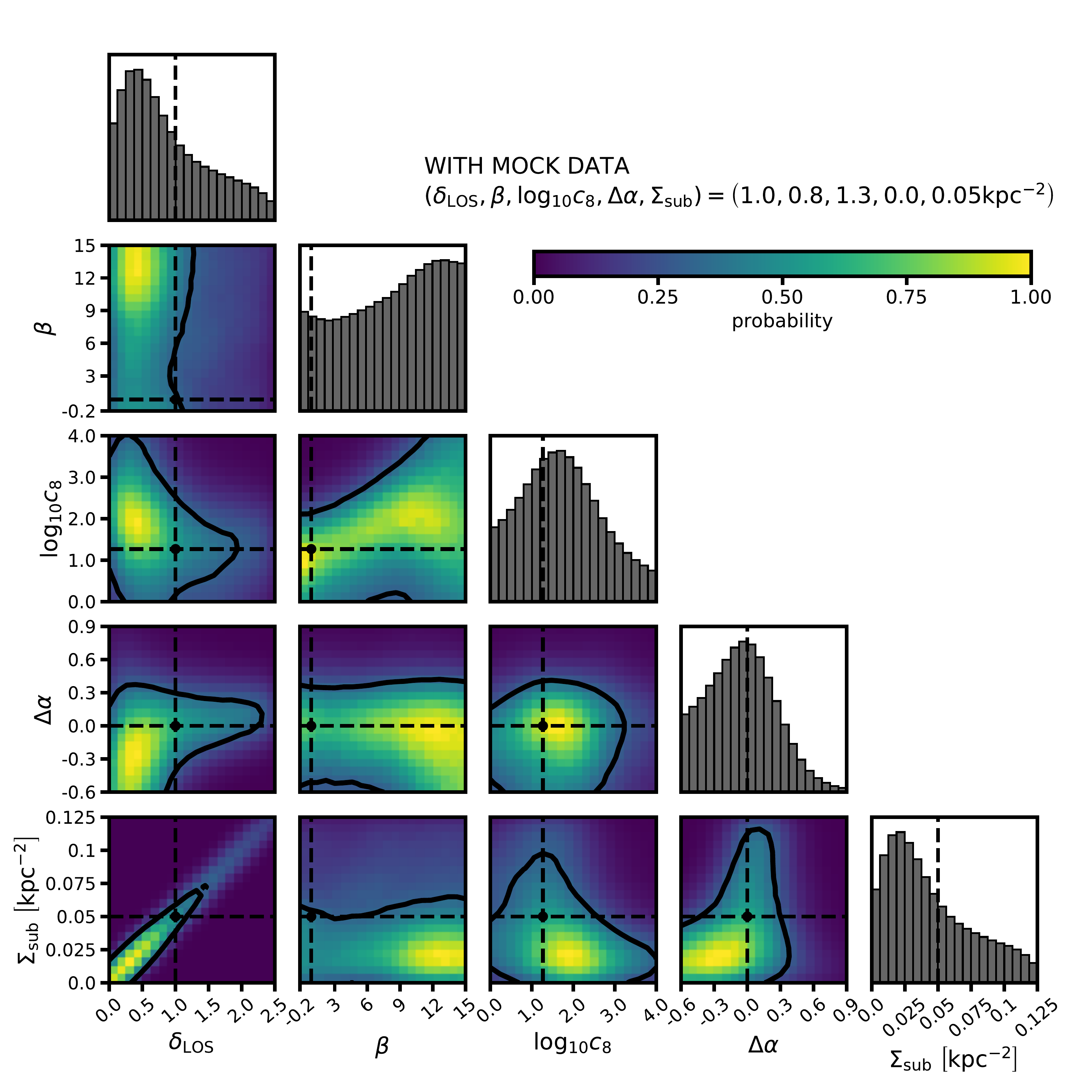}
		\includegraphics[clip,trim=0cm 0.5cm 0cm
		0cm,width=.4\textwidth,keepaspectratio,angle=0]{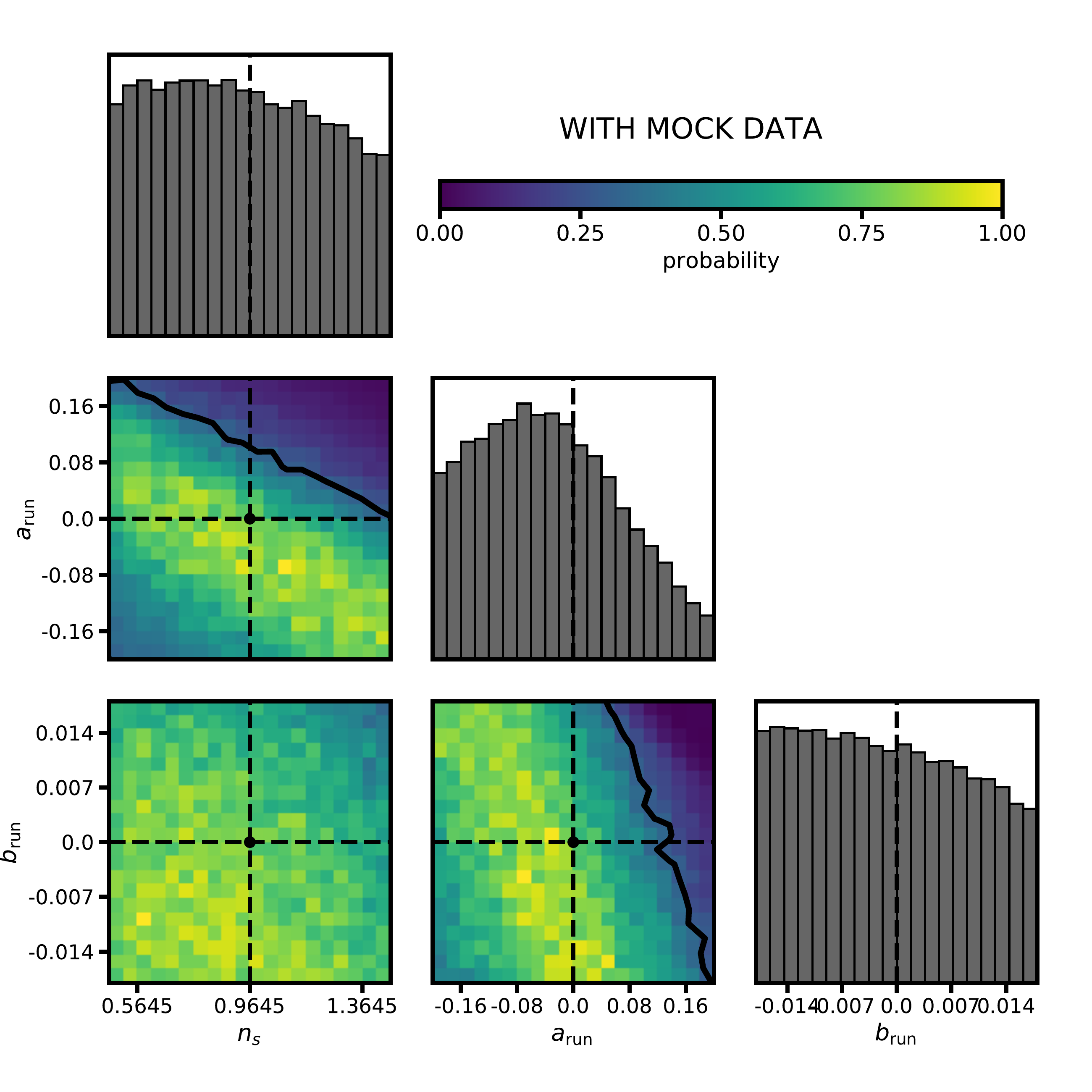}
		\includegraphics[clip,trim=0cm 0.5cm 0cm
		0cm,width=.4\textwidth,keepaspectratio,angle=0]{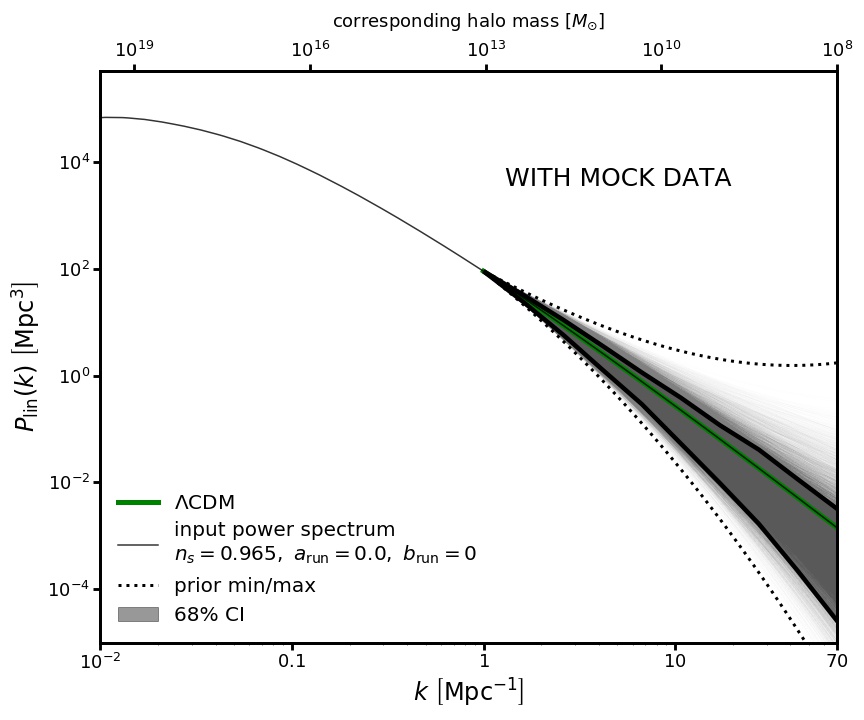}
		\caption{\label{fig:mockinf1} The inferred $\qsub$ parameters (top), the resulting constraint on the $\qp$ parameters (middle), and the reconstruction of the power spectrum (bottom), using simulated data with $n_s = 0.965$, $a_{\rm{run}} = 0.0$, and $b_{\rm{run}} = 0$, and $\left(\delta_{\rm{LOS}}, \beta, \log_{\rm{10}} c_8, \Delta \alpha, \Sigma_{\rm{sub}}\right) = \left(1.0, 0.8, 1.3, 0.0, 0.05 \rm{kpc^{-2}}\right)$. The grey shaded region shows the $68\%$ confidence interval of the inference at each $k$ scale. For reference, the green line shows the $\Lambda$CDM prediction for the power spectrum, and the solid black line enclosed by the gray shaded region shows the ``true" form of the power spectrum, as specified by the ``true" set of of $\qp$ parameters.}
	\end{figure}

\begin{figure}
		\includegraphics[clip,trim=0cm 0.5cm 0cm
		0cm,width=.45\textwidth,keepaspectratio,angle=0]{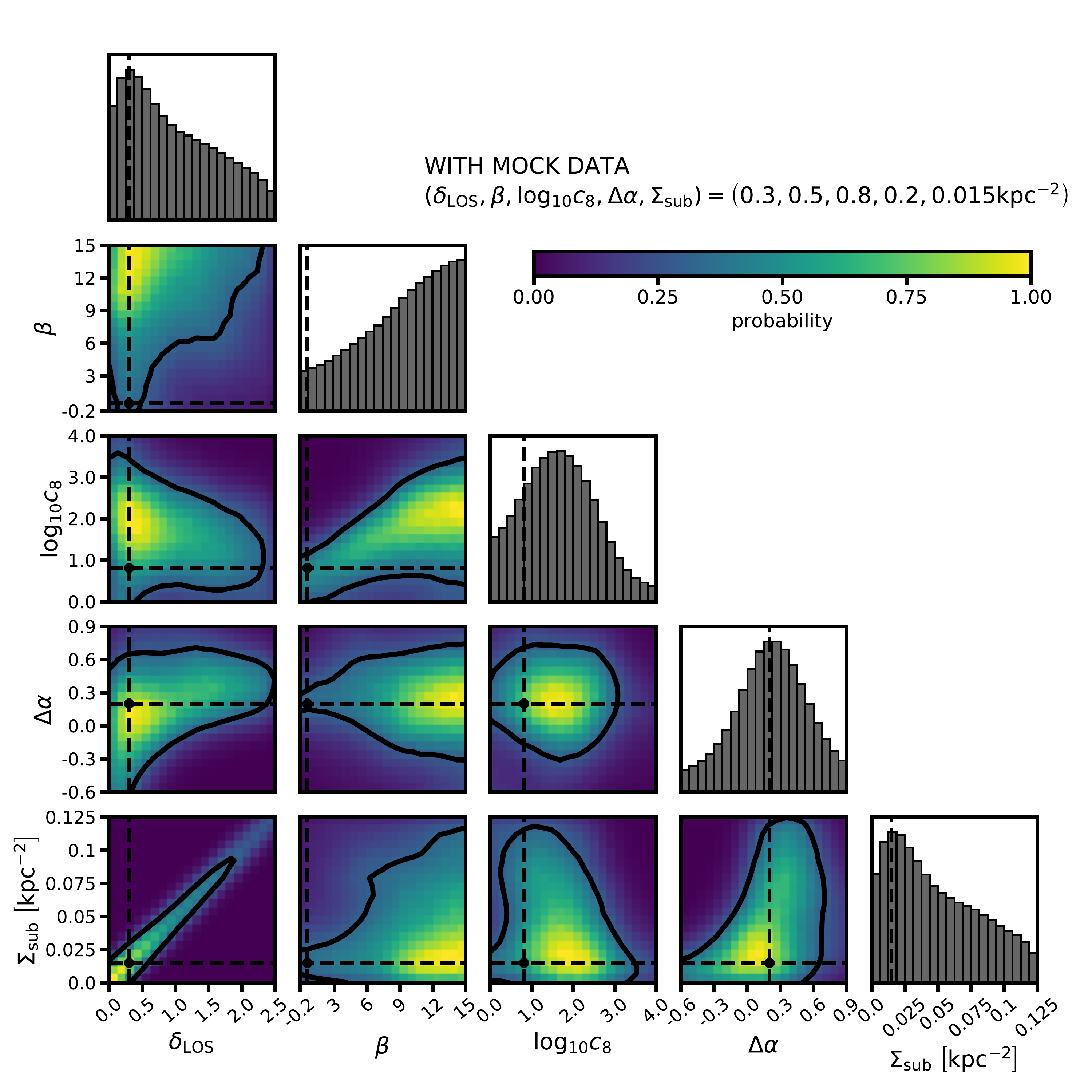}
		\includegraphics[clip,trim=0cm 0.5cm 0cm
		0cm,width=.45\textwidth,keepaspectratio,angle=0]{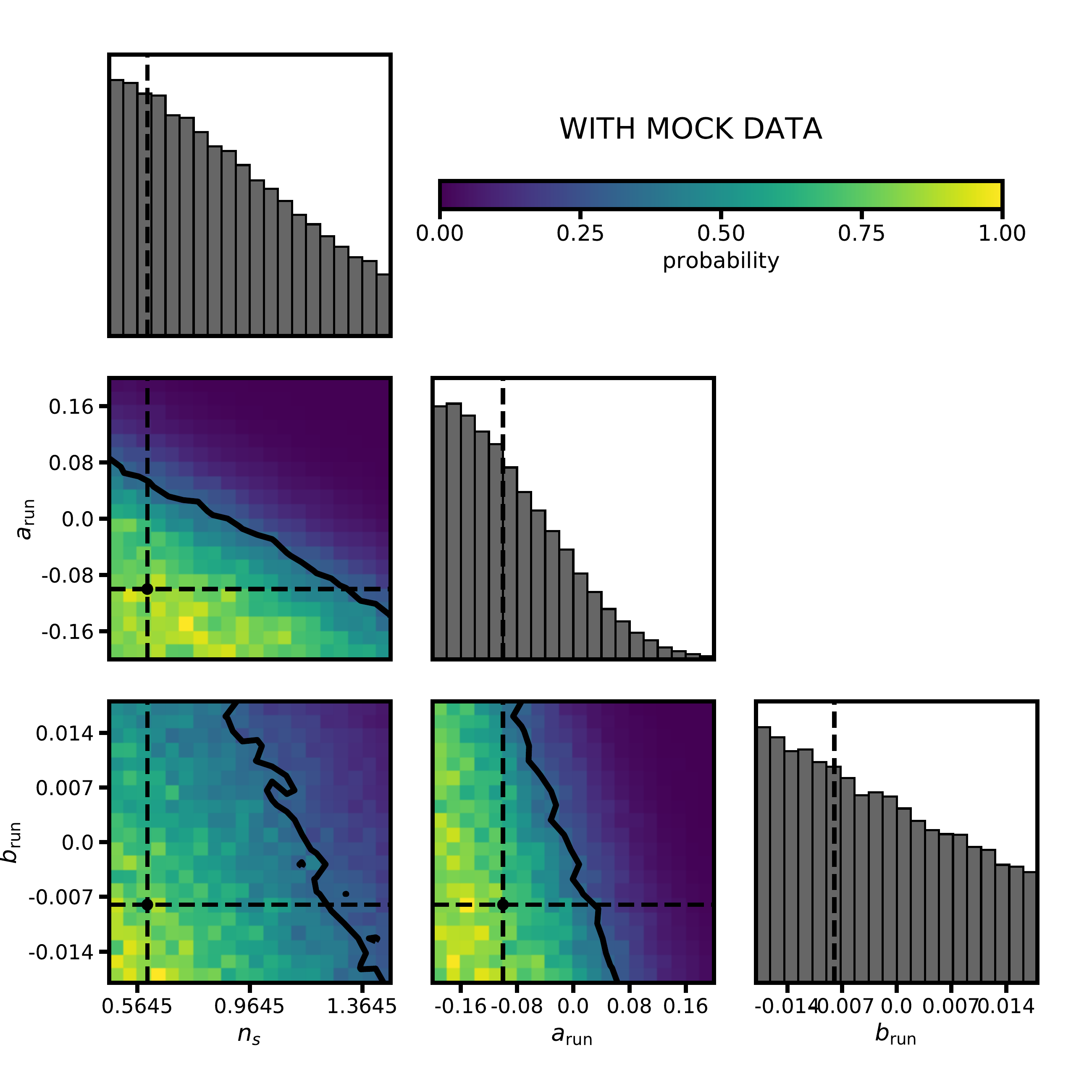}
		\includegraphics[clip,trim=0cm 0.5cm 0cm
		0cm,width=.45\textwidth,keepaspectratio,angle=0]{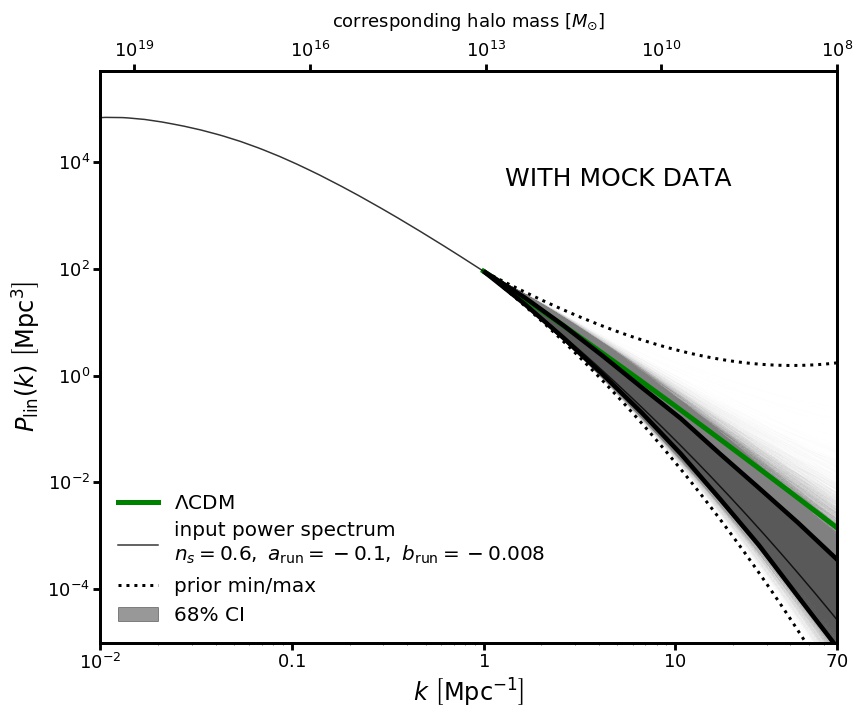}
		\caption{\label{fig:mockinf2} The same as in Figure \ref{fig:mockinf1}, but with simulated data generated from a model with $n_s = 0.6$, $a_{\rm{run}} = -0.1$, and $b_{\rm{run}} = -0.008$, and $\left(\delta_{\rm{LOS}}, \beta, \log_{\rm{10}} c_8, \Delta \alpha, \Sigma_{\rm{sub}}\right) = \left(0.3, 0.5, 0.8, 0.2, 0.015 \rm{kpc^{-2}}\right)$.}
	\end{figure}

\begin{figure}
		\includegraphics[clip,trim=0cm 0.5cm 0cm
		0cm,width=.45\textwidth,keepaspectratio,angle=0]{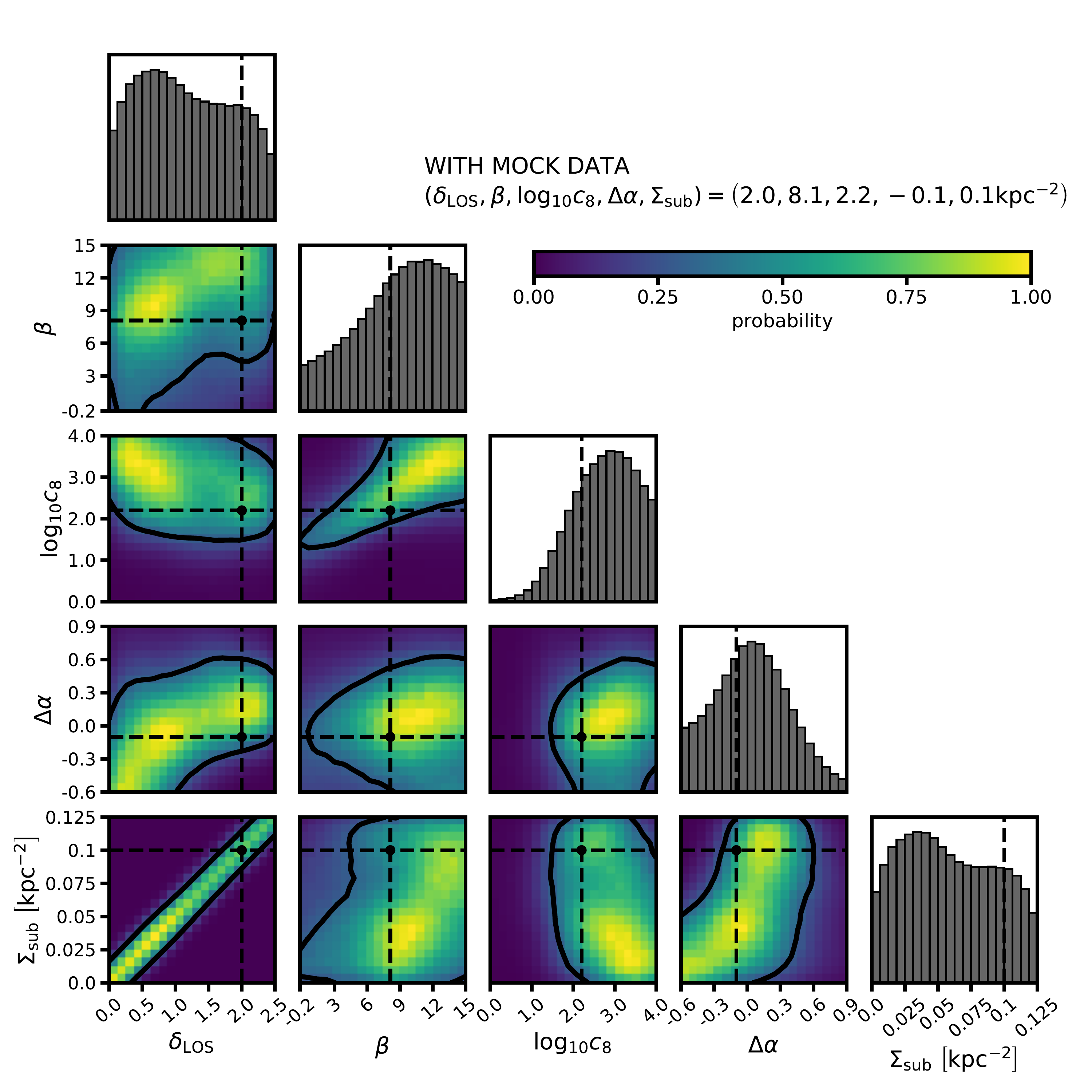}
		\includegraphics[clip,trim=0cm 0.5cm 0cm
		0cm,width=.45\textwidth,keepaspectratio,angle=0]{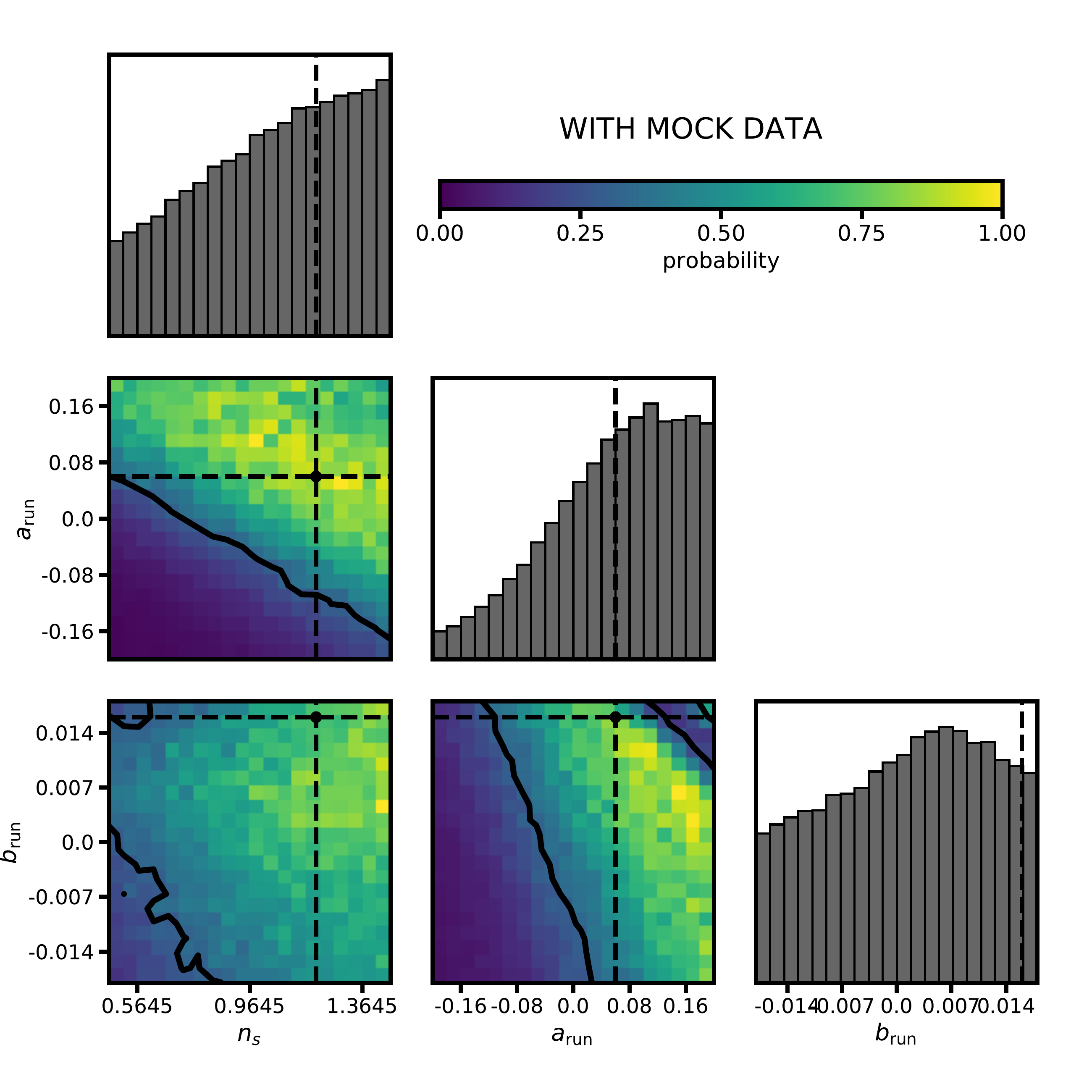}
		\includegraphics[clip,trim=0cm 0.5cm 0cm
		0cm,width=.45\textwidth,keepaspectratio,angle=0]{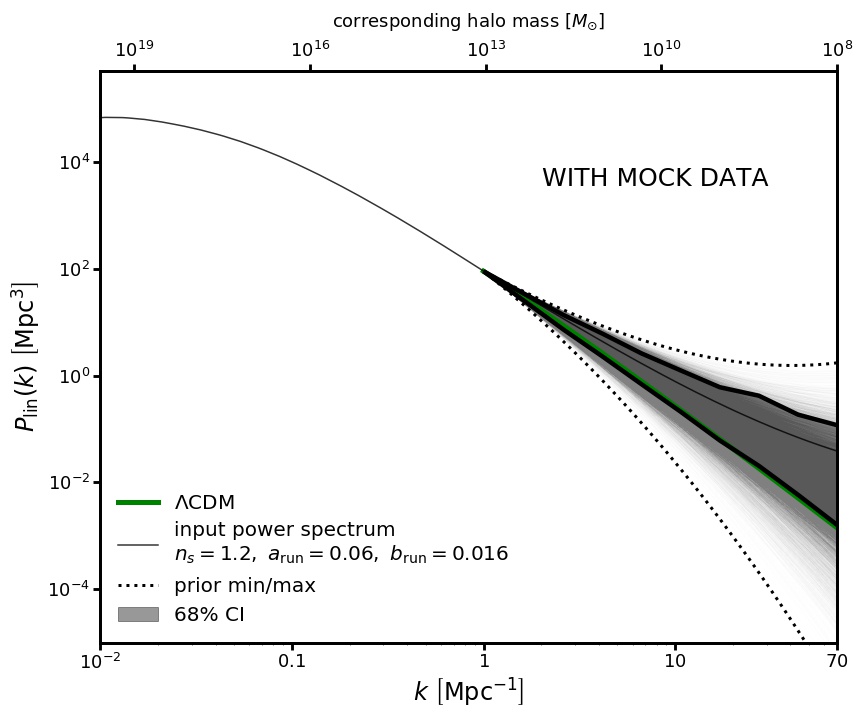}
		\caption{\label{fig:mockinf3} The same as in Figures \ref{fig:mockinf1} and \ref{fig:mockinf2}, but with simulated data generated from a model with $n_s = 1.3$, $a_{\rm{run}} = 0.06$, and $b_{\rm{run}} = 0.016$, and $\left(\delta_{\rm{LOS}}, \beta, \log_{\rm{10}} c_8, \Delta \alpha, \Sigma_{\rm{sub}}\right) = \left(2.0, 8.1, 2.2, -0.1, 0.1 \rm{kpc^{-2}}\right)$.}
	\end{figure}

	The top, middle, and bottom panels of Figures \ref{fig:mockinf1}, \ref{fig:mockinf2}, \ref{fig:mockinf3} show the resulting posterior probability distributions of the $\qsub$ parameters given the mock data, the corresponding constraints on the $\qp$ parameters, and the resulting reconstruction of $P\left(k\right)$, for the three sets of simulated data. In each case, we accurately recover (to within the $68\%$ confidence interval), each set of model parameters. 
	
	\section{Model performance}
	\label{app:goodnessoffit}
	In this appendix, we assess the goodness-of-fit of the model in terms of $\qsub$ parameters we use to interpret the data. As hyper-parameters do not have a one-to-one correspondence to data, diagnostics for goodness-of-fit are typically computed by examining the summary statistics of the data computed under the model, and comparing them with summaries computed from data generated from the model by sampling the prior \citep{Gelman++13}. 
	
\begin{figure}
		\includegraphics[clip,trim=0cm 0.5cm 0cm
		0cm,width=.45\textwidth,keepaspectratio,angle=0]{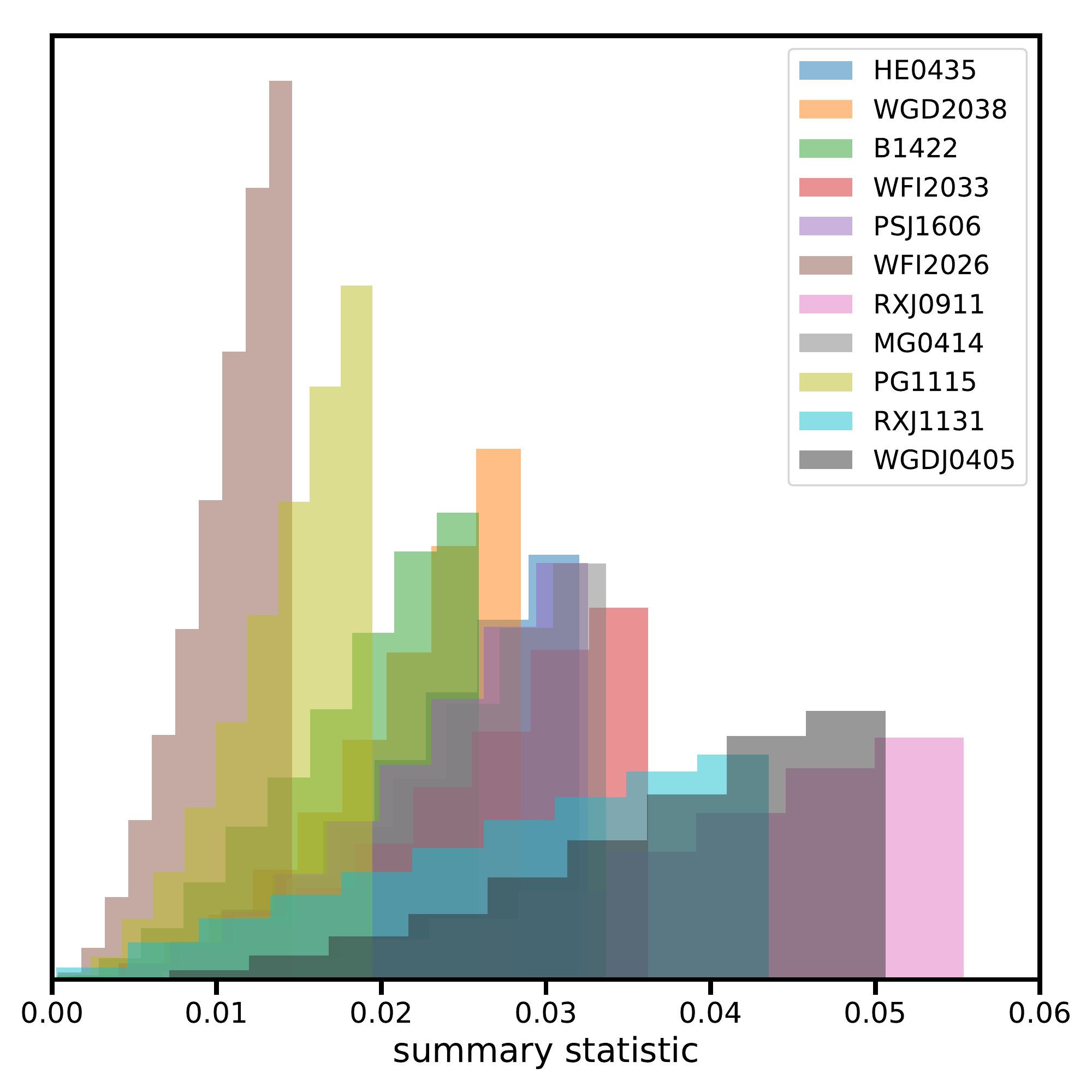}
		\caption{\label{fig:summarystatisticdistributions} Distributions of the accepted summary statistics for each lens in our sample, given the data of the $n$th lens, and the model $\qsub$.}
	\end{figure}

	First, we can check whether or not the model used to analyze the data can reproduce the data. Figure \ref{fig:summarystatisticdistributions} shows the distribution of accepted summary statistics for the eleven lenses in our sample. The hyper-parameters corresponding to these statistics populate the distributions shown in Figures \ref{fig:lensinginferencefull} and \ref{fig:lensinginferenceprior}. The summary statistic is defined as the metric distance between an observed flux ratio and a flux ratio predicted from the model (Equation \ref{eqn:summary}). To give a sense for how well the model can fit the data, we can consider typical sets of flux ratios, and ask what value of $S$ would result from a relative difference of $x_i$ between the model flux ratio, ${f}_{\rm{model(i)}}$, and the observed flux ratio, ${f}_{\rm{obs(i)}}$. For what follows, we will use notation $f_{\rm{model(i)}} = (1-x_i) f_{\rm{obs(i)}}$. As each set of image magnifications or flux ratios usually carries with it a comparable relative uncertainty, to simplify the discussion we will set each uncertainty $x_i$ to the same value, $x$. 
	
	To give a sense for what values of $S$ indicate a good fit to the data, we consider two representative cases. First, if each image has a comparable magnification, their flux ratios will all be approximately the same, i.e. $f_{\rm{obs(1)}} =  f_{\rm{obs(2)}} =  f_{\rm{obs(3)}} = 1$, and $S = \sqrt{3}x \approx 1.7x$. On the other hand, some lenses have a merging pair of highly magnified images, and two de-magnified images or $f_1 = 1.0$, $f_2 = f_3 = 0.3$, giving $S = 1.1 x$. The typical uncertainty for a flux ratio measurement in our sample is 5$\%$, so we can take $x = 0.05$, resulting in $S = 0.09$ and $S = 0.06$ for the two cases. Comparing the distributions of accepted summary statistics, we can see that all accepted samples have $S < 0.06$, indicating that we can generate simulated datasets in the forward model that match the data to a higher degree of precision than the measurement uncertainties\footnote{As a reminder, we handle measurement uncertainties in our analysis by adding the measurement uncertainties to the data simulated in the forward model before computing the summary statistics.}. 
	
	Another way to assess the adequacy of our model in terms of $\qsub$ is to compare the distribution of accepted summary statistics with the distribution of summary statistics we would obtain with a perfect generating model for the data. We can determine how the distribution of summary statistics would appear with a perfect model by applying the same rejection criterion (accept the best 3,500 summaries) to statistics computed with data generated from the model. To facilitate comparison between different distributions of accepted summary statistics, we will label the $i$th distribution of statistics by its median $s_i$, and make histograms of the distributions $p\left({\bf{S}} |{\bf{D}}, M\right)$ and $p\left({\bf{S}} |{\bf{\tilde{D}}}, M\right)$, where ${\boldsymbol{S}}$ is the set of medians $s_i$ , ${\bf{D}}$ is the observed data we used in our analysis, and ${\bf{\tilde{D}}}$ represents simulated data from the model $M$. 
	
\begin{figure}
		\includegraphics[clip,trim=0cm 0.5cm 0cm
		0cm,width=.45\textwidth,keepaspectratio,angle=0]{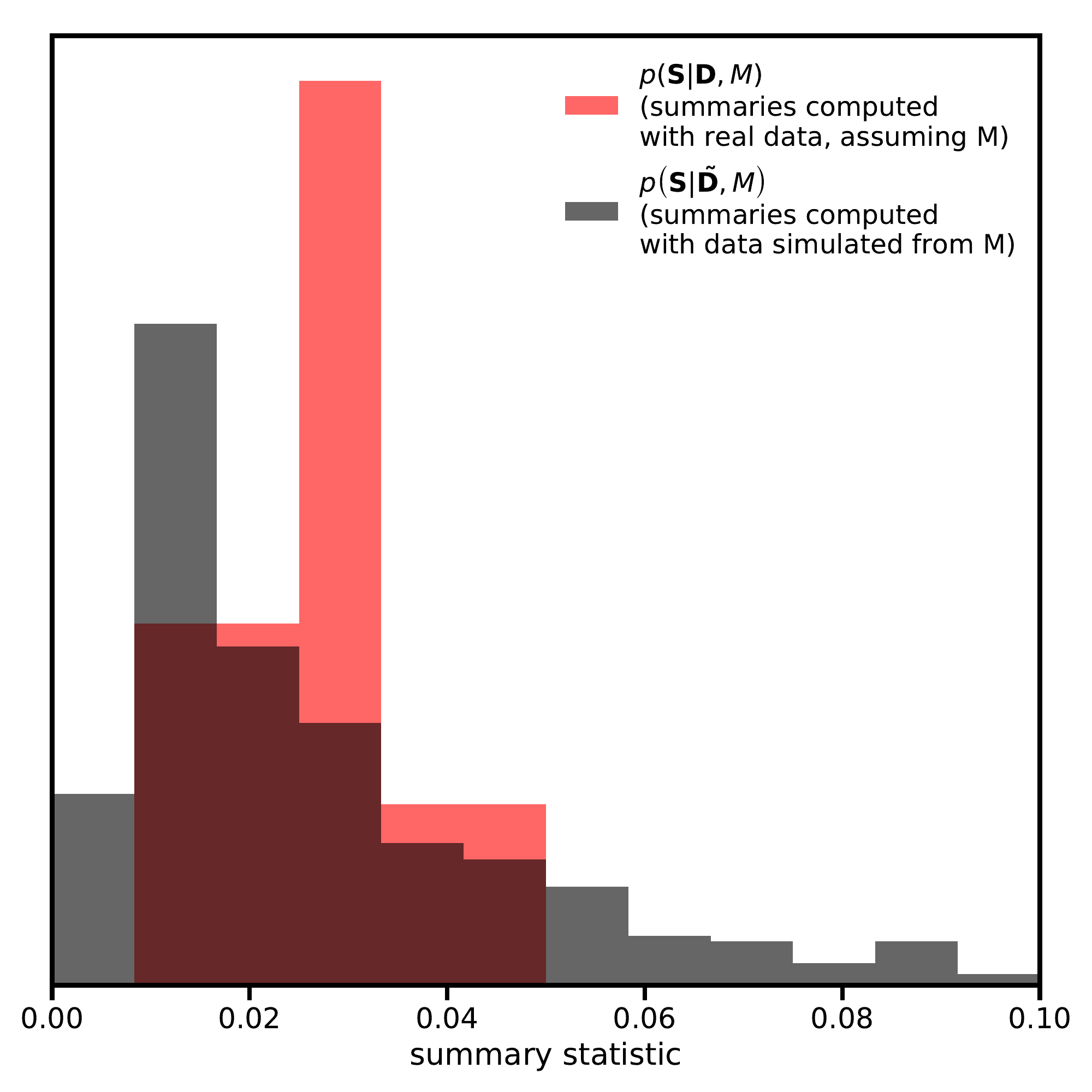}
		\caption{\label{fig:mediandistributions} Probability distributions for the statistic ${\bf{S}}$, defined as the set of medians of the distribution of summary statistics for the $n$th lens. The black distribution shows the distribution of ${\bf{S}}$ for the data in our sample, assuming the model $M$ defined by $\qsub$ parameters. The red distribution shows the distribution of ${\bf{S}}$ obtained by replacing the real data with mock data generated from the model, or how the distribution of ${\bf{S}}$ would appear with a perfect model for the data.}
	\end{figure}

	Figure \ref{fig:mediandistributions} shows, in red, the distribution of ${\bf{S}}$ for the summary statistics computed with respect to the observed data, under the model specified by $\qsub$. The black distribution in Figure \ref{fig:mediandistributions} shows distributions of $\bf{S}$ for simulated data generated from the prior predictive distribution of $\qsub$; in other words, the red distribution is the distribution of summary statistics we would acquire if we had a perfect generating model for the data. Applying a Kolmogorov-Smirnov test to the two distributions, we find a p-value of 0.21. This means that we cannot reject the null hypothesis that each set of summary statistics was generated from the same model. Put differently, with the current sample size of lenses, we cannot distinguish between a perfect model for the data, and the model with which we interpret the data. Thus, the model we use provides a sufficiently good description of the data to interpret them, given the current sample size and uncertainties. 
	
\bsp	
\label{lastpage}
\end{document}